%% file: romulusC_udgs.tex
\documentclass[useAMS,usenatbib,twocolumn]{mnras}
\usepackage[utf8]{inputenc}
\usepackage{natbib,graphics,epsfig}
\usepackage{amsmath}
\usepackage{amssymb}
\usepackage{textcomp}
\usepackage{graphicx}
\usepackage{color}
\usepackage{txfonts}
\usepackage{float,subfig}
\input{rgb}

\definecolor{drkgrn}{rgb}{0,0.5,0}

\definecolor{myblue}{rgb}{0.,0.4,0.7}

\usepackage{graphicx}
\usepackage{color}
\input{rgb}
\usepackage[T1]{fontenc}
\usepackage{aecompl}

\begin{document}

%\title{Off the Beaten Path: A New Approach to Realistically Model The Orbital Decay of Supermassive Black 
%Holes in Galaxy Formation Simulations}
\title[UDGs in RomulusC]{The Formation of Ultra-Diffuse Galaxies in the RomulusC Galaxy Cluster Simulation}
%\author{M.\ Tremmel$^{1}$, F.\  Governato$^{1}$, M.\ Volonteri$^{2,3}$, T.\ R.\ Quinn$^{1}$}
\author[M. Tremmel et al.]{M. ~Tremmel$^{1}$\thanks{email: michael.tremmel@yale.edu},
A.~C.~Wright$^{2}$,
A.~M.~Brooks$^{2}$,
F.~Munshi$^{3}$,
D.~Nagai$^{4}$,
T.~R.~Quinn$^{5}$
\\
$^1$Yale Center for Astronomy \& Astrophysics, Physics Department, P.O. Box 208120, New Haven, CT 06520, USA\\
$^2$Department of Physics \& Astronomy, Rutgers, The State University of New Jersey, 136 Frelinghuysen Road, Piscataway, NJ 08854, USA\\
$^3$Department of Physics \& Astronomy, University of Oklahoma 440 W. Brooks St., Norman, OK 73019\\
$^4$ Physics Department, Yale University, P.O. Box 208120, New Haven, CT 06520, USA\\
$^5$ Astronomy Department, University of Washington, Box 351580, Seattle, WA, 98195-1580}

%\email{mjt29@uw.edu}

\pagerange{\pageref{firstpage}--\pageref{lastpage}} \pubyear{2019}

\maketitle

\label{firstpage}

\begin{abstract}

We study the origins of 122 ultra-diffuse galaxies (UDGs) in the {\sc RomulusC} zoom-in cosmological simulation of a galaxy cluster (M$_{200} = 1.15\times10^{14}$ M$_{\odot}$), one of the only such simulations capable of resolving the evolution and structure of dwarf galaxies (M$_{\star} < 10^9$ M$_{\odot}$). We find broad agreement with observed cluster UDGs and predict that they are not separate from the overall cluster dwarf population. UDGs in cluster environments form primarily from dwarf galaxies that experienced early cluster in-fall and subsequent quenching due to ram pressure. The ensuing dimming of these dwarf galaxies due to passive stellar evolution results in a population of very low surface brightness galaxies that are otherwise typical dwarfs. UDGs and non-UDGs alike are affected by tidal interactions with the cluster potential. Tidal stripping of dark matter, as well as mass loss from stellar evolution, results in the adiabatic expansion of stars, particularly in the lowest mass dwarfs. High mass dwarf galaxies show signatures of tidal heating while low mass dwarfs that survive until $z=0$ typically have not experienced such impulsive interactions. There is little difference between UDGs and non-UDGs in terms of their dark matter halos, stellar morphology, colors, and location within the cluster. In most respects cluster UDG and non-UDGs alike are similar to isolated dwarf galaxies, except for the fact that they are typically quenched.\\
\\
\\
\\
%although \textbf{cluster dwarfs lack the population of the smallest dwarf galaxies found in our sample of simulated isolated galaxies. We attribute this to a lack of galaxies formining with low angular momentum gas in the cluster environment, which in the field is responsible for these more compact dwarfs.}

%they are typically larger in size. This is due, in part, to the fact that cluster dwarf galaxies grow from higher angular momentum gas compared to isolated dwarf galaxies.

%we find evidence that cluster dwarf galaxies grow from higher angular momentum gas, which contributes to their larger effective radii compared to isolated dwarf galaxies. %We find that central surface brightness is the main limiting factor for UDG classification at high mass while effective radius determines UDG classification at low mass. Cluster UDGs and non-UDGs form stars from higher angular momentum gas compared to isolated dwarfs from the {\sc Romulus25} simulation, contributing to a lack of compact dwarf galaxies relative to the field. \\

\end{abstract}


\section{Introduction}

%\mjt{Useful citations/suggestions from emails with Pavel Mancera Piña: PMP 2019 (UDGs look like extension of dwarf pop), mention PMP 2018 in conclusions, mention agreement on lack of high n systems with observations from Román\&Trujillo 17a, Mancera Piña+19, Venhola+17, Cohen+18. Mention lack of UDGs in central 0.1 R200 in agreement with observations (PMP 2018). This last one comes with major caveats for sims so don't make a big deal of it.}

%\mjt{useful citations/suggestions from email with Chris Conselice: Concelice 2017 for summary of previous works looking at diffuse galaxies from 80s, Concelice 2003 as well. Broaden intro to include these earlier contributions. Also cite Wttmann 2017. Unlike Chris I think that these earlier works show that there is a population well outside of the "typical" dwarfs (Impey 1988 for example).}

The population of dwarf galaxies in cluster environments has been known to include uniquely low surface brightness objects compared to low mass galaxies in the local Universe \citep[e.g.][]{sandage84, impey88, conselice03}. While their properties point to a unique formation mechanism compared to local dwarfs, a significant limitation in understanding the evolution of dwarf galaxies in such dense environments has been their elusiveness to detection. Recent advancements in detecting low surface brightness structures, led by observatories such as Dragonfly and Subaru, have revealed that this population of very diffuse, low luminosity galaxies are numerous. These galaxies, recently categorized as `ultra-diffuse' galaxies (UDGs), are characterized by their large effective radii and very low surface brightness. They are found primarily in cluster and group environments \citep{PvD15, vdBurg16, romantrujillo17, wittmann17, greco18, mancerapina18}, with significant populations detected in the Coma, Virgo, and Fornax clusters \citep{mihos15, koda15, munos15,mowla17}, as well as lower mass galaxy groups and even Milky Way-mass halos \citep{romantrujillo17b, mancerapina18}. In some cases, these UDGs have very unique properties, such as large numbers of globular clusters for their stellar mass \citep{PvD17} or dynamical masses that are dominated by stars out to large radii \citep{PvD18, PvD19}. The question of whether such galaxies can form in $\Lambda$CDM and what processes are required to produce them may have important implications to galaxy formation theory \citep[e.g.][]{dicintio17b}, and potentially even the nature of dark matter \citep[e.g.][]{wasserman19}.

Idealised simulations \citep[e.g.][]{yozin15, chowdhury19}, analytic and semi-analytic models \citep[e.g.][]{amoriscoloeb16, ogiya18, rong19, carleton19}, and cosmological simulations \citep[e.g.][]{dicintio17b, chan18, jiang19, liao19, martin19} have all been utilized in attempts to understand the origin of UDGs and their properties, often coming to different conclusions. A dynamical origin to UDGs has been proposed by several groups, which would have UDGs forming from tidal heating and/or stripping due to interactions with the galaxy cluster environment \citep[e.g.][]{ogiya18, jiang19, liao19}. However, recent observations of UDGs have shown a lack of tidal features and estimate their current tidal radii to be larger than 7 kpc \citep{mowla17}, though such observations do not rule out tidal heating which can puff up a galaxy without the features associated with tidal stripping. Another possibility is that UDGs reside in halos with particularly high spin \citep{amoriscoloeb16}, and indeed cosmological simulations of isolated galaxies have shown that angular momentum is an important component to forming low surface brightness galaxies \citep{dicintio19}. Cosmological simulations have also shown that bursty feedback from supernovae (SN), which lead to the formation of cored dark matter profiles \citep{G10,G12,PG12,PG13,dicintio14}, also lead to extended, diffuse stellar distributions and galaxies with UDG-like properties \citep{dicintio17b}. In this scenario, UDG formation does not require a dense environment, although the environment is likely important for reproducing other observed properties such as colors, gas content (or lack thereof), and stellar ages \citep{chan18}.

Self-consistently modeling UDG formation in dense environments is challenging for cosmological simulations because it requires simulating a massive halo, central galaxy, and circum-galactic/intracluster gas with enough resolution elements to study the internal structure of low mass galaxies. In general, smaller halos have been more accessible, such as Milky Way-mass halos \citep{liao19} or low-mass groups \citep{jiang19}. These simulations have found that UDGS can form either as a result of tidal interactions with their host halo and ram pressure stripping, or develop UDG-like properties while still relatively isolated due to bursty star formation and supernovae feedback, as found in \citet{dicintio17b}, or by residing in high-spin halos.

%\citet{liao19} simulate Milky Way-mass halos to reproduce observed abundances of \textbf{UDGs} in similar mass systems and find a combination of formation scenarios: tidal interactions can puff up an initially compact galaxy, while UDGs can form in isolation prior to in-fall. %The latter only occurs for galaxies within high spin dark matter halos.  
%\citet{jiang19} present results from a more massive galaxy group at significantly lower resolution \textbf{and, similarly, find two formation channels: UDGs that are classified as such prior to in-fall and UDGs that form after in-fall and the quenching of star formation. The origin of galaxies that are UDG-like prior to in-fall into a more massive halo has been found

%While \citet{liao19} attribute UDG formation prior to in-fall to the galaxies residing in high-spin dark matter halos, \citet{jiang19} find, similar to \citet{dicintio17b}, that UDGs that form in isolation do so because of bursty star formation and SN feedback.} \dn{This sentence is a bit hard to follow. It'd be easier to read if you could avoid saying who did what; e.g., "XX attributed", "YY find", and "similar to ZZ" etc. Just explain key points and include references at the end of each argument.} 
However, such experiments using cosmological simulations of cluster environments have so far been limited either to low resolution simulations able to resolve only the most massive dwarf galaxies with $\sim 1$ kpc resolution and a few hundred particles \citep{martin19} or semi-analytic galaxy evolution models applied to dark matter only simulations \citep{carleton19}. Cosmological hydrodynamic simulations that resolve the internal structure of low mass dwarf galaxies in clusters are required to better understand UDG formation.

The {\sc Romulus} simulations \citep{tremmel17} consist of a 25 Mpc-per-side uniform volume simulation ({\sc Romulus25}) as well as the {\sc RomulusC} cosmological galaxy cluster simulation \citep{tremmel19}. Both are run with state-of-the art sub-grid physics and resolution. {\sc RomulusC} is one of the highest resolution cosmological simulations ever completed of a galaxy cluster, comparable only to the recent TNG50 simulation, which also consists of a single cluster of similar mass \citep{nelson19, pillepich19}. It therefore represents an important opportunity to, for the first time, self-consistently study the evolution of dwarf galaxies (M$_{\star} < 10^9$ M$_{\odot}$) in galaxy cluster environments. Galaxies in {\sc RomulusC} with stellar and virial masses as small as $10^{7}$ M$_{\odot}$ and $3\times10^{9}$ M$_{\odot}$ respectively are resolved with $>200$ star particles and more than $\sim10^4$ dark matter particles. We are also able to use {\sc Romulus25} to self-consistently compare properties of cluster dwarf galaxies to hundreds of isolated systems simulated with the same physics and resolution.  In \S2 we provide an overview of the simulation properties and our criteria for selecting galaxies as UDGs. In \S3 we examine the $z=0$ properties of UDGs predicted by {\sc RomulusC} and compare with both observations and the overall cluster and isolated populations of dwarf galaxies. In \S4 we study the evolution of dwarf galaxies in the cluster environment and examine its role in UDG formation.  We discuss the implications and limitations of our results in \S5 and we summarize the results in \S6.

\section{The RomulusC Simulation}

{\sc RomulusC} \citep{tremmel19} is a cosmological zoom-in simulation of a small galaxy cluster with $z=0$ total virial mass of $1.5 \times 10^{14}$ M$_{\odot}$, R$_{200} = 1033$ kpc, and M$_{200} = 1.15\times10^{14}$ M$_{\odot}$\footnote{R$_{\Delta}$ is defined as the radius enclosing a mean density of $\Delta\times\rho_{crit}$, where $\rho_{crit}$ is the critical density at $z = 0$. M$_{\Delta}$ is the mass enclosed within R$_{\Delta}$.}. The initial conditions for {\sc RomulusC} were extracted using the `zoom-in' volume re-normalization technique of \citet{katz93} to define a Lagrangian region associated with the most massive $z=0$ halo of a 50 Mpc-per-side uniform volume dark matter-only simulation. The resulting `zoom-in region' was re-simulated at higher resolution with full hydrodynamic treatment using the new Tree+SPH code {\sc ChaNGa} \citep{changa15} while maintaining the gravitational influence of the evolving large-scale structure of dark matter, which is modeled with much coarser resolution.

{\sc ChaNGa} includes standard physics modules previously used in { \sc GASOLINE} \citep{wadsley04,wadsley08,wadsley17} such as a cosmic UV background \citep{HM12} including self-shielding \citep{pontzen08}, star formation, `blastwave' SN feedback \citep{Stinson06}, and low temperature metal cooling. {\sc ChaNGa} implements an updated SPH routine that uses a geometric mean density in the SPH force expression, allowing for the accurate simulation of shearing flows with Kelvin-Helmholtz instabilities \citep{ritchie01,changa15,governato15}. {\sc RomulusC} includes an updated implementation of turbulent diffusion \citep{wadsley17}, which results in a realistic intracluster medium (ICM) \citep{wadsley08, tremmel19, butsky19} and metal distributions within galaxies \citep{shen10}. Finally, 
a time-dependent artificial viscosity and an on-the-fly time-step adjustment \citep{saitoh09} system allow for more realistic treatment of weak and strong shocks \citep{wadsley17}.

{\sc RomulusC} is run with the same hydrodynamics, sub-grid physics, resolution, and cosmology as the {\sc Romulus25} simulation \citep{tremmel17}. The cosmology is $\Lambda$CDM with cosmological parameter values following the recent results from Planck \citep[$\Omega_0=0.3086$, $\Lambda=0.6914$, h$=0.6777$, $\sigma_8=0.8288$;][]{planck16}. The simulation has a Plummer equivalent force softening of $250$ pc (a spline softening of 350 pc is used, which converges to a Newtonian force at 700 pc). Unlike many similar cosmological runs, the dark matter particles are {\it oversampled} relative to gas particles, such that the simulation is run with initially $3.375$ times more dark matter particles than gas. The result is a dark matter particle mass of $3.39 \times 10^5$M$_{\odot}$ and gas particle mass of $2.12 \times 10^5$M$_{\odot}$. This will decrease numerical effects resulting from two-body relaxation and energy equi-partition, which occur when particles have significantly different masses, both of which can affect the structure of simulated galaxies \citep{ludlow19}. This increased dark matter resolution also allows for the ability to track the dynamics of supermassive black holes within galaxies \citep{tremmel15}. {\sc Romulus25} has been shown to reproduce important galaxy and supermassive black hole scaling relations \citep{tremmel17, ricarte19}.

\subsection{Sub-grid physics}

\begin{figure*}
\centering
\includegraphics[trim=10mm 10mm 0mm 10mm, clip, width=175mm]{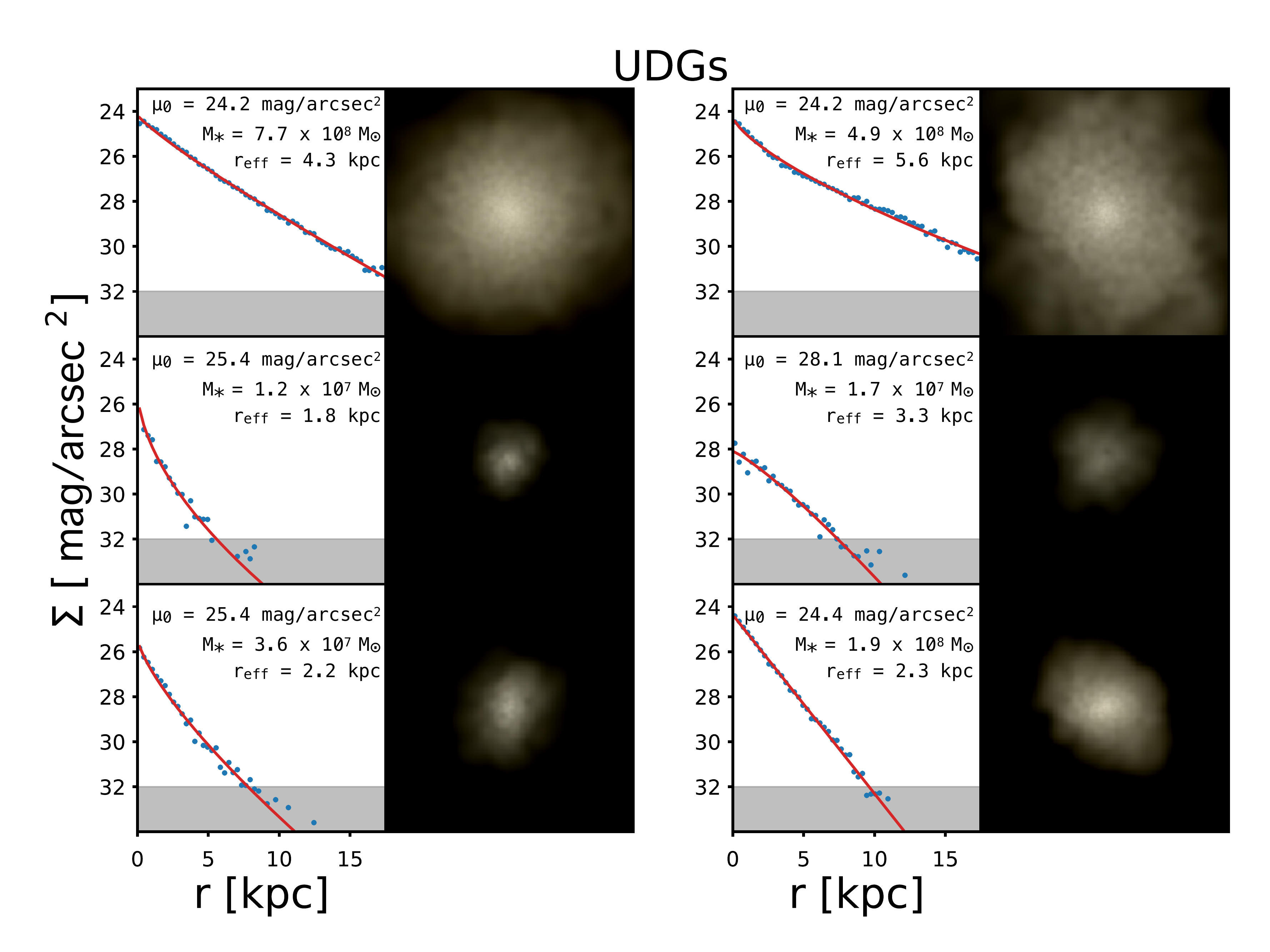}
\includegraphics[trim=10mm 80mm 0mm 10mm, clip, width=175mm]{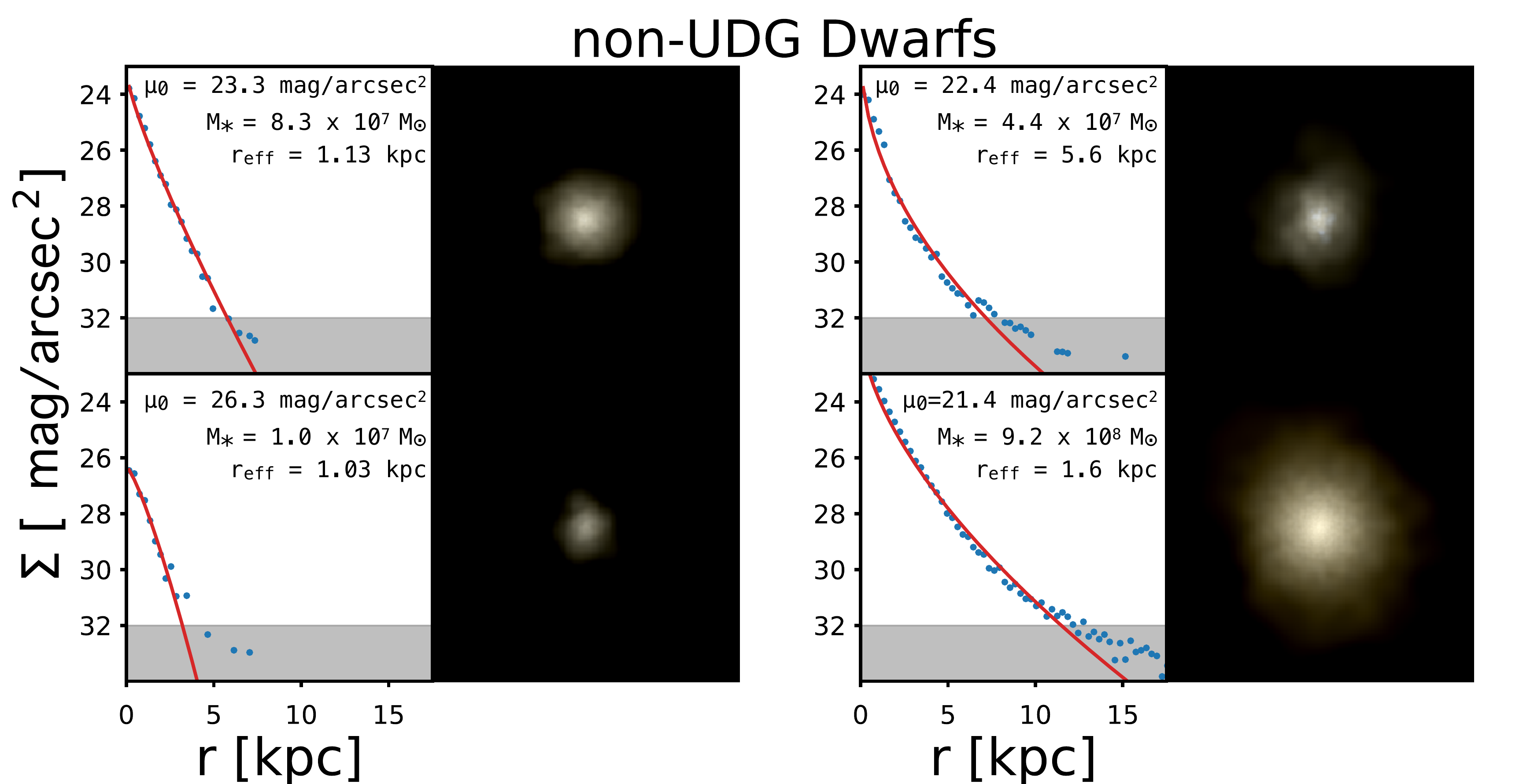}
\caption{{\sc Example UDGs and non-UDGs selected from RomulusC}. (Top) Surface brightness profiles, corresponding Sersic fits, and UVJ images of six example UDGs extracted from {\sc RomulusC} and (Bottom) the same information for four non-UDG dwarf galaxies (M$_{\star} < 10^9$ M$_{\odot}$) from the same simulation. Both images and surface brightness profiles are calculated assuming a face-on view of the galaxies (i.e. a line of sight looking down the angular momentum vector of the galaxy). The UDG examples include the ones with the most stellar mass (top left), largest effective radius (top right), largest Sersic index (1.74, middle left), lowest central surface brightness (middle right), farthest $z=0$ distance to cluster center (1489 kpc, bottom left), and closest distance to cluster center (153 kpc, bottom right). For the non-UDG dwarfs, the top examples fail both size and surface brightness criteria for being a UDG. The bottom left is the non-UDG dwarf galaxy with the smallest effective radius and on the bottom right is the non-UDG dwarf with the highest central surface brightness.}
\label{rogue_gallery}
\end{figure*}

\subsubsection{Star formation and gas cooling}

Star formation and associated feedback from supernovae are crucial processes that require sub-grid models in cosmological simulations like {\sc RomulusC}.  As in previous work \citep{Stinson06} for runs at this resolution, star formation (SF) is regulated with parameters that encode star formation efficiency in dense gas, the coupling of SN energy to the ISM, and the physical conditions required for star formation:

\begin{enumerate}
\setlength\itemsep{1em}
\item The normalization of the SF efficiency, c$_{\star} = 0.15$, and formation timescale, $\Delta t = 10^6$ yr, are both used to calculate the probability $p$ of creating a star particle from a gas particle that has a dynamical time $t_{\mathrm{dyn}}$

\begin{equation}
p =\frac{m_{gas}}{m_{star}}(1 - e^{-c_{\star} \Delta t /t_{\mathrm{dyn}}}).
\end{equation}
 
\item The fraction of SN energy coupled to the ISM, $\epsilon_{\mathrm{SN}} = 0.75$.

\item The minimum density, n$_{\star} = 0.2$ cm$^{-3}$, and maximum temperature, T$_\star = 10^4$ K, thresholds beyond which cold gas is allowed to form stars.
 
 \end{enumerate}

Star particles form with a mass of $6\times10^4$ M$_{\odot}$, or 30\% the initial gas particle mass. We assume a Kroupa IMF \citep{kroupa2001} with associated metal yields and SN rates. Feedback from SN uses the `blastwave' implementation \citep{Stinson06}, with thermal energy injection and a cooling shutoff period approximating the `blastwave' phase of SN ejecta when cooling is inefficient. Passive evolution of stellar populations are accounted for directly in the simulation via mass loss due to stellar winds and supernovae ejecta. In post-processing we also account for the gradual dimming of older stellar populations due to the loss of more massive, brighter stars which is included in the population synthesis models used to generate our stellar emission tables.

Gas cooling at low temperatures is regulated by metal abundance as in \citet{eris11} as well as SPH hydrodynamics that includes both thermal and metal diffusion as described in \citet{shen10} and \citet{governato15}. An important limitation of the {\sc Romulus} simulations is the lack of cooling from high temperature metal lines. For low mass galaxies such as those we focus on in this work, metal line cooling is sub-dominant due to both low metallicity and low virial temperature of the dark matter halos hosting the galaxies. However, massive halos like galaxy clusters will be affected by such choices. The choice to not include metal-line cooling is outlined in more detail in \citet{tremmel19} and is due to the fact that, although {\sc RomulusC} has the highest resolution for a cosmological simulation performed at this mass scale, it is still not enough to self-consistently simulate the multiphase ISM and, in particular, molecular cooling. \citet{christensen14b} found that the inclusion of metal cooling without molecular hydrogen physics and more detailed models of star formation resulted in over-cooling in galaxies. While some simulations that are run at similar resolution to {\sc Romulus} opt to include high temperature metal line cooling and instead boost the efficiency of stellar feedback \citep{shen12,dallavecchia12,eagle15, sokolowska16,sokolowska18}, this will not necessarily result in realistic circumgalactic media nor realistic galaxies beyond bulk properties such as their final stellar and gas masses. \citet{christensen14b} find  that ISM models that include both metal lines and H$_2$ physics result in galaxies with star formation histories and outflow rates more similar to primordial cooling runs than to simulations with metal lines and no H$_2$. This will be particularly true for the low mass halos presented here, which would likely be more affected by artificially boosted SN feedback prescriptions. The structure of the ICM of the {\sc RomulusC} cluster has been shown to be consistent with observations, despite the lack of metal-line cooling \citep{tremmel19}.

\begin{figure}
\centering
\includegraphics[trim=5mm 0mm 45mm 35mm, clip, width=83mm]{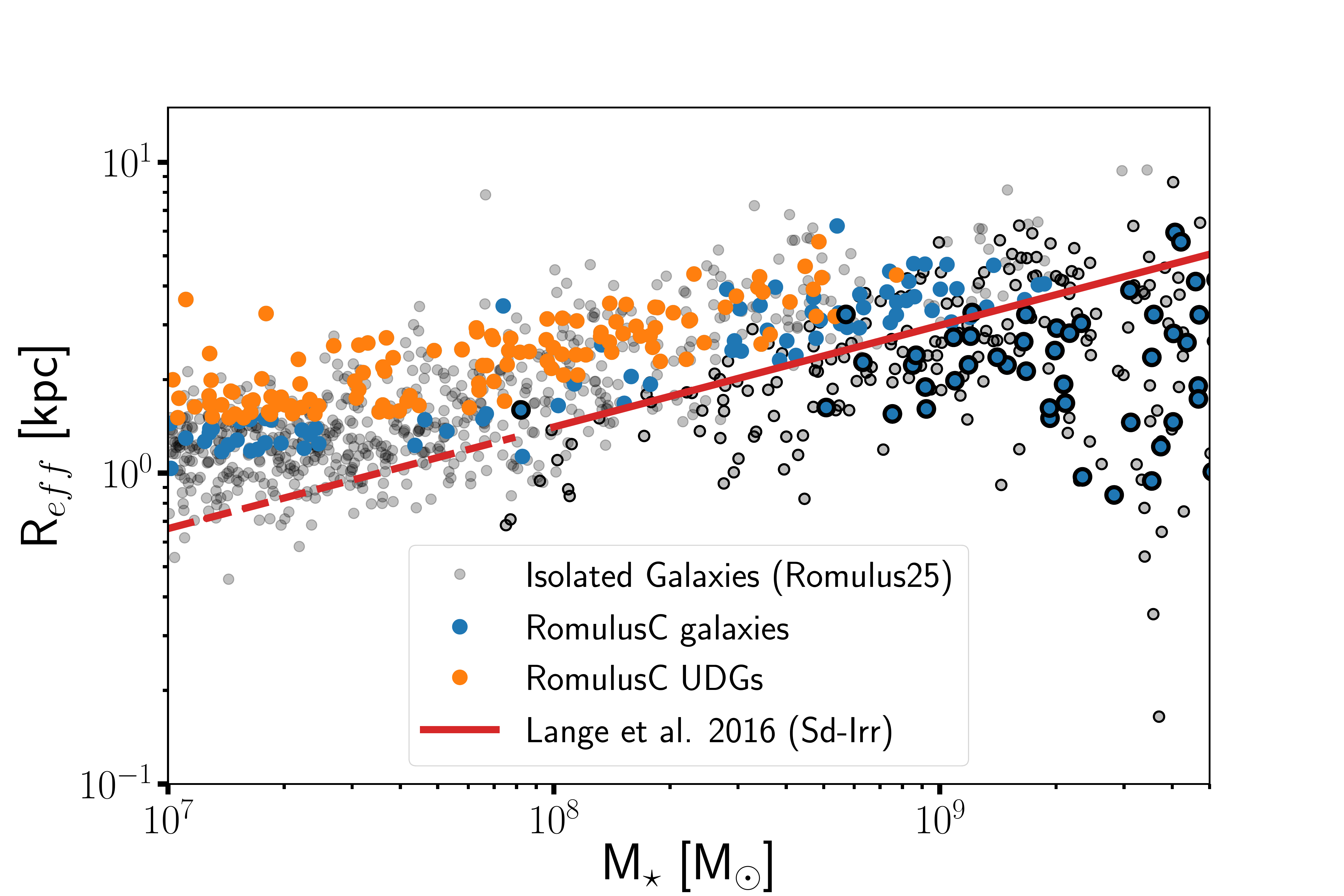}
\caption{{\sc Size mass relation in Romulus}. The g-band effective radius versus stellar mass for isolated galaxies from {\sc Romulus25} (grey) as well as cluster galaxies (blue) and UDGs (orange). In general, cluster galaxies follow the same relation as the isolated galaxies, which is slightly above the observed relation shown in red from \citet{lange16}, the dashed portion representing an extrapolation of the observed relation. This is mostly due to selection effects. The points with black borders are those with mean surface brightness $<24.5$ mag/arcsec$^2$ to approximate the selection effects of the \citet{lange16} sample. This subset of {\sc Romulus} data better matches the observed relation. While non-UDG cluster dwarfs typically follow the relation down to $\sim10^{7.5}$ M$_{\odot}$, UDGs tend to lie well above the relation. Dwarf galaxies in the cluster lack the population of more compact (R$_{eff} < 1$ kpc) galaxies that exist in isolation, particularly at low mass ($<10^{7.5}$ M$_{\odot}$). \citet{lange16} use r-band to calculate the effective radius, but we confirm that the difference is negligible were we to fit the r-band profile instead.}
\label{size_mass}
\end{figure}

\subsubsection{Black hole physics}

\begin{figure*}
\centering
\includegraphics[trim=85mm 5mm -5mm 5mm, clip, width=200mm]{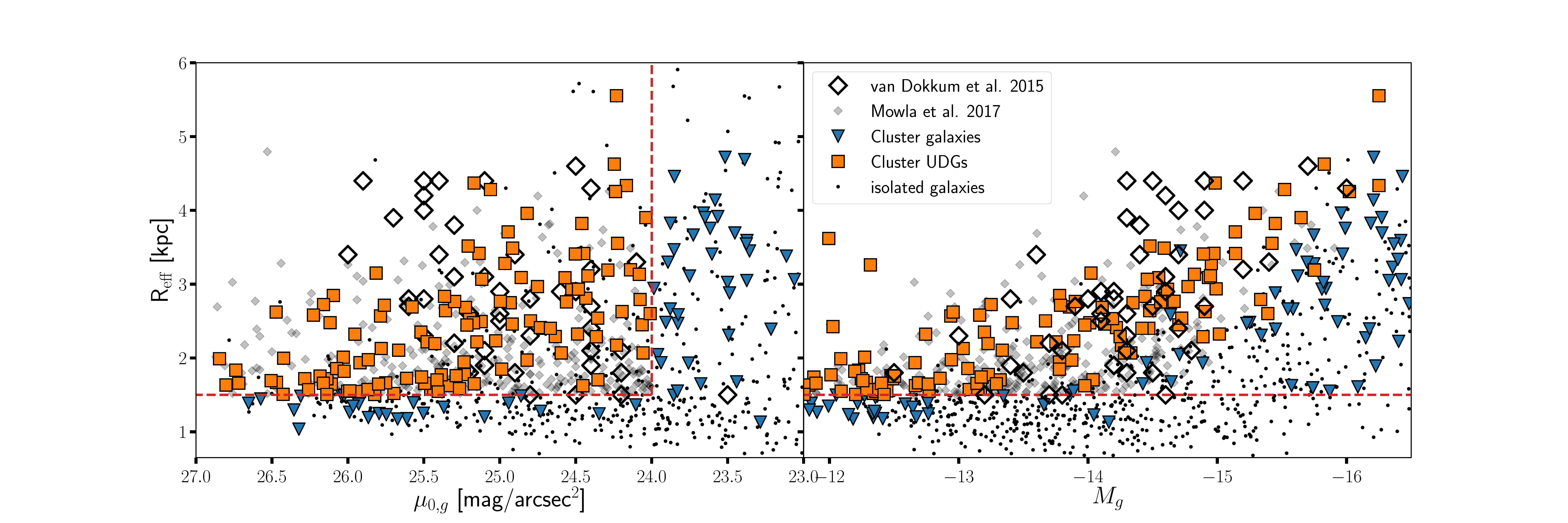}
\caption{{\sc Comparison with observed cluster UDGs.} Scatter plots of properties of simulated cluster UDGs (orange) and other non-UDG cluster galaxies (blue) relative to observed properties of UDGs in cluster environments from \citet{PvD15} and \citet{mowla17} (black and grey diamonds respectively). For the latter we convert Subaru R-band magnitude to g-band assuming $g-R = 0.5$, as described in \citet{mowla17}. Note that one of the points from \citet{PvD15} lies outside of the UDG region we define with red lines (DF21 with a central surface brightness of 23.5 mag/arcsec$^2$) and we include it here for completeness. Also shown are simulated isolated galaxies extracted from {\sc Romulus25} (black points). UDGs in {\sc RomulusC} match well in general with observed galaxies, though the simulation has fewer large ($>3$ kpc), low central surface brightness ($>25$ mag/arcsec$^2$) UDGs. UDGs in {\sc RomulusC} do not inhabit a well separated region of R$_{eff}-\mu_0-M_g$ space compared to the rest of the galaxy population, but are rather just the low surface brightness portion of a roughly continuous population.}
\label{obs_compare}
\end{figure*}

Supermassive black hole (SMBH) formation, dynamics, growth, and feedback are implemented into all {\sc Romulus} simulations and described in more detail in \citet{tremmel17}, where it is shown that low mass galaxies are generally unaffected by SMBH feedback \cite[though see][]{sharma19}. While SMBH feedback certainly affects the evolution of the brightest cluster galaxy and other massive galaxies, it only marginally changes the overall structure of the ICM, even in the cluster core \citep{tremmel19}. While in this work we verify that interaction between dwarf galaxies and their environment is the most important process to forming UDGs in cluster environments, we explore in more detail how the presence of SMBHs in dwarf galaxies may affect their final morphology and star formation history in \citet{sharma19}. We do confirm that the presence of a SMBH is not required to form a UDG, nor to keep a galaxy from becoming a UDG.

 For the sake of completeness, we will briefly review the implementation of SMBH physics in {\sc Romulus}. SMBHs are seeded in cold ($T<10^4$ K), dense ($n > 15 n_{\star}$), and pristine ($Z<10^{-4}$) gas, primarily in the first Gyr of the simulation. SMBH growth is modeled with a modified Bondi-Hoyle formalism that accounts for angular momentum supported gas. A fraction (0.2\%) of the accreted mass is transferred to surrounding gas particles as thermal energy and the cooling of particles receiving this energy is turned off for the length of the SMBH's timestep (generally $<10^5$ yr). Accretion onto SMBHs is Eddington limited, assuming a radiative efficiency of 10\%. SMBHs are allowed to move freely within their host galaxies, with unresolved dynamical friction accounted for in a sub-grid model \citep{tremmel15}, resulting in SMBH mergers that can be significantly delayed with respect to galaxy mergers \citep{tremmel18,tremmel18b}.

\subsubsection{Sub-grid parameter optimization}

Free parameters within our sub-grid models for star formation, SN feedback, and SMBH growth and feedback, were optimized using dozens of zoom-in cosmological simulations run with the same resolution as {\sc Romulus}, as described in \citet{tremmel17}. The parameter combination that resulted in galaxies that best matched four roughly orthogonal empirical relations were selected. The relations used were: 1) stellar mass-halo mass \citep{moster13}, 2) SMBH mass-stellar mass \citep{schramm2013}, 3) HI mass-stellar mass \citep[using ALFALFA data, see][]{cannon11,AlfAlfaHaynes11}, and 4) bulge-to-total ratio versus specific angular momentum \citep{obreschkow14}. Calibration was only performed on isolated galaxies at $z=0$. Therefore, the evolution of galaxy properties (morphology, star formation history, etc) with redshift, as well as $z=0$ properties of satellite galaxies, are purely predictions from the simulation and have in no way been optimized to match any observations.

\subsection{Halo and Galaxy Extraction}

\begin{figure*}
\centering
\includegraphics[trim=40mm 0mm 60mm 40mm, clip, width=170mm]{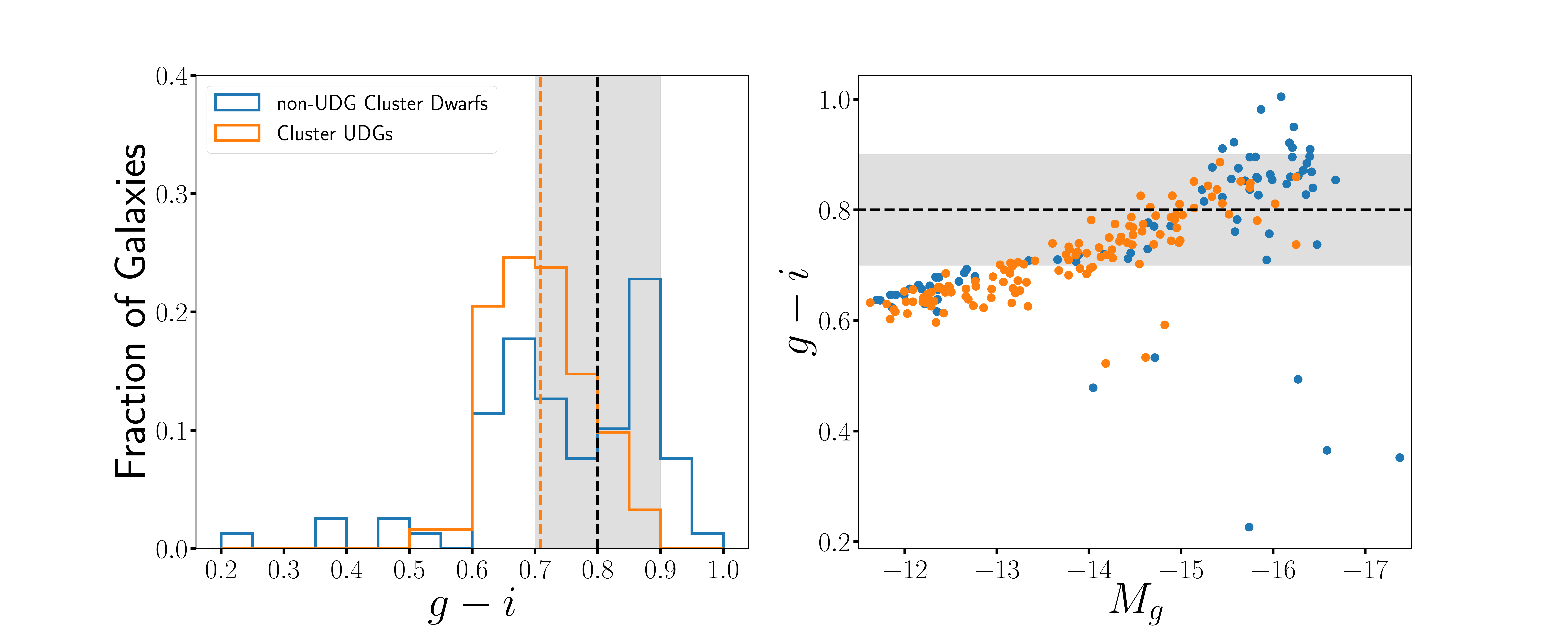}
\caption{{\sc Colors of Cluster Dwarf Galaxies in RomulusC}. {\it Left:} The Distribution of $g-i$ colors of UDG (orange) and non-UDG (blue) dwarf galaxies in {\sc RomulusC}. {\it Right:} Dwarf galaxy $g-i$ colors for cluster UDGs and non-UDGs (same colors) as a function of total g-band magnitude. The dashed line and grey region represent the mean $g-i$ color from \citet{PvD15} and associated standard deviation. As discussed in \citet{tremmel19}, the overwhelming majority of dwarf galaxies in {\sc RomulusC} are quenched and red. The colors of UDGs are similar to those in observations of Coma UDGs. UDGs and non-UDGs have similar colors at a given magnitude. The fact that our UDG population has slightly bluer colors, with an average of 0.7 compared to 0.8 from the \citet{PvD15} sample, is due to the lower mass, dimmer galaxies included in our simulation. In \citet{PvD15} there is only one galaxy with $M_g>-13$. Several UDGs in the more complete \citet{mowla17} sample exist at such low magnitudes, but $g-i$ colors have not been included for those galaxies yet. Were we to disregard these lower mass galaxies, the average UDG color would be more similar to the \citet{PvD15} sample.}
\label{udg_color}
\end{figure*}

Dark matter halos, sub-halos, and all of their baryonic content, including central galaxies, are extracted using the Amiga Halo Finder \citep{knollmann09}. AHF utilizes a spherical top-hat collapse technique to define the virial radius (R$_{\mathrm{vir}}$) and mass (M$_{\mathrm{vir}}$) of each halo and sub-halo. In our analysis we also include M$_{200}$ and R$_{200}$, the mass and radius containing an average overdensity of 200 times the critical density of the universe at that halo's redshift. We only consider a halo or sub-halo resolved if it has a virial mass of at least $3\times 10^9$ M$_{\odot}$, meaning that it contains at least $\sim10^4$ dark matter particles. The centers of halos are calculated using the shrinking spheres approach \citep{power03}, which consistently traces the centers of the central galaxies within each halo. By nature of being a zoom-in simulation, the main halo and surrounding region in {\sc RomulusC} are simulated at high resolution while the rest of the cosmological volume is sampled at much lower mass resolution. Galaxies near the boundaries of the simulation can become `contaminated' with low resolution elements. We avoid including such galaxies in our analysis by requiring each galaxy to have less than $5\%$ of its dark matter particles be contaminated by low resolution elements. We only include galaxies that, at $z=0$, exist well within our zoom-in region, no more than 1.5 Mpc from the center of the cluster, while the zoom-in region extends out to $\sim2$ Mpc.

Following \citet{munshi13}, when comparing our results with observations we apply a factor of 0.6 and 1.25 to the stellar and halo masses in order to provide a more accurate representation of what the inferred values would be given typical observational  and abundance matching techniques. According to \citet{munshi13} these corrections are rather conservative for galaxies in the mass range we focus on in this Paper. Such a correction has also been justified by recent observations comparing more advanced techniques to typical photometric estimates of stellar mass \citep{leja19}. We note that the results of this paper are insensitive to this choice, as they do not play a role in determining the structural properties of the galaxies (effective radius, central surface brightness, etc) and will only affect the (observed) stellar masses we predict from the simulation.

In our analysis we define dwarf galaxies to be any resolved galaxy with a post-correction stellar mass below $10^9$ M$_{\odot}$. % (i.e. after these corrections are taken into account).
In addition to galaxies from {\sc RomulusC}, we also compare our cluster dwarf galaxies to the population of isolated dwarf galaxies in the {\sc Romulus25} cosmological simulation. As discussed in \S2, both simulations are run with the same resolution, sub-grid physics, and cosmology. When selecting for isolated galaxies, we select only galaxies that do not lie within the virial radius of another halo with larger virial mass. We also employ an additional criterion motivated by the results from \citet{geha12}, where each isolated dwarf galaxy must be at least 1.5 Mpc away from any galaxy with stellar mass greater than $2.5\times10^{10}$ M$_{\odot}$. Dwarf galaxies at closer distances to such massive galaxies are likely to have been affected by their environment and should not be considered isolated. Galaxies and their host halos are extracted from the simulation in the same way as in {\sc RomulusC}.

\subsection{Selection of UDGs in RomulusC}

\begin{table}
\centering
\caption{Number of total dwarf galaxies and UDGs in different mass bins in {\sc RomulusC}. The errors in UDG fraction are Poisson errors.}
\begin{tabular}{cccc}
\hline
log(M$_{\star}$ [M$_{\odot}$]) & N$_{\mathrm{total}}$ & N$_{\mathrm{udg}}$ & UDG Fraction\\
\hline
$7-7.5$ & 64 & 41 & $0.65\pm0.1$\\
$7.5-8$ & 45 & 37 & $0.82\pm0.14$\\
$8-8.5$ & 43 & 31 & $0.72\pm0.13$\\
$8.5-9$ & 49 & 13 & $0.36\pm0.1$\\
\hline
\end{tabular}
\label{frac_table}
\end{table}

\begin{figure*}
\centering
\includegraphics[trim=35mm 32mm 25mm 60mm, clip, width=190mm]{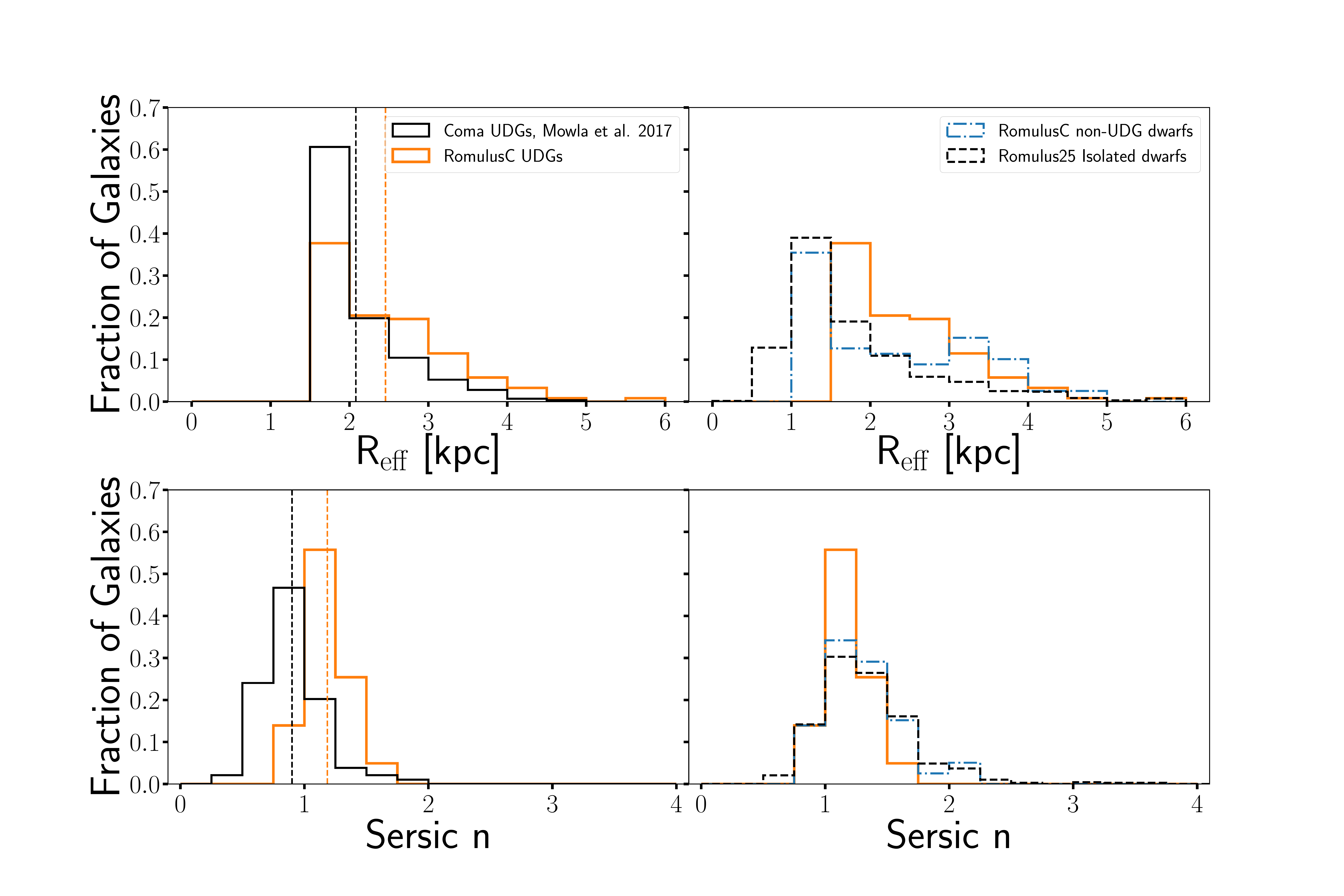}
\caption{{\sc Morphological Properties of UDGs in Romulus}. The distribution of effective radii (top) and Sersic index $n$ (bottom) for UDGs (orange, solid lines) in {\sc RomulusC}. On the left, we compare the distributions for UDGs in {\sc RomulusC} to those in the \citet{mowla17} sample in Coma (black, solid). The average values of both distributions are shown as vertical dashed lines. Overall, we find good agreement between the {\sc RomulusC} UDGs and observed galaxies, though the simulated UDGs are more likely to be larger and more centrally concentrated (higher Sersic $n$). On the right we compare {\sc RomulusC} cluster UDGs to non-UDG dwarfs in {\sc RomulusC} (blue, dot-dashed lines), as well as isolated dwarf galaxies in {\sc Romulus25} (black, dashed lines). We find that the Sersic indices are the same for all galaxy populations, while cluster dwarfs are more likely to have large effective radii ($>2$ kpc) compared to isolated dwarfs. We also do not see any dwarf galaxies below 1 kpc in effective radius in the cluster environment, while they exist in the field.}
\label{morph_compare}
\end{figure*}

For each resolved, uncontaminated galaxy in RomulusC at $z=0$, we generate surface brightness profiles in V and B bands due to the stellar populations, where stellar emission is calculated using tables generated from population synthesis models \citep[http://stev.oapd.inaf.it/cgi-bin/cmd;][]{marigo08,girardi10}. This will naturally account for the passive evolution of stellar populations, which result in a dimming over time as the most massive stars evolve and explode as SN. From Johnson V and B bands, we generate g-band surface brightness profiles following the conversion from \citet{jester05}. The profiles are integrated within circular annuli, assuming only stars within each given dark matter halo contribute and absorption from gas and dust is unimportant. This latter assumption is safe for dwarf galaxies, which typically have low metallicities \citep[e.g.][]{vanZee00}. The profiles are then fit to a single Sersic profile \citep{sersic63} of the following form:

\begin{equation}
\Sigma(r) = \mu_e + 2.5(0868 n - 0.142) \left (\left [\frac{r}{\mathrm{R}_{eff}} \right ]^{1/n} - 1 \right).
\label{sersic_fit_eqn}
\end{equation}

We fit each galaxy using a least squares fit on Sersic index, $n$, effective radius, R$_{eff}$, and the surface brightness at the effective radius, $\mu_e$. In order to both match the angular resolution of typical UDG observations \citep[e.g.][]{PvD15, mowla17} and avoid fitting to structure below the resolution limit of the simulation, we create and then fit our surface brightness profiles using radial bins of 300 pc. We fit out to the first radial bin with surface brightness dimmer than 32 mag/arcsec$^2$, a typical limit for the most sensitive low surface brightness observations \citep[e.g.][]{PvD14}. We place a further limitation on the radial bins used in this fit that they extend no further than the closest companion galaxy. At this resolution, it is difficult to resolve satellite galaxies at such low masses so this is a constraint rarely applied. Following \citet{PvD15}, we classify a galaxy as being a UDG if it meets each of two criteria based on the single Sersic fits we generate: 1) it has an effective radius (R$_{eff}$) that is greater than 1.5 kpc, and 2) it has a central surface brightness ($\mu_0$) dimmer than 24 magnitudes/arcsec$^2$. Note that $\mu_0$ is different from $\mu_e$ that we fit to in Eqn.~\ref{sersic_fit_eqn}. Rather, this is the value of the surface brightness from the Sersic fit (Eqn.~\ref{sersic_fit_eqn}) evaluated at $r = 0$.

Each profile and Sersic fit is generated as if the observer is looking at the galaxy face-on. Galaxy re-orientations are performed based on the angular momentum of gas particles within the inner 5 kpc of the halo or, when there are less than 100 such particles, star particles within the inner 5 kpc. The rationale behind this is that we want to capture all galaxies that could potentially be categorized as a UDG and, by orienting in this way, we maximize the effective size and minimize the surface brightness of all of our galaxies. This way, when we select UDGs, we will be selecting all galaxies that would potentially be categorized as such were they to be viewed at random inclinations. %In future work, we will explore the effects of orientation and chosen band-pass in categorizing galaxies as UDGs, as well as perform a full GalFit analysis to determine surface brightness profiles and calculate axes ratios in a more detailed way. \amb{I don't think I'd include this last sentence about future work. Or move it to Section 6.}

We find a total of 122 galaxies that fit our adopted definition of UDG within 1.5 Mpc of the center of the cluster at $z=0$ and 80 within R$_{200}$. Example images, surface brightness profiles, and Sersic fits are shown in Figure~\ref{rogue_gallery}. This predicted abundance is a factor of $\sim2-3$ higher than many estimates for low mass clusters such as {\sc RomulusC} \citep{vdBurg17,romantrujillo17b, mancerapina18}, though in agreement with other cosmological simulation results \citep{liao19,jiang19}. We do, however, stress that this is an upper limit as we do not account for the effect of random orientation which can make galaxies appear less diffuse. If galaxies have a low axes ratio, doing a full GalFit analysis may also decrease some effective radii. Some numerical effects may also artificially boost our population of UDGs. We discuss all of this further in \S5.

\section{Properties of UDGs in RomulusC}
Table~\ref{frac_table} shows the fraction of galaxies that we categorize as UDGs as a function of galaxy stellar mass. Low mass galaxies are more likely to be UDGs and all of our UDGs are less than $10^9$ M$_{\odot}$ in stellar mass. UDGs in {\sc RomulusC} have stellar masses as high as $7.72\times10^8$ M$_{\odot}$ and total halo virial masses as high as $3.5\times10^{10}$ M$_{\odot}$ at $z=0$ (halo masses are typically smaller than at in-fall, as discussed further in \S3.3), corresponding to $4.4 \times10^{10}$ M$_{\odot}$ in dark matter only simulations \citep{munshi13}. As described in \S2.2, in the following analysis we compare to non-UDG dwarf galaxies, which we define as any resolved galaxy with stellar mass below $10^9$ M$_{\odot}$. In this section we describe the observable properties of the UDGs in our sample in terms of morphology and color, comparing both to the ambient simulated dwarf galaxy population in the cluster and field environments, as well as observations of cluster UDGs. We also examine their position and kinematics within the cluster environment in {\sc RomulusC}.

%\amb{This information feels a bit adrift.  A histogram or something would be more useful.} \mjt{I don't really want to add another plot if I can avoid it}

\subsection{UDG morphology and colors}

Figure ~\ref{size_mass} shows the size-mass relation for {\sc RomulusC} galaxies as well as isolated dwarf galaxies from the {\sc Romulus25} simulation. In terms of their size, UDGs are not a unique population compared to non-UDGs. Rather, they represent, particularly at stellar masses below $\sim2\times10^8$ M$_{\odot}$, the larger galaxies among the overall distribution of cluster dwarf galaxies. Cluster galaxies follow the same relation as the isolated galaxies, although they lack the smallest galaxies at any given stellar mass. This is contrary to observations that find significant populations of dwarf galaxies in clusters with effective radii below 1 kpc \citep[e.g.][]{gavazzi05,eigenthaler18, venhola19}.

The isolated galaxy population is consistent with the empirical relation from \citet{lange16} at stellar masses above $10^{7.5}$ M$_{\odot}$. Although biased slightly high on average, the {\sc Romulus25} isolated galaxies are within a factor of two of the empirical relation. %\textbf{This discrepancy goes away if we limit our simulated sample to what would be detectable in the GAMA survey, corresponding roughly to an effective surface brightness brighter than 24.5 mag/arcsec$^2$.} 
Non-UDG cluster galaxies match well to the \citet{lange16} results while the UDGs are above it. At the lowest masses (M$_{\star}<10^{7.5}$M$_{\odot}$) the deviation from the (extrapolated) empirical relation for both cluster and isolated dwarf galaxies is more significant. The observations from \citet{lange16} are insensitive to the lowest surface brightness galaxies and when we account for this our results match much better to the empirical relation (thick outlined points in the figure). These low surface brightness dwarfs contribute to the slight bias of our simulated galaxies toward higher effective radius in both environments. For galaxies with M$_{\star}<10^{7.5}$ M$_{\odot}$ resolution effects could be important in determining their larger sizes, as we discuss further in \S5.

%Cluster galaxies follow the same relation as the isolated galaxies, although they lack the smallest galaxies at any given stellar mass. This is contrary to observations that find  compact dwarf galaxies in the Coma, Virgo, and Fornax clusters with effective radii below 1 kpc \citep[e.g.][]{gavazzi05,eigenthaler18, venhola19}. %In terms of their size, UDGs are not a unique population compared to non-UDGs. Rather, they represent, particularly at stellar masses below $\sim2\times10^8$ M$_{\odot}$, the larger galaxies among the overall distribution of dwarf galaxies. 

%We note that it is also likely that the observations from \citet{lange16} are insensitive to the lowest surface brightness galaxies, which we do not exclude from our simulated sample. This may contribute to why our simulated dwarf galaxies are more likely to be above the observed relation in both field and cluster environments. The discrepancy is worse for lower mass galaxies (M$_{\star}<10^{7.5}$ M$_{\odot}$) which, as we discuss in \S6, is likely due to limited spatial resolution.

\begin{figure}
\centering
\includegraphics[trim=3mm 2mm 45mm 30mm, clip, width=83mm]{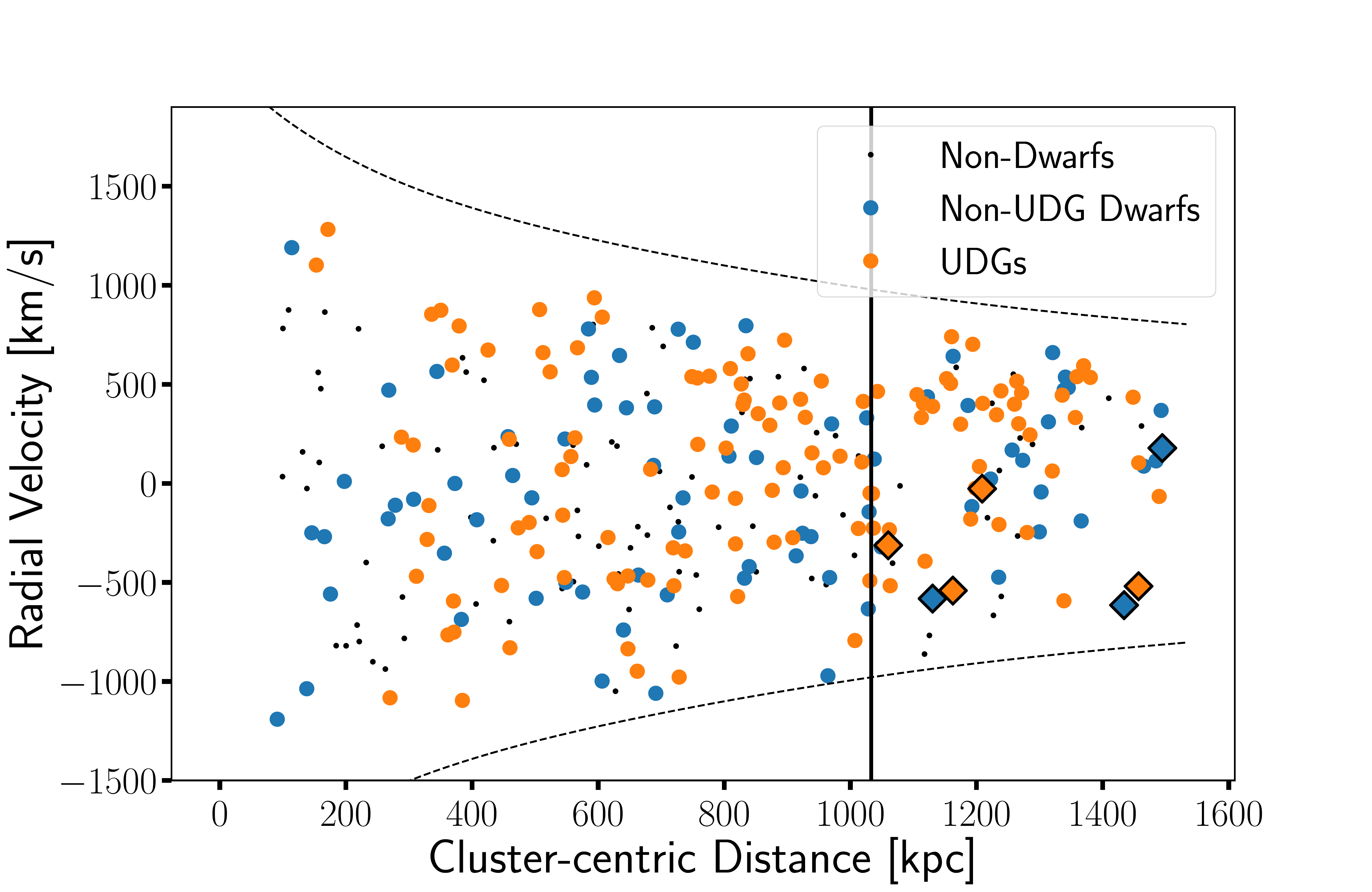}
\caption{{\sc Distance and Radial velocity for Cluster Galaxies}. UDGs (orange), non-UDG dwarf galaxies (blue), and more massive galaxies (black) all inhabit the same regions in phase space within the cluster environment at $z=0$. The solid line denotes R$_{200}$ of the cluster at $z=0$ and the dashed curves denote the cluster's escape velocity. The diamond points represent galaxies that have not yet fallen inward of R$_{200}$. Several galaxies are splashback galaxies that have moved back outside R$_{200}$ after their initial in-fall.}
\label{pos_vel}
\end{figure}

Figure~\ref{obs_compare} plots the effective radius, central surface brightness, and total g-band magnitude ($M_{g}$) for simulated cluster galaxies from {\sc RomulusC} and isolated galaxies from {\sc Romulus25} alongside the observed sample of Coma cluster UDGs from \citet{PvD15} and \citet{mowla17}. Cluster UDGs do not inhabit a unique region in R$_{eff}-\mu_0-$M$_{g}$ space compared to cluster and isolated dwarf galaxies, though there are more large, low surface brightness, low-luminosity galaxies in the cluster compared to the field. There is an overall good agreement between the observed UDGs and those in {\sc RomulusC}. The most noticeable, though minor, difference is the lack of galaxies in {\sc RomulusC} with low ($\mu_0>25$ mag/arcsec$^2$) central surface brightness and $R_{\mathrm{eff}} > 3$ kpc. %There is also a higher abundance of smaller (R$_{eff}$<2 kpc) UDGs with central surface brightness $>25$ mag/arcsec$^2$ and M$_g$ > -13 in {\sc RomulusC}, resulting in a slightly flatter relationship between R$_{eff}$ and M$_g$ compared to observations. 

Figure~\ref{udg_color} shows the distribution of UDG and non-UDG dwarf galaxy colors in {\sc RomulusC}. The typical dwarf galaxy colors in {\sc RomulusC} are red ($g-i > 0.6$) down to a g-band magnitude of -12, consistent with observed cluster dwarf galaxies \citep{roediger17,eigenthaler18}. Dwarf galaxies we classify as UDGs and non-UDGs have similar colors, with dimmer, lower mass galaxies typically being bluer, also consistent with observations \citep{boselli14, roediger17,eigenthaler18}. The $g-i$ colors are slightly bluer than the average UDG color presented in \citet{PvD15} ($\langle g-i \rangle = 0.8\pm0.1$), but this difference is due to the inclusion of lower luminosity, bluer galaxies from {\sc RomulusC}. The {\sc RomulusC} UDG population deviates from the typical color of observed UDGs below $M_g\sim-13$, where there is only one UDG in the \citet{PvD15} sample. %The extended sample of \citet{mowla17} does have lower luminosity galaxies but does not include $g-i$ colors.

The right-hand panels of Figure~\ref{morph_compare} compare the distribution in R$_{eff}$ and Sersic index of cluster UDG and non-UDG dwarf galaxies, as well as isolated galaxies from {\sc Romulus25}. Other than the fact that UDGs are not allowed to have small effective radii by definition and are therefore more likely to have larger sizes, cluster UDGs do not look significantly different from non-UDGs or field galaxies in terms of R$_{eff}$ and are essentially identical in terms of their Sersic index distribution. As mentioned above, we see that {\sc RomulusC} lacks dwarf galaxies with R$_{eff} < 1$ kpc that exist in isolation.  Overall, these results indicate that UDGs are not a unique population of galaxies, but rather a sub-set of the ambient dwarf galaxy population in our simulations. Further, their morphology is not significantly altered by the cluster environment compared to isolated galaxies. In the left hand panels, we also compare to observed UDG morphologies from \citet{mowla17}. While we find overall good agreement with observed cluster UDGs, {\sc RomulusC} UDGs have slightly larger sizes and larger Sersic indices. Despite their slightly larger Sersic indices compared with the \citet{mowla17} sample, the lack of UDGs with $n>2$ is consistent with a wide range of observations \citep{romantrujillo17,venhola17,cohen18,mancerapina19}.

\subsection{Location, dynamics, and infall time}

\begin{figure}
\centering
\includegraphics[trim=15mm 30mm 30mm 60mm, clip, width=80mm]{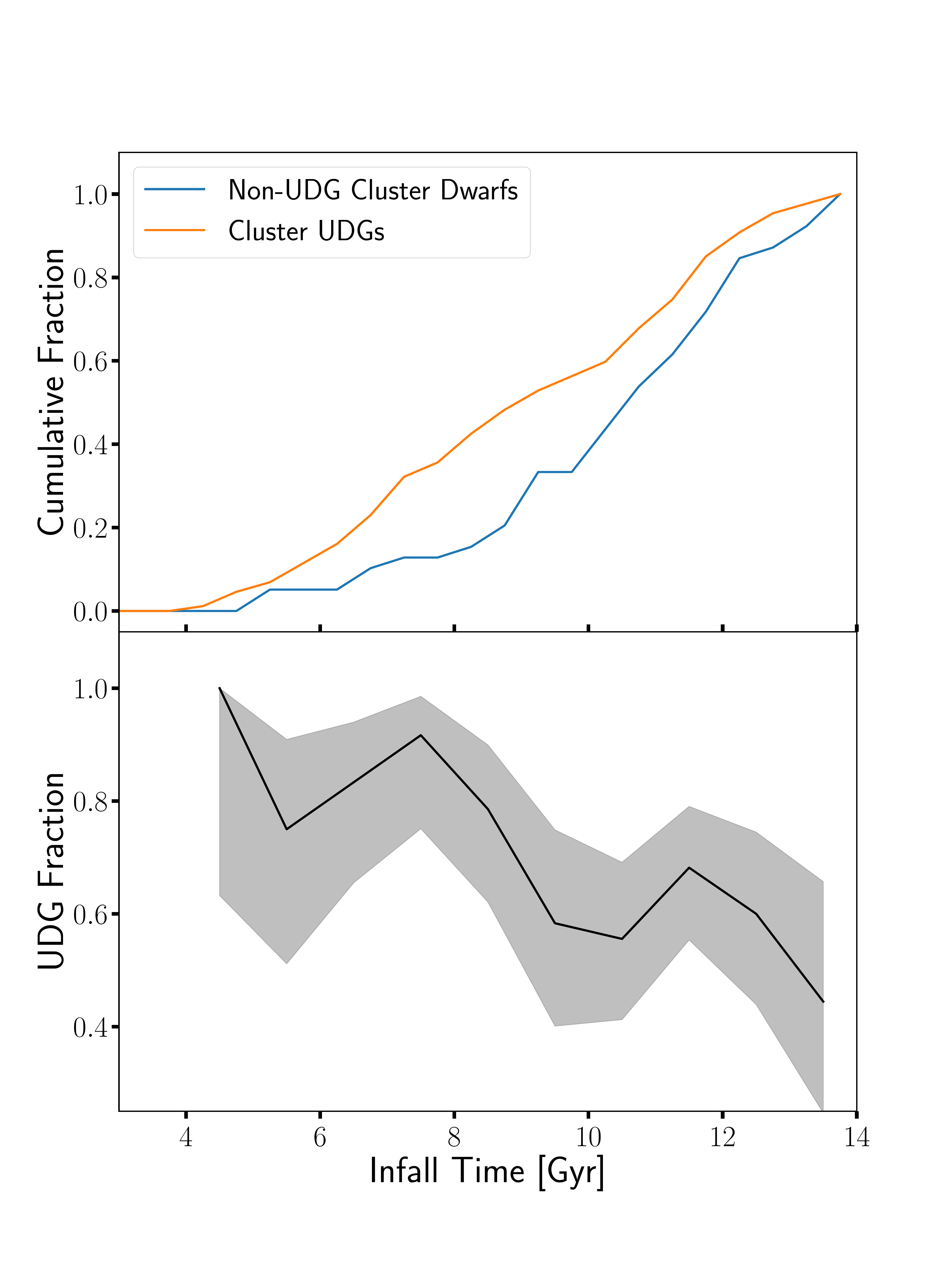}
\caption{{\sc Galaxy In-fall Times}. \textit{Top:} The cumulative distribution of in-fall times for UDGs (orange) and non-UDG dwarf galaxies (blue). \textit{Bottom:} The fraction of dwarf galaxies that are UDGs at $z=0$ as a function of in-fall time. The grey shaded region represents the 68\%  binomial confidence interval \citep{cameron11}. Low mass galaxies crossing R$_{200}$ at $z>0.5$ are about two times more likely to be classified as UDGs by $z=0$ than galaxies that in-fall at $z<0.5$.}
\label{tinfall}
\end{figure}

Figure~\ref{pos_vel} plots the cluster-centric distance and radial velocity of cluster galaxies. The small number of dwarf galaxies that have yet to cross R$_{200}$ by $z=0$ are shown as diamonds. There is also a significant population of `splashback' galaxies that have already crossed R$_{200}$ but their orbits have taken them back outside of it. Most of these galaxies happen to be moving outward (positive radial velocities), but some are on their way back inward. While there is overall little difference in phase space location between UDGs and non-UDGs, interestingly there is a dearth of UDGs within $\sim200$ kpc from cluster center (or 0.2 R$_{200}$). This agrees with observational results from \citet{mancerapina18} and may point to the fact that UDGs are more likely to become tidally disrupted at small cluster-centric distances. However, we stress that limitations in halo finding and artificial satellite disruption due to resolution effects (as discussed in \S5.4) would need to be addressed in more detail before any direct conclusions can be made from the simulations, which we leave to future work.

There is little difference between the phase space location of UDGs, non-UDGs, and more massive galaxies, but there is a difference between their in-fall times, defined as the time a galaxy first crosses R$_{200}$ of the cluster. The upper panel in Figure ~\ref{tinfall} plots the distribution of in-fall times for UDG and non-UDG dwarf galaxies in {\sc RomulusC}. The small number of dwarfs that have yet to cross R$_{200}$ are not included. UDGs are more likely to have fallen in at earlier times. The lower panel in Figure ~\ref{tinfall} shows the fraction of galaxies that are UDGs at $z=0$ as a function of in-fall time. More than 80\% of the dwarf galaxies that in-fall before $z\sim0.5$ become UDGs by $z=0$. This falls to $\sim50\%$ for galaxies that have fallen into the cluster more recently.

\begin{figure}
\centering
\includegraphics[trim=15mm 30mm 30mm 60mm, clip, width=80mm]{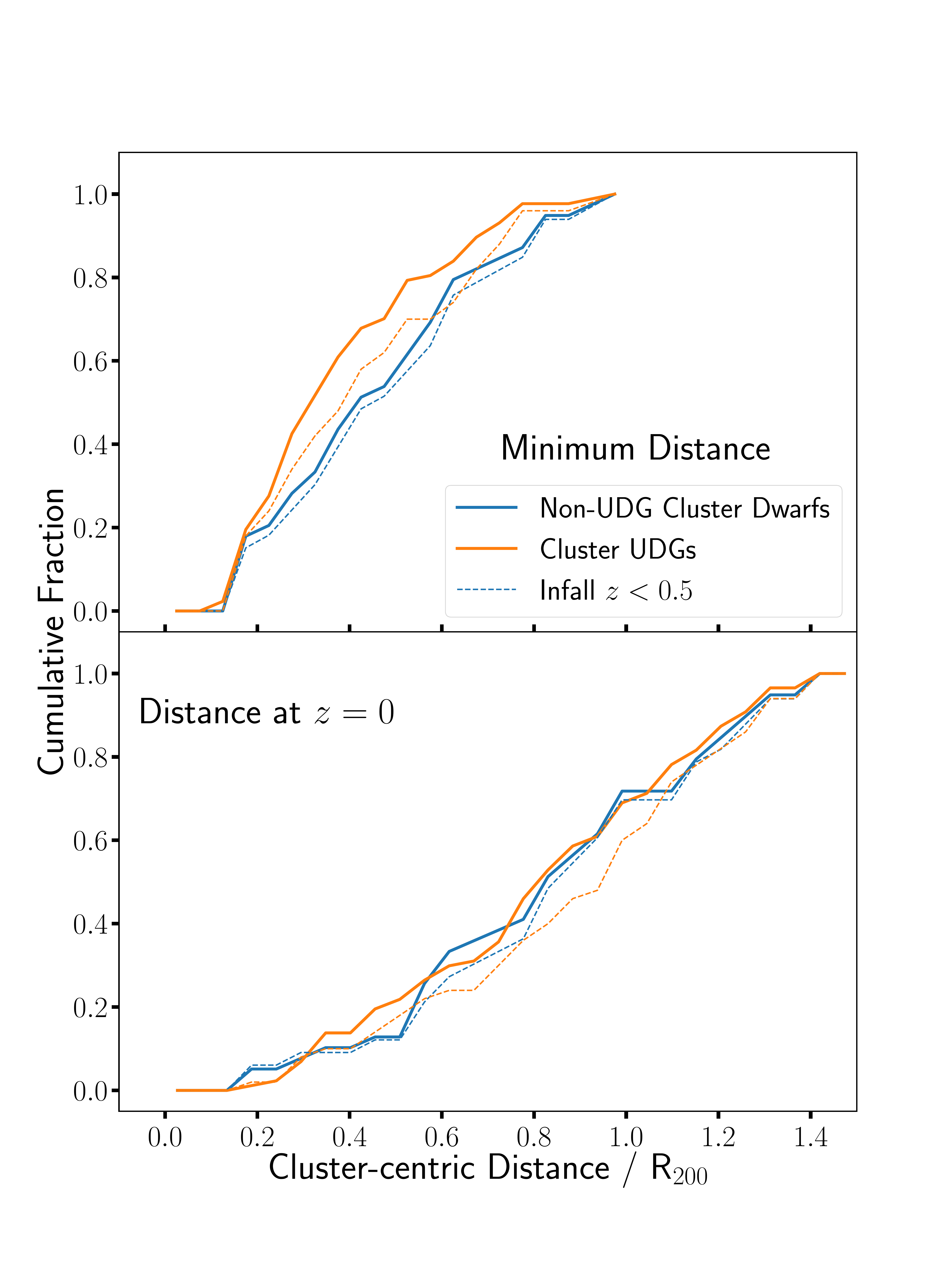}
\caption{{\sc Final and Minimum Cluster-centric Distances for Dwarf Galaxies}. The distribution of final (bottom) and minimum (top) distances each UDG (orange) and non-UDG (blue) dwarf galaxy attains relative to cluster center in {\sc RomulusC}. We derive the minimum distance relative to R$_{200}$ as calculated for the cluster progenitor at each snapshot. The overall distribution at $z=0$ is the same for both populations while UDGs have gotten preferentially closer to the center of the cluster throughout their orbital evolution. This difference can be attributed to the earlier in-fall times of UDGs compared to non-UDGs (see Figure~\ref{tinfall}), as the distributions look similar if we only consider the $z<0.5$ in-fall populations (dashed lines). However, it is true that UDGs have preferentially felt higher densities, not only because they get closer to the center, but also because they do so earlier when the (proto-)cluster is more dense.}
\label{distance}
\end{figure}

Figure~\ref{distance} plots the distribution of both $z=0$ cluster-centric distances and the minimum distance attained by cluster dwarf galaxies. Again, only those that have already crossed R$_{200}$ are included. The final distance distributions are nearly identical, but there is a difference in the minimum distance to cluster center attained throughout their orbital evolution. The difference is relatively small and is mostly due to the different in-fall times, as the difference is diminished if we consider only galaxies with in-fall at $z<0.5$. However this does mean that UDGs have typically interacted with a denser environment as the (proto-)cluster is denser at high redshift. Still, we note that not all UDGs have passed close to cluster center ($\sim30\%$ of UDGs have never passed closer than 0.5 R$_{200}$). Further, UDGs and non-UDGs  that have in-fall times at lower redshift have similar minimum cluster-centric distances.

\subsection{Dark matter halos}

\begin{figure}
\centering
\includegraphics[trim=10mm 0mm 45mm 30mm, clip, width=83mm]{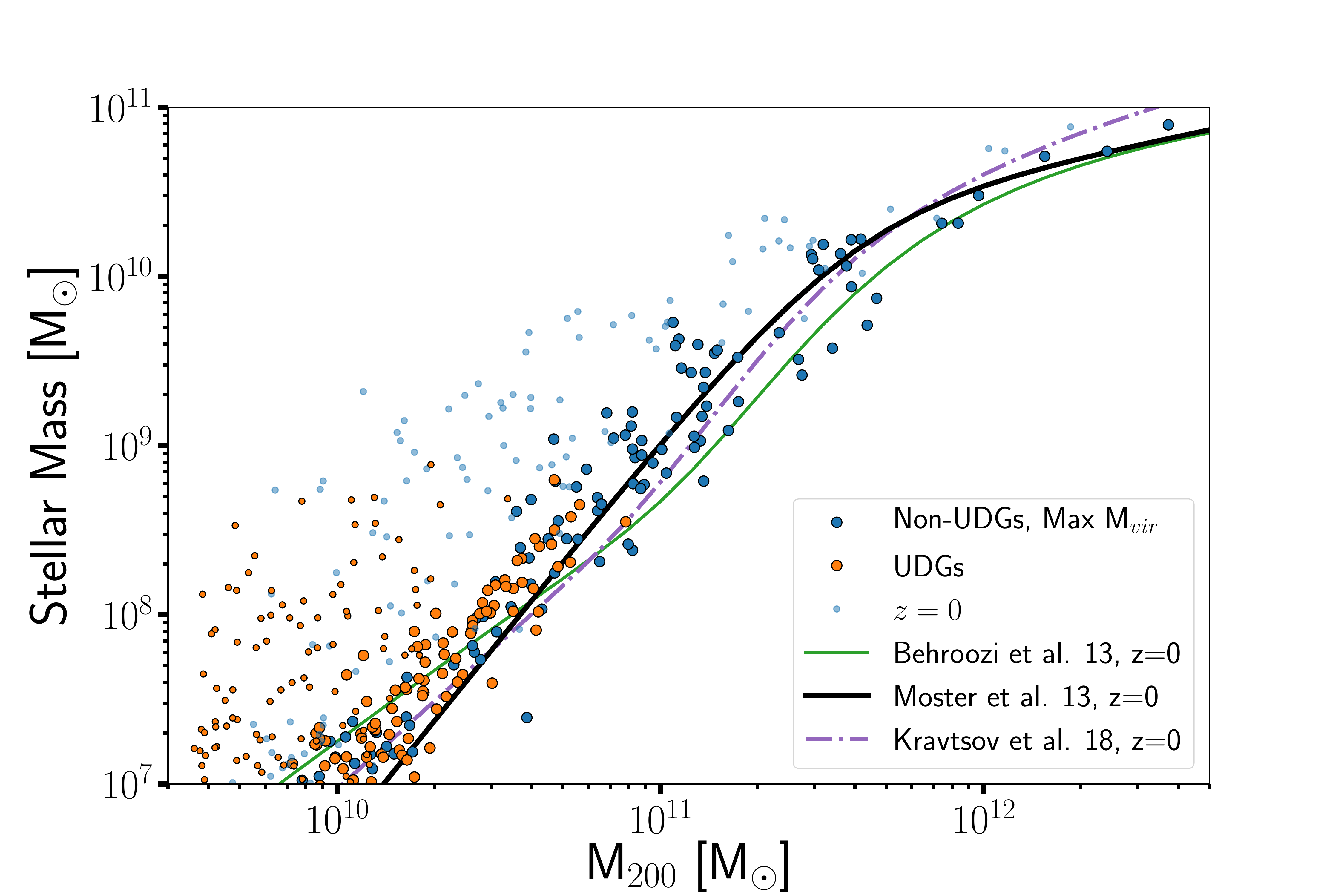}
\caption{{\sc Stellar Mass -- Halo Mass Relation for Cluster Galaxies}. The small points are the properties of the galaxies and their M$_{200}$ values at $z=0$. The large points represent the stellar and M$_{200}$ values at the time their peak halo mass has been reached. Overplotted are three abundance matching results from \citet[][green]{behroozi13}, \citet[][black]{moster13}, and \citet[][magenta, dot-dash]{kravtsov18}. The {\sc Romulus} simulations were optimized to fit the \citet{moster13} relation at $z=0$. In this respect, UDG and non-UDG dwarf galaxies have normal halo masses for their stellar mass at in-fall.  Deviations from the relation at $z=0$ are due to tidal stripping of mass after in-fall. Importantly, this tidal stripping mostly affects the outskirts of the dark matter halos of cluster member galaxies rather than their stellar masses. Stripping can also remove dark matter particles from the core that have radial orbits \citep{zolotov12}, however the cuspy dark matter profile is still intact at $z=0$. All of our UDGs are dwarf galaxies in low mass halos, with the most massive UDG-hosting halo in our sample being just below $10^{11}$ M$_{\odot}$ at its peak.}
\label{smhm}
\end{figure}

Figure~\ref{smhm} plots the relationship between stellar mass and halo mass at $z=0$ (small points) for {\sc RomulusC} galaxies. Plotted in larger points is the relation for these same galaxies at the time they reach their peak halo mass. Prior to in-fall, cluster galaxies are consistent with the same abundance matching relations as isolated galaxies in {\sc Romulus25} \citep{tremmel17}. While these galaxies often in-fall much earlier than $z=0$, we expect the stellar mass halo mass relation to remain relatively constant with redshift \citep{behroozi13}. Because the {\sc Romulus} simulations were calibrated to match the \citet{moster13} relation for isolated $z=0$ galaxies, this result is not purely a prediction of the simulation. What it shows is that, prior to in-fall, UDGs and non-UDGs alike had halo masses consistent with the isolated galaxy population of {\sc Romulus25}. Following in-fall, halos lose dark matter from tidal stripping, but the central galaxy remains unaffected, pushing the galaxies to the left of the abundance matching relations. 

%\amb{rachel is making Ray include more updated SMHM relations, like Behroozi 2018.  Maybe you should consider adding some newere ones, too.} \mjt{looking at the figure in Behroozi 2019, the newer empirical models from Behroozi and Moster for example have lower stellar mass for low mass halos... I think this would make our results look artificially weird, given that we used the earlier abundance matching results to constrain our model... given that I only want these to train the eye rather than test the SMHM relation, I'm weary to put in these results.} %Following in-fall, the cluster halos lose the outskirts of their dark matter halos to tidal stripping, but the central galaxy remains nearly constant in stellar mass, pushing the galaxies to the left of the abundance matching relations. 

At the time of peak halo mass, UDGs in {\sc RomulusC} have stellar masses as high as $6.3\times10^8$ M$_{\odot}$ and halo virial masses as high as $6.5\times10^{10}$ M$_{\odot}$, corresponding to a mass of $8.2\times10^{10}$ M$_{\odot}$ in a dark matter only simulation \citep{munshi13}. While it is often difficult to make concrete observational conclusions about the halo mass of UDGs, Dragonfly 44 is thought to occupy a relatively massive halo based on its large velocity dispersions \citep{PvD16}, though current estimates place its most likely dynamical mass to be not much beyond $10^{11}$ M$_{\odot}$ \citep{PvD19}. The large populations of globular clusters observed in several UDGs \citep{PvD17} has been used as an argument that they come from more massive halos than their stellar masses would normally imply. Our results do not support this picture, as both UDGs and non-UDGs inhabit average mass DM halos for their stellar mass at the time of in-fall. 

Other UDGs have been reported to have much lower dynamical masses than expected, with dark matter sub-dominant compared to stars in terms of total mass, or potentially missing all together \citep{PvD18b, PvD19b}. We do not attempt to extract such halos from the simulation, as we place strict boundaries on which halos we consider well resolved based on the number of dark matter particles they have. It is also unclear whether our halo finder would be properly optimized to find such systems. 

It is possible that these observed galaxies do have dark matter, but heavily cored dark matter profiles. Our galaxies do not form heavily cored dark matter profiles due to their resolution being too low (see \S5 for more discussion on this). There is some decrease to the dark matter density in the centers of cluster satellite halos due to tidal stripping (see \S4.1), but not enough to form resolved cores. The typical dark matter density profile inside 1 kpc for cluster dwarf galaxies is cuspy with a slope of approximately $-1.5$. This is in contrast with recent results from cosmological simulations presented by \citet{dicintio17b} that isolated UDGs have significantly cored dark matter profiles. This is because the same mechanism that causes cored dark matter profiles \citep[rapid, repeated outflows from bursty star formation and supernovae feedback;][]{PG12,PG13} results in the puffing up of the stellar mass as well. Not only do we not see evidence of dark matter cores, but we also see no evidence of star formation histories being any more bursty for UDGs versus non-UDGs. We stress that puffing up both DM and stellar mass distributions via supernovae feedback requires higher resolution than what we have in {\sc Romulus} \citep[][ see discussion in \S5]{dutton19}. So, while this is a feature of our simulation resolution rather than a true prediction, our results show that the formation of UDGs can occur without bursty supernovae feedback causing both stars and dark matter to expand. Rather the formation channel at play in {\sc Romulus} must be unique from what is predicted for isolated UDGs in \citet{dicintio17b}.

\section{The Origin of UDGs in Galaxy Clusters}

\begin{figure*}
\centering
\includegraphics[trim=20mm 60mm 5mm 5mm, clip, width=190mm]{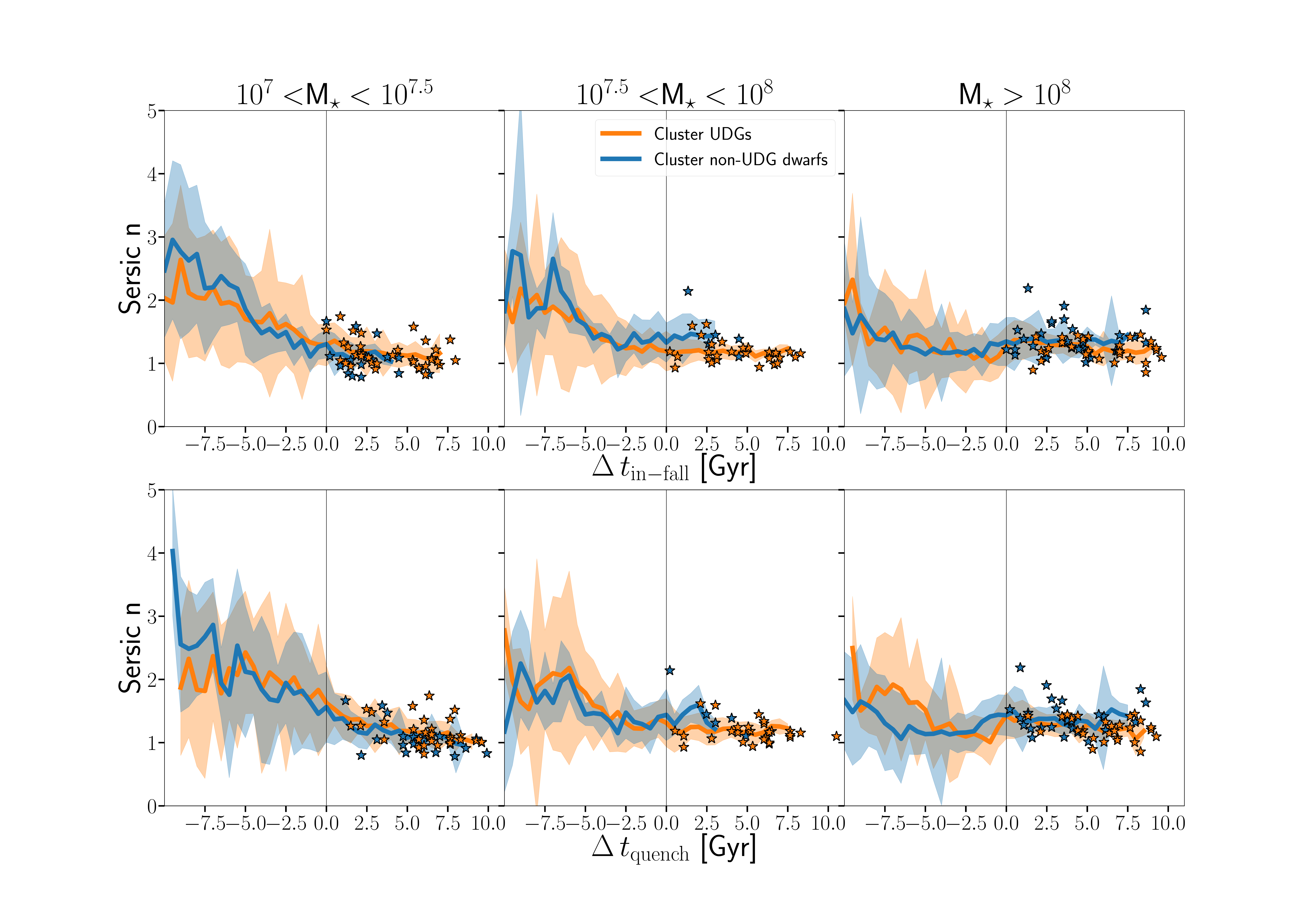}
\caption{{\sc Evolution in Sersic index for cluster dwarf galaxies}. This figure shows the evolution of the Sersic index, $n$, over cosmic time for cluster dwarfs. The average evolutionary track for UDGs (thick orange line) and non-UDGs (thick blue line) in {\sc RomulusC}, only including galaxies that have fallen inward of R$_{200}$ and can be traced back to at least 2 Gyrs after the Big Bang. The stars represent the $z=0$ values. The average is only calculated at times with at least 3 galaxies in that group. The times are plotted relative to the in-fall time (top) and quenching time (bottom). The shaded regions show the standard deviation at each time. Galaxies are more concentrated at earlier times, with $n$ typically falling from $\sim$2 to $\sim1$. There is no apparent change in evolution associated with either quenching or falling into the cluster.}
\label{sersic_evol}
\end{figure*}

In this section we explore the origin of UDGs in galaxy clusters by tracking {\sc RomulusC} galaxies back in time and examining how their key morphological properties change as they interact with the cluster environment. Because UDGs in our simulation span two orders of magnitude in stellar mass, in order to ensure that we control for any differences in evolution as a function of galaxy mass we split galaxies into three stellar mass bins: $10^7 < M_{\star} < 10^{7.5}$, $10^{7.5} < M_{\star} < 10^{8}$, and $10^{8}$ < $M_{\star} < 10^{9}$ M$_{\odot}$. Each bin contains approximately one-third of the total UDG population. %By binning in this way we can ensure that we control for any differences in evolution as a function of stellar mass. %\amb{You haven't really taken advantage of these mass bins in later discussions.  I think the formation probably is different between the highest and lowest stellar mass UDGs, but that's not emphasized.}

\subsection{Cluster In-fall and Quenching}

\begin{figure*}
\centering
\includegraphics[trim=15mm 60mm 5mm 5mm, clip, width=190mm]{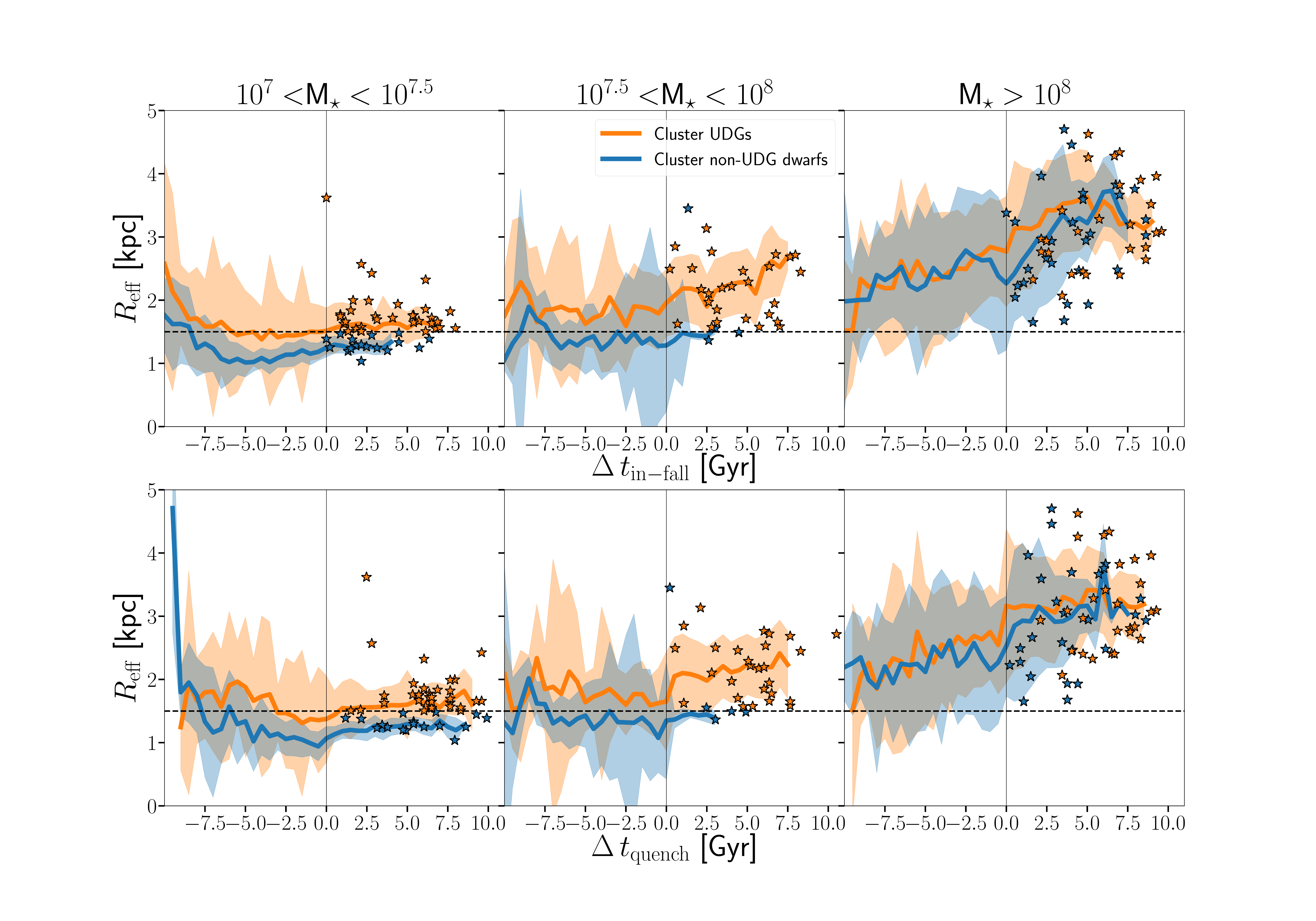}
\caption{{\sc Evolution in effective radius for cluster dwarf galaxies}. Similar to Figure~\ref{reff_evol} but tracking the evolution in the effective radius, R$_{eff}$, for cluster dwarf galaxies. Solid lines represent the average evolutionary tracks for times with at least three galaxies and the shaded regions represent the standard deviation at each time. Galaxies with M$_{\star}<10^8$ M$_{\odot}$ are separated into UDG or non-UDG mostly based on their final effective radius. UDGs tend to have larger radii long before in-fall into the cluster or quenching. High mass galaxies show no difference in evolution between UDGs and non-UDGs. There is a tendency for galaxies to increase in effective radius after in-fall and/or quenching. This is in part due to decreasing the central dark matter densities from tidal stripping. In general the UDG status of a galaxy is not contingent upon this effect.}
\label{reff_evol}
\end{figure*}

\begin{figure*}
\centering
\includegraphics[trim=20mm 60mm 5mm 5mm, clip, width=190mm]{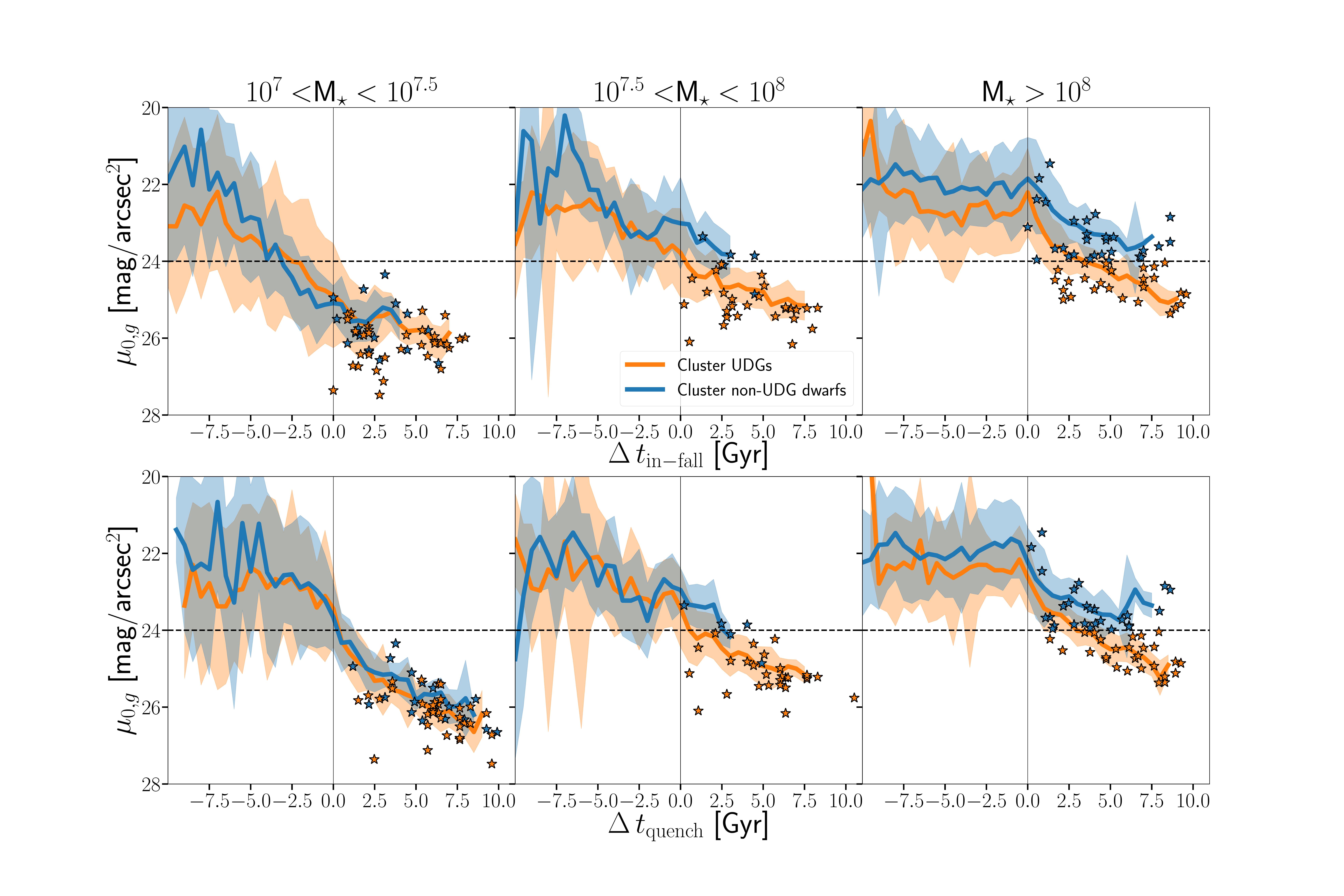}
\caption{{\sc Evolution in $\mu_0$ for cluster dwarf galaxies.} Similar to Figure~\ref{reff_evol} but tracking the evolution in the central surface brightness, $\mu_0$, for cluster dwarf galaxies. Solid lines represent the average evolutionary tracks for times with at least three galaxies and the shaded regions represent the standard deviation at each time. There is a significant evolution in $\mu_0$ with time. This evolution is more closely related to the quenching of star formation than to in-fall into the cluster, though often both are related. Once a galaxy quenches, the passive evolution of its stellar population causes the central surface brightness to drop by orders of magnitude over the course of the next few Gyrs. The low mass bin shows most clearly that this is more related to quenching than in-fall into the cluster. Many low mass galaxies quench long before they cross R$_{200}$. The difference between the leftmost plots on the top and bottom show that a much clearer trend and coherent change in evolution occurs when normalizing the time to $t_{\mathrm{quench}}$ rather than $t_{\mathrm{in-fall}}$.}
\label{mu0_evol}
\end{figure*}

We track several properties back in cosmic time and plot their evolution relative to t$_{\mathrm{in-fall}}$, the first time when each galaxy crosses R$_{200}$, and t$_{\mathrm{quench}}$, the time at which star formation ceases in the inner 1 kpc of each galaxy. Figures~\ref{sersic_evol},~\ref{reff_evol}, and~\ref{mu0_evol} plot such evolutionary tracks for Sersic index ($n$), R$_{eff}$, and central surface brightness ($\mu_0$) respectively. The thick blue and orange lines represent the mean values at each time for non-UDGs and UDGs (at $z=0$), respectively, and the shaded regions the standard deviation at any given time. We present the average time evolution in units of Gyrs, but have also examined the results were we to normalize by the dynamical time at R$_{200}$ following the same methodology as \citet{wang20} and \citet{jiang16}. We find no meaningful difference in the behavior when using this normalization.%The thin lines represent individual galaxies and the stars the values at $z=0$. The mean values are only calculated for times that include at least 3 galaxies. 

Some galaxies fail to track back in time, due to the halo finder failing to recover their progenitor halo during a step. This can happen either due to a close interaction with the cluster center or with another cluster member galaxy that causes the halo finder to be unable to separate the two halos. The galaxies included on these plots are those that trace back to at least $t=2$ Gyrs in simulation time and prior to both in-fall and quenching. These criteria remove $\sim20-30\%$ of dwarf galaxies in each bin. We stress that at earlier times the galaxies are much less massive and so are resolved with fewer particles, making some of the earliest morphological properties uncertain, particularly at our lowest stellar mass bin.%, but does not affect our conclusions. %\amb{In my own manual tracing, I have found you can usually pick up the galaxy again before the encounter.  Not always, but often.  This reads like you throw all those out and don't pick up any at higher z.  Is that right?}
%\mjt{yes, but this is all automated so we would have to do bridging between non adjacent steps... in my experience using BHs as particle tracers to `fix' this issue, it sometimes takes several steps back before the galaxy reappears... this is too difficult to automate for so many galaxies right now.}

%\mjt{\textbf{[CURRENTLY REWORKING THIS SECTION]}}

For Sersic $n$ (Figure~\ref{sersic_evol}) there is little difference in the mean evolution between UDG and non-UDG dwarf galaxies in all mass bins. There is a gradual decrease in $n$ from $\sim2$ at early times to $\sim1$ that is most apparent at low mass. %As time goes on and star formation declines in the centers of these galaxies, the Sersic index becomes lower.

The effective radius (Figure~\ref{reff_evol}) of {\sc RomulusC} dwarfs tends to increase by $\sim20-30\%$ following the quenching of star formation (bottom figures). This increase is likely due to a combination of tidal heating, tidal and ram pressure stripping, and stellar mass loss due to passive stellar evolution. In the highest mass bin, many galaxies that become UDGs already fit our R$_{eff}$ criteria for UDG classification well before quenching or cluster in-fall. At the lowest mass bin, where this size evolution is most important in determining each galaxy's classification as a UDG at $z=0$, UDG progenitors already have typically larger effective radii compared to their non-UDG counterparts long before in-fall or quenching. However, the relatively small increase in size following the quenching of star formation helps many of these low mass galaxies fit our UDG criteria.% Importantly, the progenitor galaxies to UDGs have typically larger effective radii compared to their non-UDG counterparts well before cluster in-fall.}

Substantial evolution is seen in the central g-band surface brightness, $\mu_{0,g}$, for both UDG and non-UDG galaxies in the cluster as shown by Figure~\ref{mu0_evol}. The highest mass dwarf galaxies have much higher surface brightness prior to in-fall into the cluster. The same is true for lower mass dwarf galaxies but many, particularly those in the lowest mass bin, begin to see substantial change well before in-fall. More enlightening is the evolution in $\mu_{0,g}$ with respect to t$_{\mathrm{quench}}$ (lower panels in Figure~\ref{mu0_evol}). Prior to quenching the central surface brightness is relatively constant, but following the shutdown of star formation $\mu_0$ declines steadily with time. This decline corresponds to as much as 1-3 magnitudes/arcsec$^2$ by $z=0$. For all UDGs at all masses this evolution is important for bringing them below the central surface brightness threshold set by our selection criteria. For the lowest mass bins, every galaxy falls below this threshold by $z=0$ (which is why the galaxy size becomes the most important criteria for determining their status as a UDG). For the highest mass bin the opposite is true: all galaxies are large enough to be UDGs so the central surface brightness is the determining factor in their final classification.

%The effective radius (Figure~\ref{reff_evol}) tends to increase with time, particularly at larger masses. Following in-fall into the cluster, the effective radius increases by a typical value of $\sim20-30\%$ compared to its value at in-fall with the most extreme cases increasing by $\sim50\%$ (the same is true when taken relative to t$_{\mathrm{quench}}$). For high mass galaxies, the evolution in size with time is most apparent, but also the least important as all galaxies in the simulation with stellar masses above $10^8$ M$_{\odot}$ have effective radii large enough for UDG classification long before in-fall or quenching. Both UDGs and non-UDGs at $z=0$ have similar evolutionary histories with respect to effective radius at the highest masses. For lower mass galaxies, the effective radius is a critical factor in determining their $z=0$ status as UDGs. For the two lower mass bins, both UDG and non-UDG galaxies increase their effective radius over time, but UDGs are often large enough to be considered UDGs long before they quench or in-fall past R$_{200}$. At any given time galaxies that will become UDGs are more likely to have larger radii compared to galaxies that will not become UDGs in the two lower mass bins.

\begin{figure*}
\centering
\includegraphics[trim=20mm 60mm 5mm 5mm, clip, width=190mm]{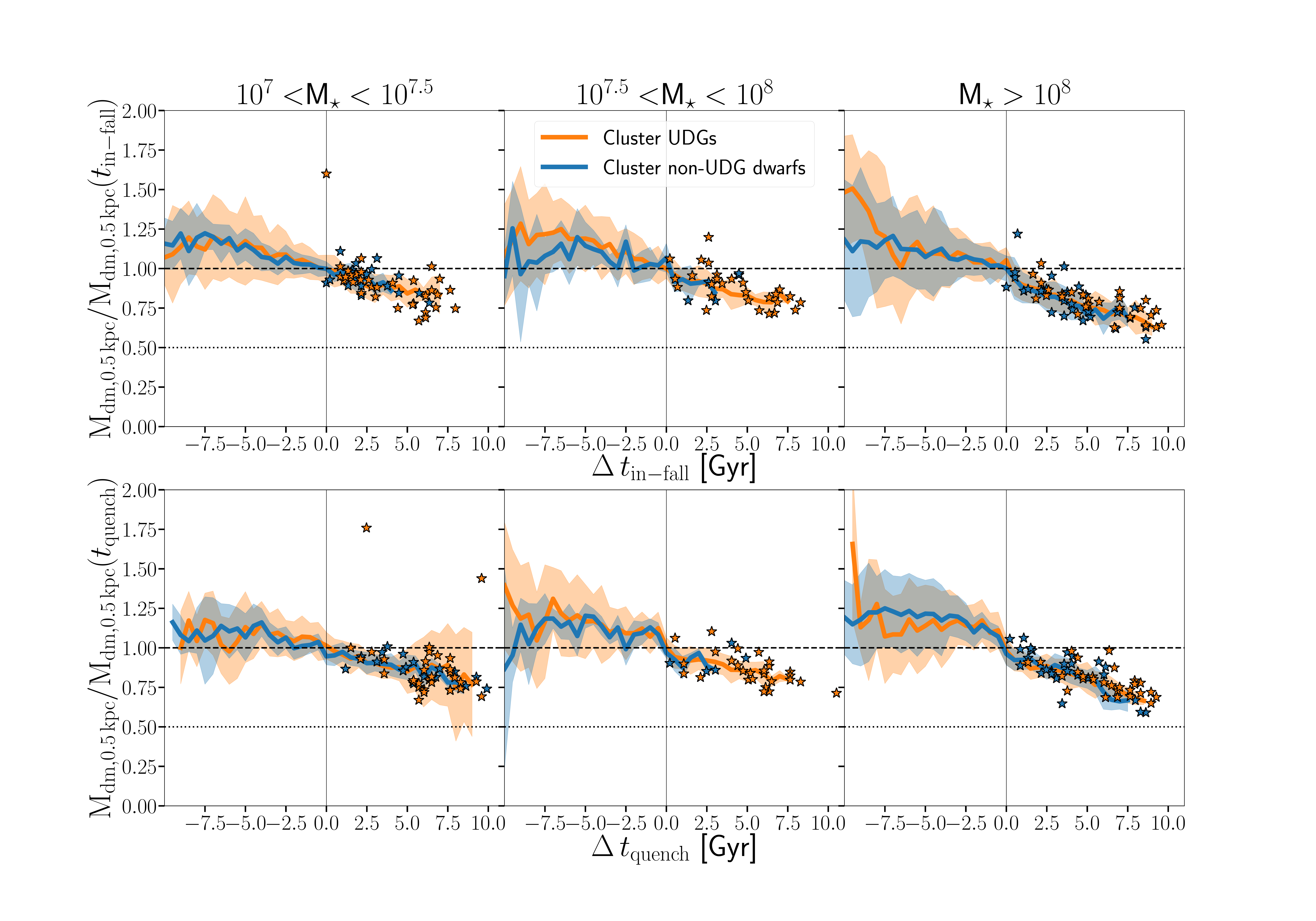}
\caption{{\sc Evolution in the central dark matter density in cluster dwarf galaxies}. Similar to Figures~\ref{sersic_evol},~\ref{reff_evol}, and~\ref{mu0_evol}, but following the evolution of dark matter density within the inner 0.5 kpc of cluster dwarf galaxies. Solid lines represent the average evolutionary tracks for times with at least three galaxies and the shaded regions represent the standard deviation at each time. The decrease in central density is an effect seen in previous simulations that results from tidal stripping. A significant amount of mass within dark matter cores are on radial orbits that can take them out to large distances where they are stripped away \citep{zolotov12}.}
\label{dm_den_evol}
\end{figure*}

\begin{figure*}
\centering
\includegraphics[trim=20mm 60mm 5mm 5mm, clip, width=190mm]{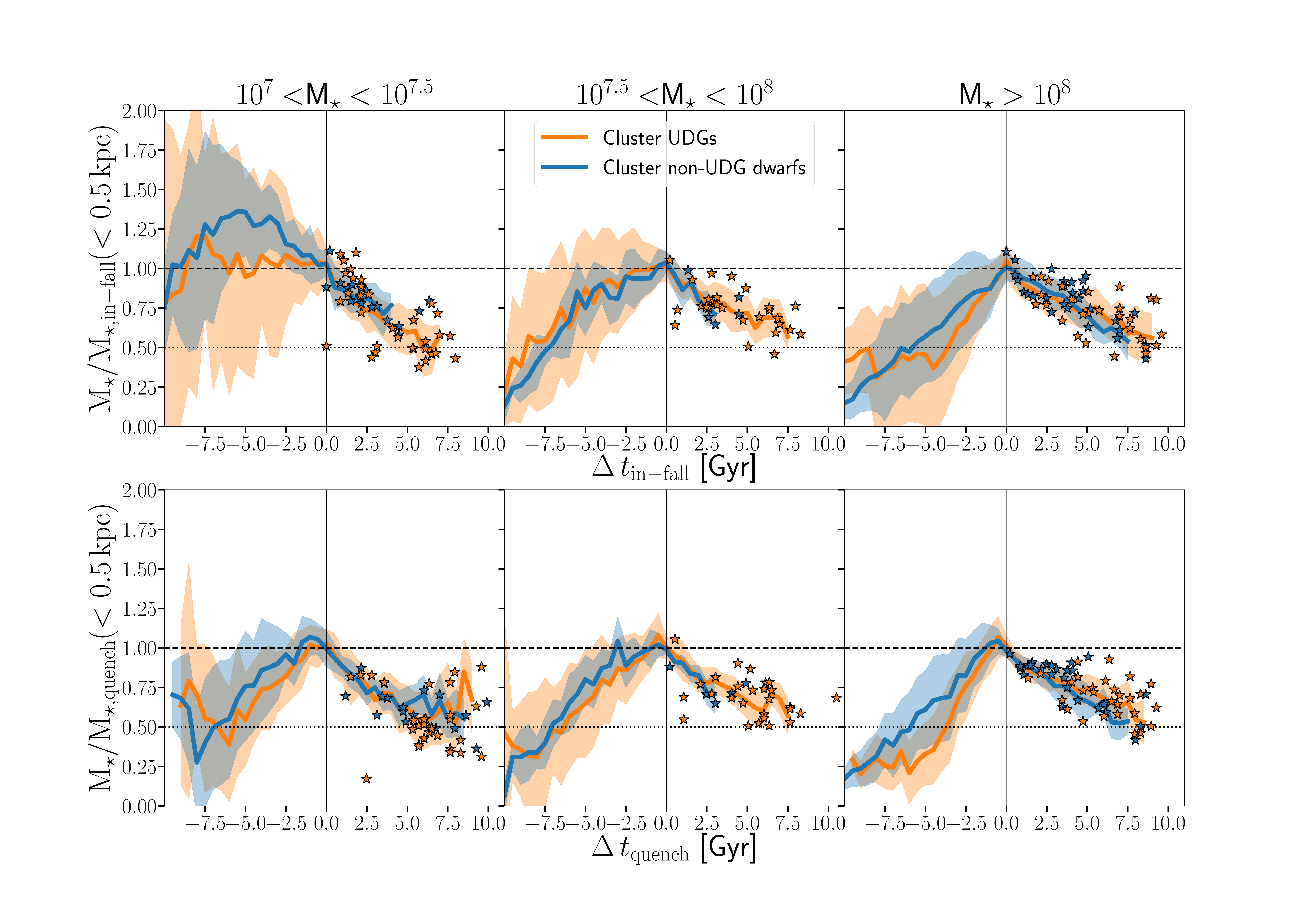}
\caption{{\sc Evolution in the central stellar density in cluster dwarf galaxies}. Similar to Figure~\ref{dm_den_evol} but for the central stellar density inside 0.5 kpc. Solid lines represent the average evolutionary tracks for times with at least three galaxies and the shaded regions represent the standard deviation at each time. Prior to quenching the stellar mass in the center is increasing, as that is where a significant portion of the star formation takes place. After quenching and in-fall into the cluster the stellar density declines by typically $\sim25-50\%$ and as much as 75\%. As stars age in the centers of galaxies that have no star formation, they lose mass due to stellar winds, which are typically 30-40\% of their original mass. Therefore a significant portion of this decline in central stellar density can be attributed to such wind mass loss. Once the galaxy is quenched and fully within the cluster environment, there is no additional star formation and the gas lost to winds is quickly stripped away, rather than recycled to form new stars. Additional effects that contribute to this decrease are the stripping of both gas and dark matter from galaxy centers due to ram pressure and tidal stripping respectively.}
\label{star_den_evol}
\end{figure*}

%\textbf{It is possible that the low surface brightness of cluster UDGs is caused by a decreasing central stellar density. 
%\amb{I would make this one paragraph that focuses on Figure 13 alone, then a new one below focusing on Figure 14.}
In Figure~\ref{dm_den_evol} we explore the evolution of dark matter in the central 0.5 kpc of cluster dwarf galaxies. Dark matter content in the centers of cluster dwarfs decreases by as much as $\sim25\%$ by $z=0$. The evolution in dark matter is due to tidal stripping as the halos interact with the cluster potential, which is why this decline typically begins close to cluster in-fall. This tidal stripping can affect the dark matter in the center of the halo where dark matter particles often have more radial orbits that take them out to larger distances where they are susceptible to these tidal effects \citep{zolotov12}. While this is not extreme enough to form dark matter cores in {\sc RomulusC}, it is able to decrease the density an appreciable amount. Figure~\ref{star_den_evol} shows the evolution in stellar mass in the inner 0.5 kpc of the same galaxies in {\sc RomulusC}. Following cluster in-fall and quenching, the stellar mass decreases. A significant part of this is likely due to stellar evolution, where $\sim20-30\%$ of the mass is removed from stars in the form of winds and SN ejecta. In addition to this, we expect adiabatic expansion to take place in response to the rapid removal of gas and dark matter due to ram pressure and tidal forces respectively \citep{arraki14}.

\begin{figure*}
\centering
\includegraphics[trim=10mm 5mm -25mm 5mm, clip, width=170mm]{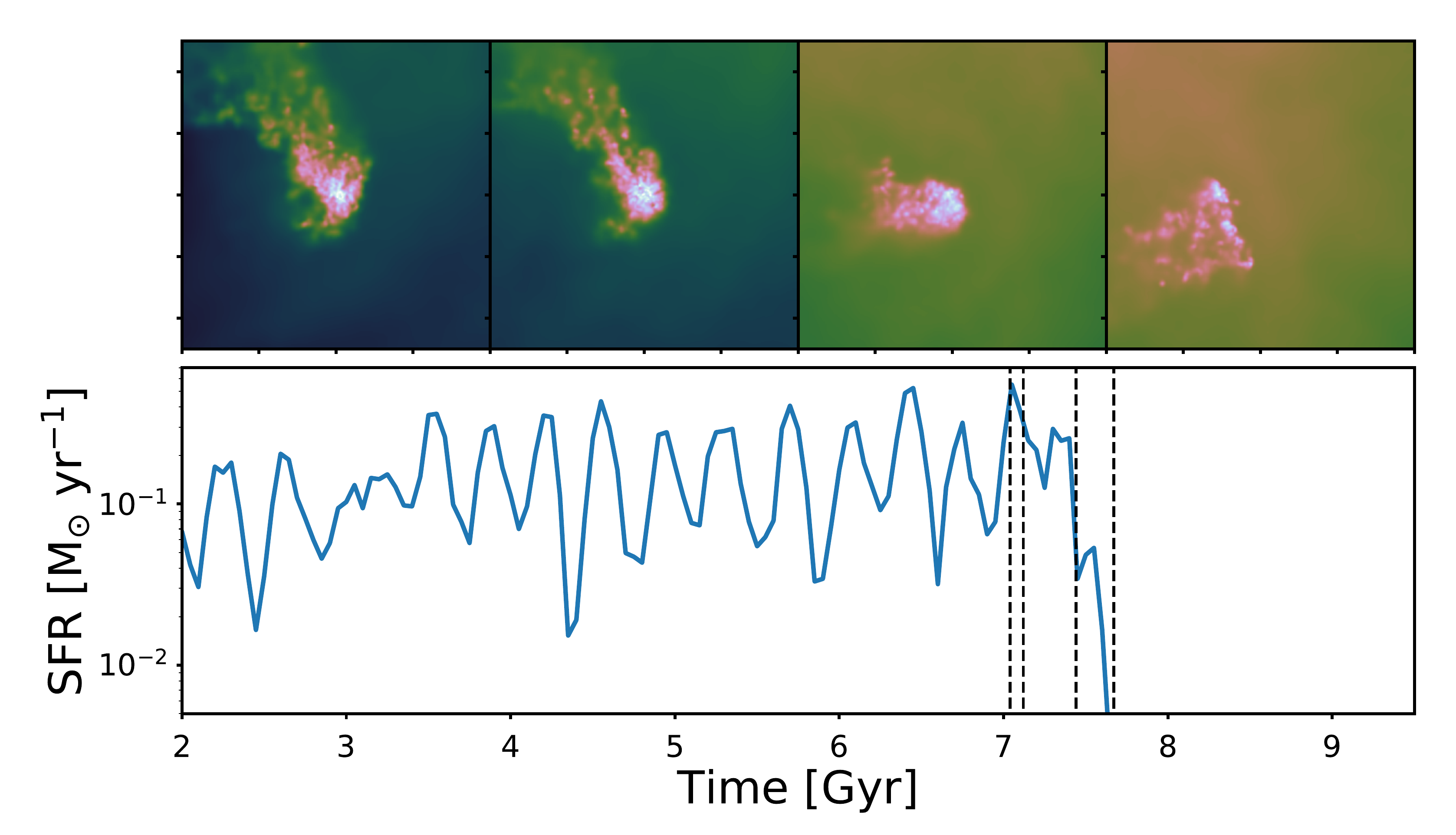}
\includegraphics[trim=18mm 0mm 5mm 25mm, clip, width=170mm]{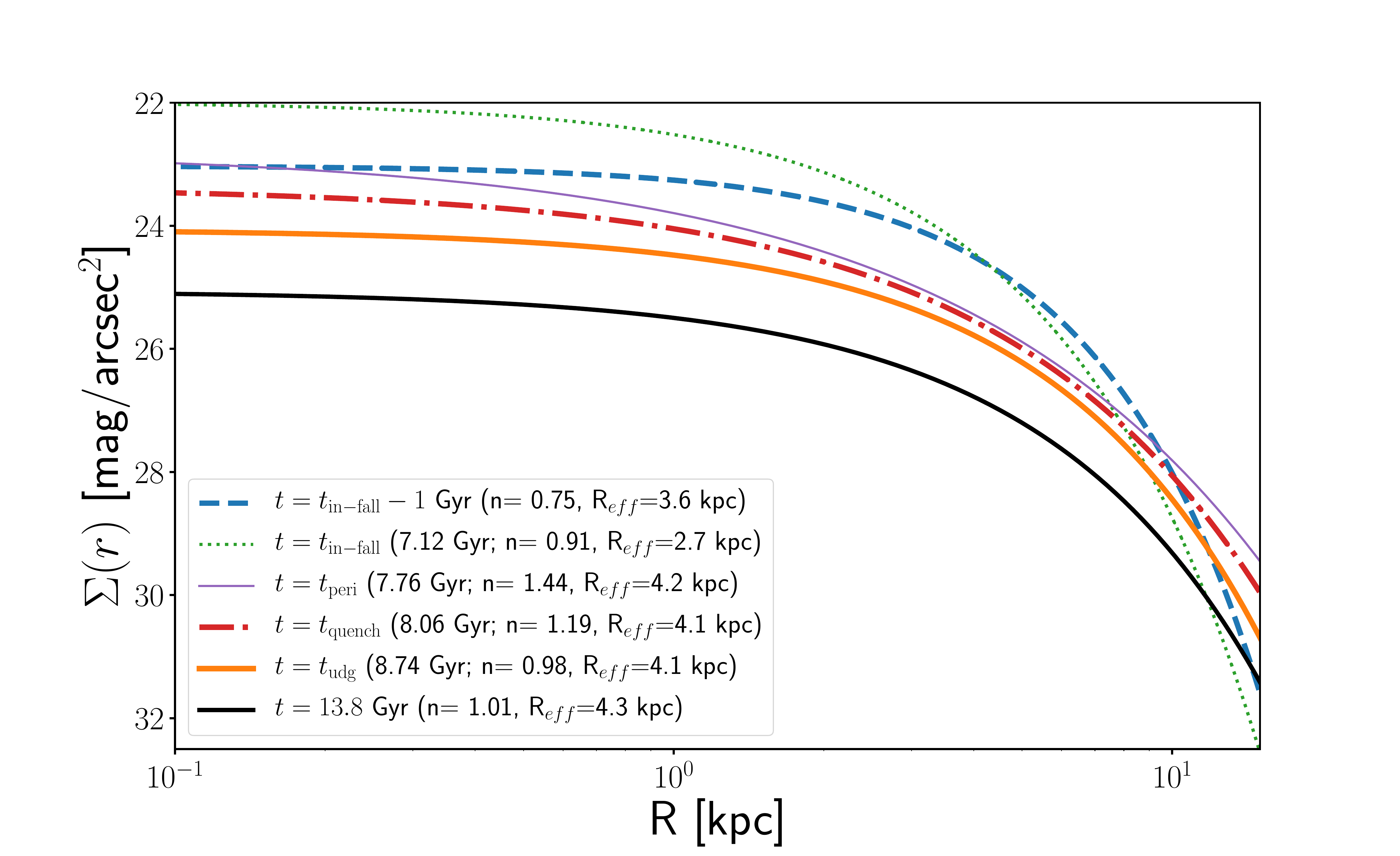}
\caption{{\sc Ram Pressure, Quenching, and the Passive Evolution of Surface Brightness}. Here we show the evolution of an example UDG selected from {\sc RomulusC} for its $z=0$ properties. This example galaxy is a relatively massive (M$_{\star}(z=0) = 10^{8.53}$ M$_{\odot}$) system with otherwise typical morphology for UDGs at this mass in the simulation ($n = 1.01$, $R_{\mathrm{eff}} = 4.3$ kpc, and $\mu_{0,g} = 25$ mag/arcsec$^2$). The top panel shows four snapshots of the gas column density in the galaxy (with the colors on a scale from $3\times10^{19}$ to $5\times10^{21}$ m$_p$/cm$^{2}$) as it is crossing R$_{200}$ for the first time. Tails forming from ram pressure stripped gas are seen in all four panels. The middle panel shows the star formation history of the galaxy as a function of time, with the four snapshots marked as vertical dashed lines. The first snapshot after in-fall, which corresponds to the leftmost image in the top panel, occurs at 7.12 Gyr and within 1 Gyr star formation is quenched in the galaxy, with the last remnants of dense gas being pushed away by ram pressure. The bottom panel plots the surface brightness profile of the galaxy at five different times: 1 Gyr before in-fall (blue, dashed), at in-fall (green, dotted), at pericenter (thin, purple), the time star formation fully quenches (red, dot-dash), the time when the galaxy would first be considered a UDG (orange, solid), and the profile at $z=0$ (black, solid). %\acw{Going in ``rainbow'' order might make interpreting this a little more intuitive.} 
The inner surface brightness profile increases just after in-fall, likely due to the compression from ram pressure pushing more gas to the center to form stars. At pericenter the galaxy increases its effective radius due to tidal effects and soon after star formation is quenched. Following quenching, because the galaxy is no longer forming new stars, the surface brightness continues to drop roughly evenly across all scales until the galaxy would be considered a UDG. As the galaxy passively evolves, the surface brightness decreases further and the galaxy remains ultra diffuse at $z=0$.}
\label{ram_pressure}
\end{figure*}

While the stellar mass in the center of these galaxies does decrease, it doesn't do so by a substantial enough amount to account for the several magnitude decrease in central surface brightness shown in Figure~\ref{mu0_evol}. It can therefore be concluded that the decrease in central surface brightness is rather the result of passive stellar evolution. Figure~\ref{ram_pressure} shows how this process works in more detail for the most massive UDG in our sample. On the top we show the gas column density at four snapshots as the galaxy first crosses inward of R$_{200}$. Below that, we show the evolution of its star formation rate, which declines over the next Gyr and fully quenches at $8.06$ Gyr, approximately 1 Gyr after the initial in-fall. On the bottom we show the evolution of the surface brightness profile at six different times. Before in-fall (blue, dashed line), the galaxy is star forming with a central surface brightness 1 magnitude brighter than the UDG threshold. As the galaxy crosses R$_{200}$ (green, dotted line) ram pressure begins to quench star formation. However, while ram pressure begins to remove gas from the outskirts of the galaxy, it can also compress gas toward the center
%in the outskirts, but it also compresses gas toward the center of the galaxy
\citep{fujita99,bekki03,kronberger08,du19, steyrleithner20}, resulting in a momentary increase in the central surface brightness. This effect can also be seen for UDGs in our highest mass bin in Figure~\ref{mu0_evol}. On average, the effect is less than what is seen in Figure~\ref{ram_pressure}. %We do not attempt to explain here why this increase in central surface brightness occurs around the time the galaxy crosses R$_{200}$ specifically. Such analysis is beyond the scope of this paper and would require a more thorough analysis on ram pressure, which we leave to future work. 
Indeed an in-fall radius of R$_{200}$ is a somewhat arbitrary choice and we leave to future work more detailed investigations of environmental effects on galaxy properties and their dependence on cluster-centric radius. In general, it is reasonable to say that beyond R$_{200}$ galaxies are experiencing significant environment effects. Such influence is not limited to galaxies within R$_{200}$, as lower mass galaxies have their gas removed by ram pressure well outside R$_{200}$. The more massive galaxies like the one in Figure~\ref{ram_pressure}, however, are able to maintain their gas supply long enough to feel the stronger ram pressure within R$_{200}$ including the increase in star formation activity from ram pressure-induced compression before completely quenching. 
By the time the galaxy has reached pericenter (thin, magenta line), star formation has been quenched everywhere but the very center and tidal heating has caused the effective radius to increase rapidly (see \S4.2). The high ram pressure felt by the galaxy at pericenter passage is enough to fully quench it (red, dot-dashed line), leaving the stellar population to age and dim over time. This passive evolution results in a decrease in the surface brightness, eventually causing the already large galaxy to become very low surface brightness such that it would be considered a UDG (orange, solid line) $\sim700$ Myr after quenching and $\sim1$ Gyr after pericenter passage. This passive dimming continues with roughly constant effective radius and Sersic index until $z=0$ (black, solid line).

\begin{figure}
\centering
\includegraphics[trim=20mm 3mm 15mm 25mm, clip, width=90mm]{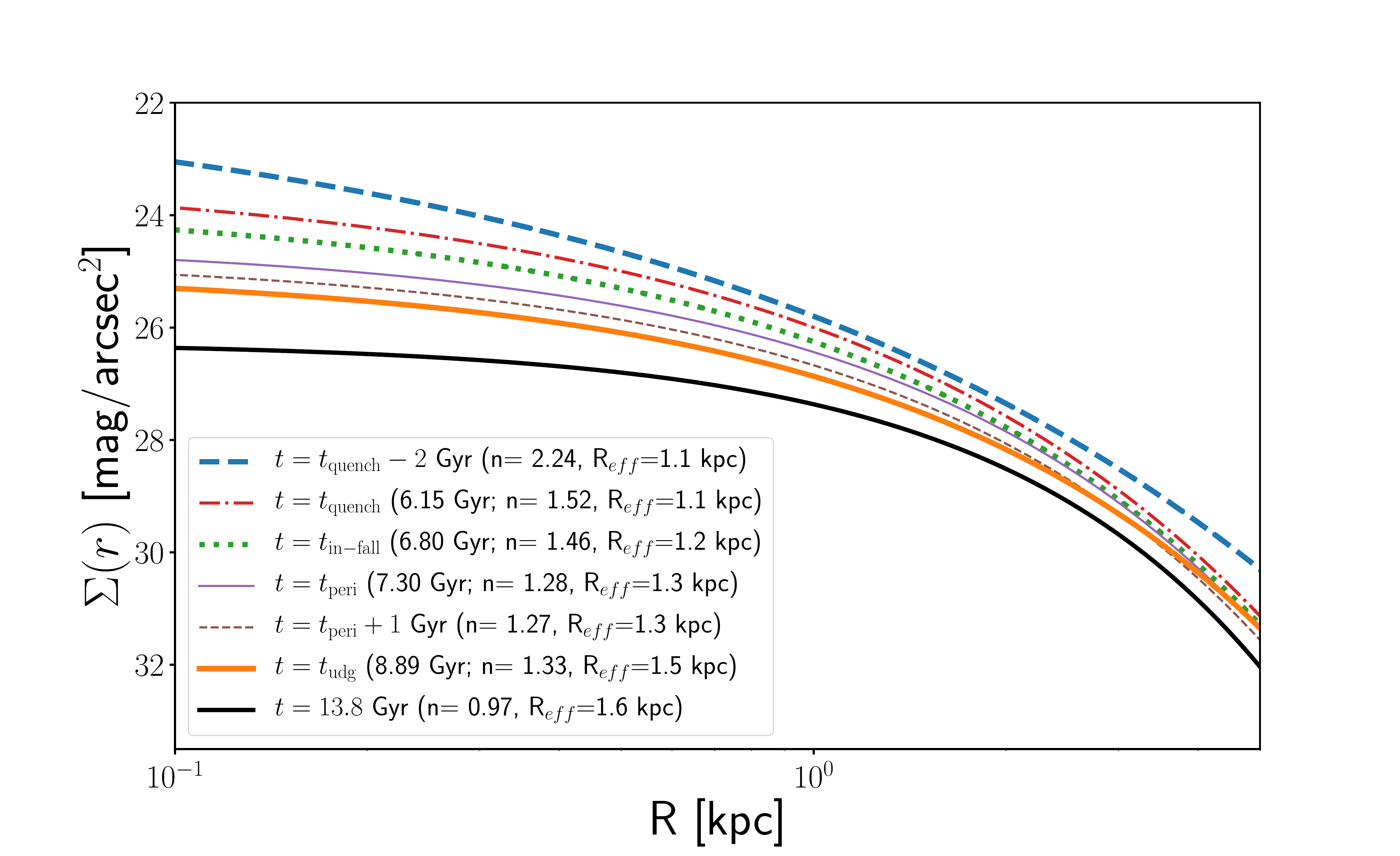}
\caption{{\sc Surface Brightness Evolution for a Low Mass UDG.} Similar to the bottom panel of figure~\ref{ram_pressure} but for a low mass UDG with $z=0$ stellar mass of $1.6\times10^7$ M$_{\odot}$. Unlike the higher mass example, this galaxy quenches $\sim650$ Myr prior to cluster in-fall. After this, the surface brightness decreases continuously through time and the effective radius increases. Also unlike the more massive example, this galaxy experiences less tidal heating at pericenter. Rather than a sharp increase, the effective radius gradually increases by $\sim36\%$ over the course of $\sim2.5$ Gyr following quenching as the galaxy adiabatically expands in response to the loss of both dark matter (tidal stripping) and stellar mass (winds and ram pressure), as shown in Figures~\ref{dm_den_evol} and~\ref{star_den_evol}. This evolution slows down with time as the younger, massive stars leave the main sequence.}
\label{sbprofile2}
\end{figure}

Figure~\ref{sbprofile2} shows another example of the evolution of the surface brightness profile, this time for a low mass dwarf galaxy (M$_{\star} = 1.6\times10^7$ M$_{\odot}$) that is both too bright and too small to be considered a UDG by our definition at in-fall. Low mass galaxies typically quench more easily and do not require larger amounts of ram pressure at pericenter. Rather, in this example the galaxy quenches several hundred Myr prior to cluster in-fall. A couple Gyrs prior to quenching (blue, dashed line) the galaxy fails both central surface brightness and effective radius criteria for being a UDG. As star formation quenches, the surface brightness decreases and the size begins to gradually increase over time. At pericenter (thin, purple line) the galaxy does not undergo a dramatic increase in effective radius as seen in the more massive galaxy. %\amb{A larger radius for pericenter could also lead to less effect, so can you compare the radii that each has pericenter?}\mjt{I don't really understand what you mean. What I hoped to convey with this figure is that the radius is gradually increasing beginning at the time of quenching. It is the rate of change of effective radius. This galaxy is smaller because it has a smaller mass.} \amb{I mean that maybe larger dwarfs have smaller pericenters (due to dynamical friction acting more on them), which is what makes tidal heating more relevant for them than the smaller guys. In that case, comparing Figures 15 and 16 might not be a fair comparison.  Keeping the new figure is good, I'm just not sure it's an apples-to-apples comparison with Fig 15.  I may be wrong, of course, since you've showed UDGs as a group tend to have smaller pericenters than non-UDGs, and you show below that size evolution in the low mass guys is fairly gradual.  Maybe make Figure 8 for the different mass bins and see if there's any trends?} 
Rather, the galaxy continues to gradually increase in effective radius until it becomes large enough and dim enough to be considered a UDG (orange line). In total, the radius increases by 36\% between quenching and 2.7 Gyr later when it would be considered a UDG. During the next $\sim5$ Gyr the effective radius  increases by less than 10\% (black line; $z=0$). During the few Gyr following quenching, stars will lose mass to winds which will then be removed from the relatively shallow potential well of the low mass galaxy by ram pressure. The resulting mass loss, combined with the tidal stripping of dark matter, will result in adiabatic expansion of stars and a gradual increase in effective radius \citep{arraki14}. % \amb{I note that in this galaxy's case, it's the Reff expansion (with quenching) that makes it a UDG at z=0. Based on Figure 11, while may of these UDGs are already larger than most dwarfs, it's the time after infall/quenching that really pushes the mean over 1.5 kpc. So I agree that overall is't passive evolution that is the driving force, but that expansion really matters in this mass bin, too.}\mjt{I completely agree. showing this example is good, I agree with the referee. But here where it matters we do not see a rapid incrase in effective radius as we would expect for tidal heating at pericenter, and which we see at higher masses.}

\begin{figure}
\centering
\includegraphics[trim=20mm 3mm 15mm 25mm, clip, width=90mm]{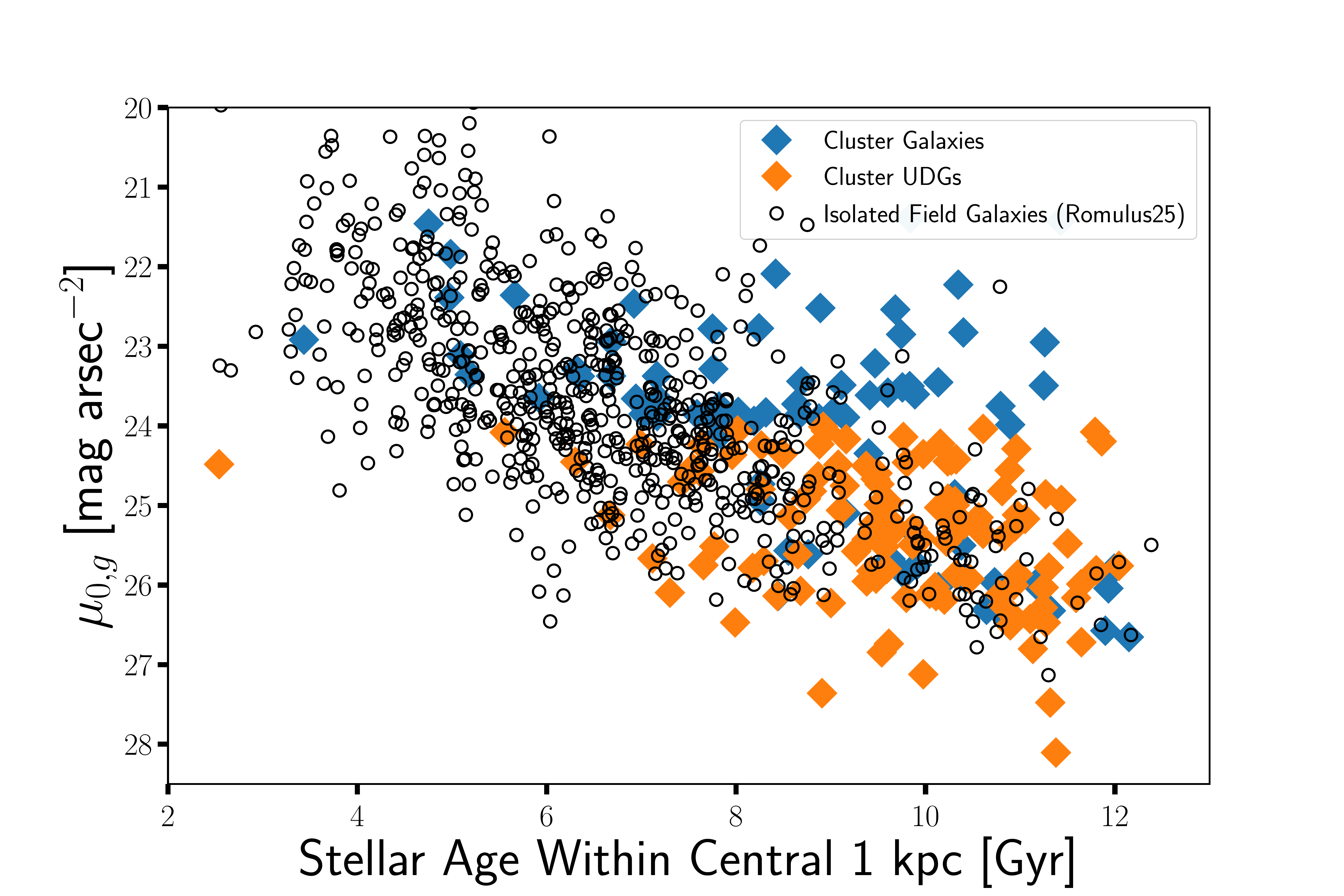}
\caption{{\sc Effect of passive evolution on central surface brightness.} The relationship between central surface brightness and the mass weighted stellar age in the inner 1 kpc of a galaxy for cluster non-UDG dwarf galaxies (blue diamonds), cluster UDGs (orange diamonds), and isolated dwarf galaxies (black open circles). There is a clear trend that galaxies with older stellar populations in their centers have lower central surface brightness. This is true for both cluster and isolated galaxies in the {\sc Romulus} simulations. Galaxies in both environments follow the same trend, though most cluster galaxies are older while isolated dwarf galaxies tend to have younger stars in their centers.}
\label{age_mu0}
\end{figure}

Passive evolution is not unique to the cluster environment, as shown in Figure~\ref{age_mu0}. Central surface brightness values of galaxies in isolation follow the same trend with stellar age as galaxies in {\sc RomulusC}. Galaxies with older stellar populations at their centers have lower central surface brightness. The cluster environment is important mainly as a way to quench star formation in low mass galaxies, which are generally star forming in isolation.

\subsection{Tidal Interactions and UDG Formation}

%\amb{I'm not so convinced by this next section.  It seems to me that, while tidal heating isn't important for the most massive bin (since they are already big), it can really contribute in the two lower mass bins.  Perhaps we should also discuss what is meant by tidal heating. Does it have to be a fast expansion at peri?  Why not the slower expansion due to mass loss after infall?}

In the previous section, as well as Figure~\ref{smhm}, we discuss how dwarf galaxies lose dark matter after cluster in-fall due to tidal stripping and that this mass loss not only occurs in the outskirts of satellite dwarf halos but also in their centers. In addition, progenitors to UDGs have preferentially earlier cluster in-fall times and get closer to cluster center relative to non-UDGs (see \S3.2). This early in-fall and close approach to cluster core may also make UDG progenitors more susceptible to tidal heating from the cluster potential. In this section we examine in more detail the effect of tidal interactions with the cluster potential, in particular the effect of a galaxy's first pericenter passage on its morphological evolution. Again, we present the average time evolution in units of Gyrs, but have also examined the results normalized by the dynamical time following \citet{wang20} and \citet{jiang16}, but find no meaningful difference when this normalization is used.

\begin{figure*}
\centering
\includegraphics[trim=20mm 50mm 5mm 5mm, clip, width=190mm]{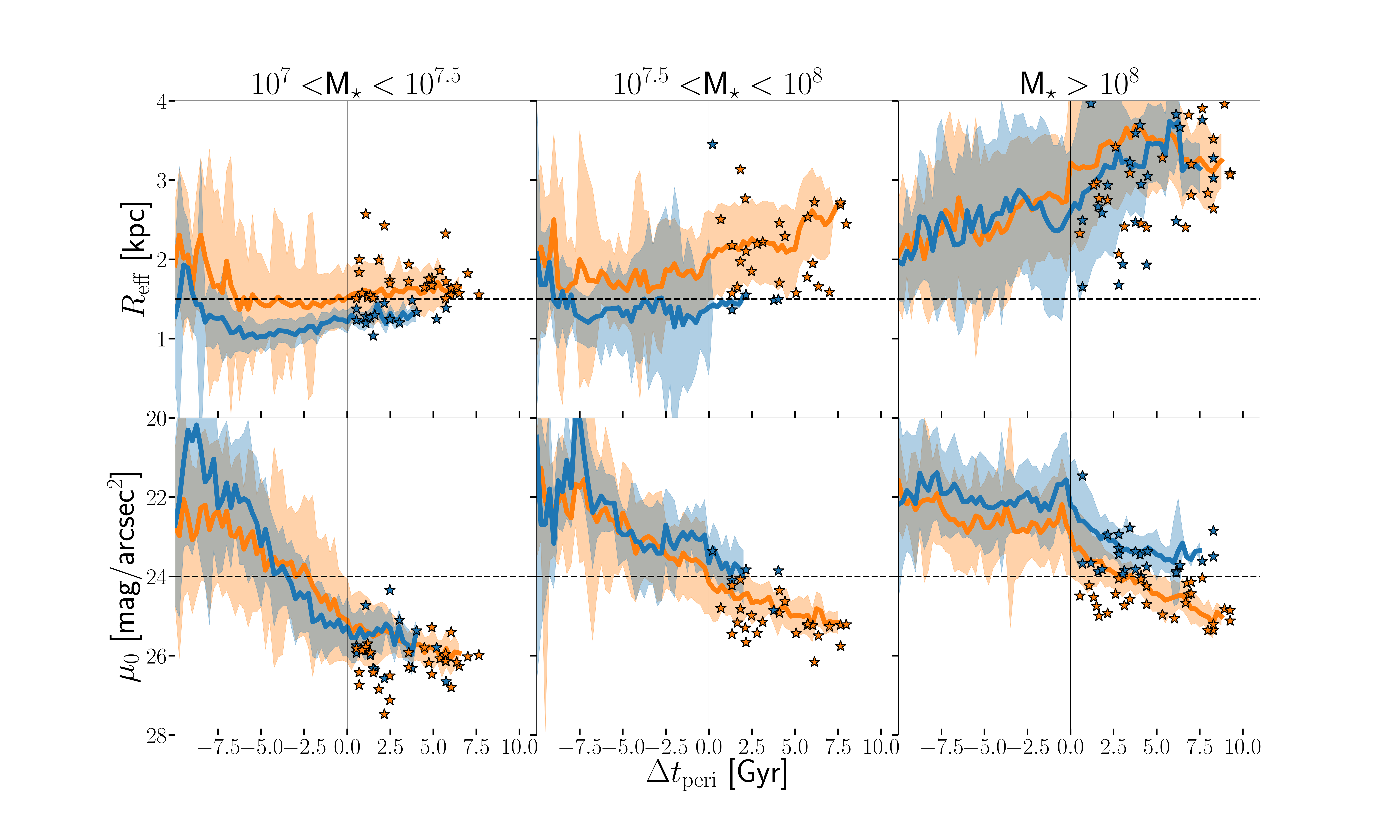}
\caption{{\sc Effect of pericenter passage on galaxy size and central surface brightness. Similar to Figures 10-14, but following the evolution of R$_{\mathrm{eff}}$ (top) and $\mu_0$ (bottom) for each galaxy relative to the time of pericenter passage. Solid lines represent the average evolutionary tracks for times with at least three galaxies and the shaded regions represent the standard deviation at each time.} At higher masses, there is evidence that tidal forces at pericenter are causing the effective radius to increase by $\sim20-30\%$ within a few hundred Myr. At low masses such a rapid change is not seen, with a gradually increasing effective radius starting Gyr prior to pericenter passage.}
\label{peri_reff_mu}
\end{figure*}

The top panels of Figure~\ref{peri_reff_mu} follow R$_{\mathrm{eff}}$ as a function of time relative to first pericenter passage. In the two more massive bins there is typically an increase in effective radius at pericenter of $\sim20-30\%$, consistent with previous results examining UDG evolution in group environments \citep{jiang19}. However, the evolution in effective radius at low masses, where the evolution is more critical to their final classification, is much more gradual, with the increase beginning around the time of quenching, rather than at pericenter passage. While the rapid increase in effective radius at higher masses is likely the result of tidal heating, the gradual increase seen at smaller masses is more consistent with adiabatic expansion in response to decreasing dark matter and stellar mass in the centers of galaxies (Figures~\ref{dm_den_evol} and~\ref{star_den_evol}).

%Dark matter is primarily due to tidal stripping \citep{zolotov12,arraki14}, while th

%is more closely tied to quenching (Figure~\ref{reff_evol}) than it is to pericenter passage. Rather than a quick increase in effective radius indicative of tidal heating, the gradual increase in size occurs mostly within the few Gyrs following quenching. This is indicative that the size evolution is indeed due to the gradual mass loss of stars and dark matter (Figures~\ref{dm_den_evol} and~\ref{star_den_evol}).}

\begin{figure*}
\centering
\includegraphics[trim=20mm 50mm 5mm 5mm, clip, width=190mm]{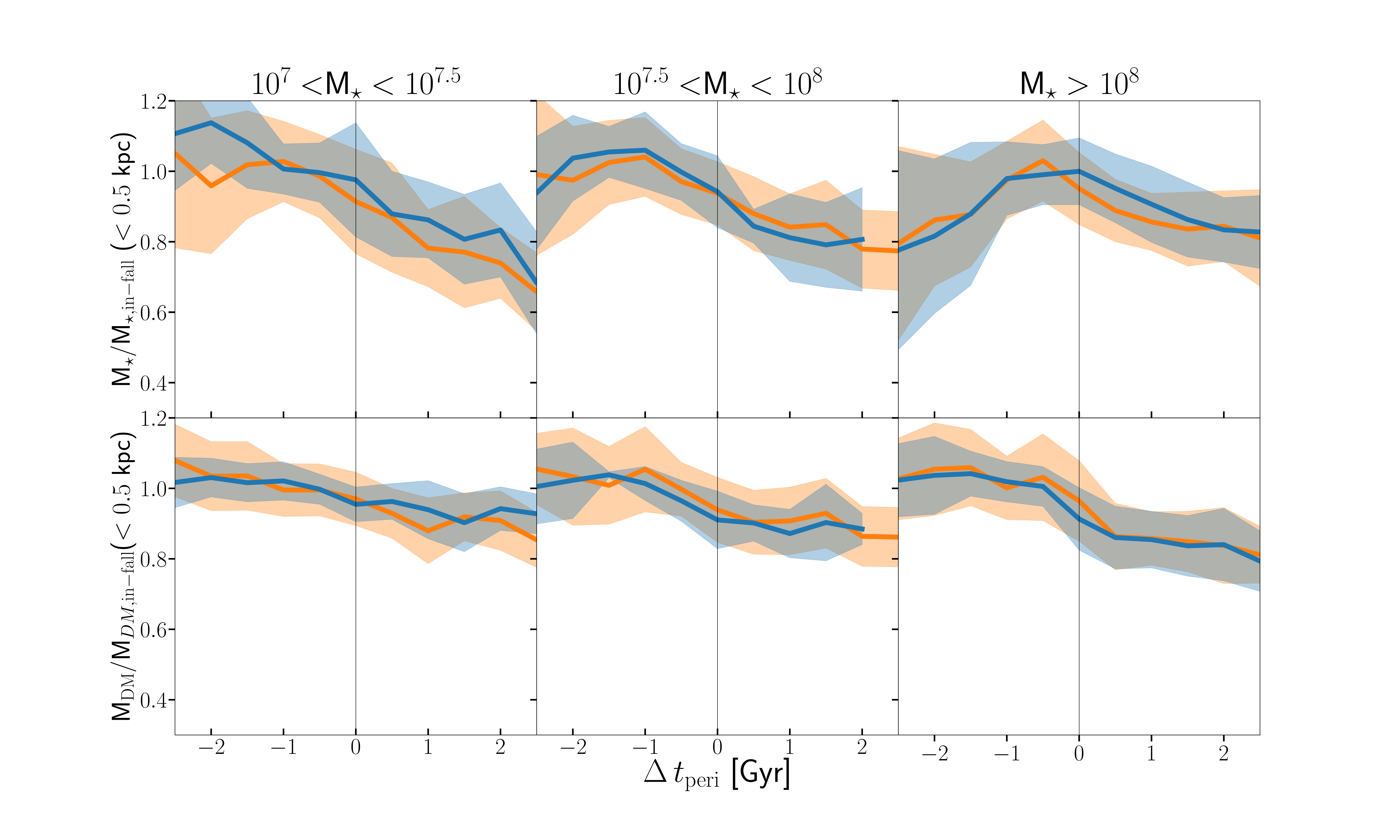}
\caption{{\sc Effect of pericenter passage on dark matter and stellar mass distributions.} Similar to Figure~\ref{peri_reff_mu}, but following the evolution of dark matter and stellar mass within inner 0.5 kpc of each galaxy relative to the time of pericenter passage. Solid lines represent the average evolutionary tracks for times with at least three galaxies and the shaded regions represent the standard deviation at each time. If significant tidal heating were taking place, we would expect to see significant re-distribution of mass within the galaxy, with the core becoming more diffuse. The values are shown with respect to the maximum values throughout the simulation. There is some evidence of a decrease for the middle mass bin UDGs and for the lowest mass bin non-UDGs around pericenter, but the effect is small relative to the scatter and there is no evidence that overall UDGs are more likely to experience strong tidal heating from the cluster potential compared to non-UDGs.}
\label{peri_mdm_mst}
\end{figure*}

Figure~\ref{peri_mdm_mst} plots the evolution in stellar and dark matter mass within 0.5 kpc of galaxy center relative to pericenter passage. Unlike the other figures, we zoom into the 5 Gyrs bracketing pericenter passage for all galaxies. At high mass, we see some evidence for a decrease in both stellar mass and dark matter mass as the galaxy experiences its first pericenter passage, indicative of both tidal heating and enhanced tidal stripping as the galaxy interacts with the higher density near the center of the cluster. At low masses, however, there is little evidence of a decrease in central density associated with pericenter passage. Rather, there is a gradual decline in both stellar mass and dark matter mass in the central 0.5 kpc over time beginning prior to pericenter passage and continuing well after.

Figure~\ref{peri_reff_mu} also shows the evolution in central surface brightness relative to pericenter passage (bottom panels). The majority of the more massive dwarfs quench close to their first pericenter passage and, consistent with Figure~\ref{mu0_evol} in the previous section, the central surface brightness begins to decline after this point. Lower mass galaxies have their central surface brightness already well in decline at the time of pericenter passage, consistent with the fact that these galaxies typically quench $>1$ Gyr prior to pericenter passage.

While the size evolution due to tidal heating helps determine the final size of the more massive dwarfs, this evolution does not in the end determine their classification as UDGs. The classification of low mass dwarfs is more sensitive to their final size, which we have shown in this section to not have strong effects, on average, from tidal heating. At no mass scale is there evidence that UDGs and non-UDGs experience different evolution in their size or central surface brightness.

\subsection{Angular momentum and galaxy sizes}

%\amb{Is this paragraph necessary?  I feel like it could be cut and we could open this section with the next paragraph.} \dn{Agree with AMB that removing this par can help tighten up this section a bit more.} We have established in the previous section that the sizes of dwarf galaxies increase slightly with time, but such processes are not substantially important for determining the UDG population at $z=0$. However, Figures~\ref{size_mass} and~\ref{morph_compare} show that, compared to field galaxies, cluster dwarfs are more likely to be large, with the cluster lacking dwarfs of sizes below 1 kpc that exist in isolation. This makes galaxies in the cluster environment inherently more likely to become UDGs. In \S5 we discuss the extent to which the lack of small galaxies at the lowest masses may be a resolution effect, but we stress that both {\sc RomulusC} and {\sc Romulus25} have the same resolution and physics and so we can self-consistently compare isolated and cluster dwarf galaxies to better understand why their sizes could be different. 

Previous work has theorized that angular momentum is important for determining the sizes of galaxies and that low surface brightness galaxies reside preferentially in halos with larger spin \citep[e.g.][]{dalcanton97, dicintio19}. A similar argument has been made with respect to the origin of UDGs \citep[e.g.][]{amoriscoloeb16}. To study the effect of angular momentum on the sizes of galaxies in cluster and field environments in the {\sc Romulus} simulations, we use the dimensionless spin parameter, $\lambda'$ \citep{bullock01a}. AHF calculates this parameter for each halo in the simulation at all times, accounting for the angular momentum of all halo particles ($\lambda'_{\mathrm{tot}}$) as well as just gas particles ($\lambda'_{\mathrm{gas}}$). This dimensionless parameter is convenient because it is redshift independent, allowing us to directly compare the spins of halos and gas at different times. However, it is also important to take into account the fact that cluster galaxies are subject to  environmental effects that strip away dark matter, as well as all gas, from low mass galaxies. It is reasonable that the final halo spins may not fully represent the effect that angular momentum has on the halo as the galaxy is forming most of its stars. In Figure~\ref{reff_evol} we show that lower mass dwarf galaxies that are larger at $z=0$ (i.e. UDGs) are also larger at earlier times. In order to capture a potential trend between angular momentum and the formation of dwarf galaxies, we use values for $\lambda'_{\mathrm{tot}}$ and $\lambda'_{\mathrm{gas}}$ calculated at $t_{50}$, the time at which 50\% of the stars have formed in the galaxy. By doing this we sample the state of the halo and potential star forming gas during a period when the galaxy is still growing and ensure that the gas and dark matter have not yet been significantly stripped by the cluster environment. 

\begin{figure*}
\centering
\includegraphics[trim=100mm 60mm 5mm 5mm, clip, width=190mm]{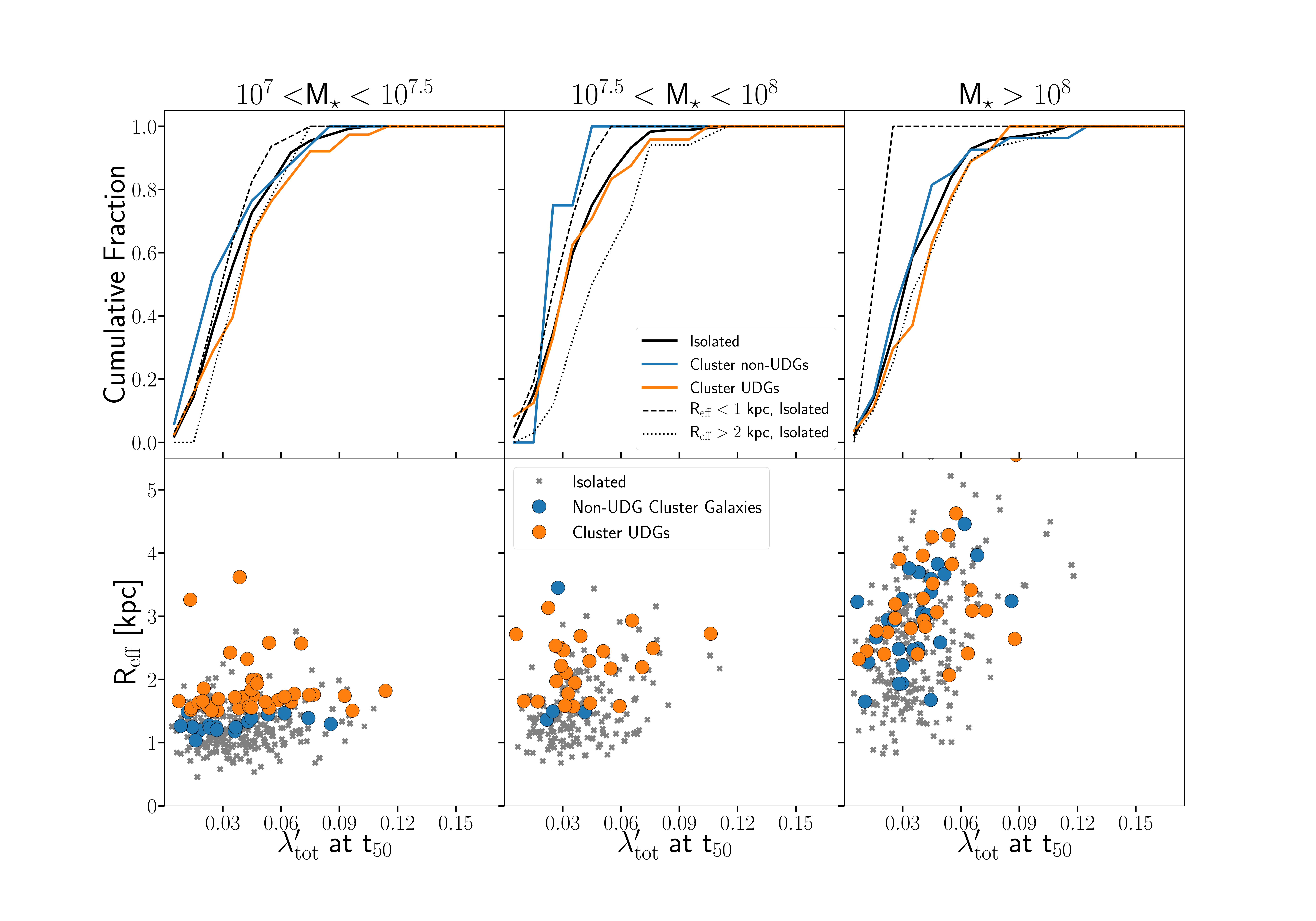}
\caption{{\sc Total spin parameter at $t_{50}$}. \textit{Top:} The distribution of total halo spin at $t_{50}$, the time at which 50\% of the stars formed in the galaxy, for non-UDG cluster dwarf galaxies (blue) and for cluster UDGs (orange). Also plotted for each mass bin are the spin distributions for isolated dwarf galaxies from {\sc Romulus25} (black, solid). Additionally, we plot the distribution of spin for isolated galaxies that, at $z=0$, have large (R$_{eff}>2$ kpc) and small (R$_{eff}<1$ kpc) sizes. There is not a significant difference in the spin distributions of UDGs and non-UDGs and both are similar to the overall spin distribution of isolated dwarf galaxies. \textit{Bottom:} Effective radius as a function of total spin parameter at $t_{50}$ for non-UDG cluster galaxies (blue), UDGs (orange), and isolated dwarf galaxies (grey). At low mass there is no significant dependence of final size on the spin parameter at $t_{50}$, but there is at higher masses.}
\label{lambda_tot}
\end{figure*}

The top panels of Figure~\ref{lambda_tot} plot the distributions of $\lambda'_{\mathrm{tot}}$ at $t_{50}$ for non-UDG cluster dwarfs (blue) and for UDGs (orange) in our three mass bins. We also show the distribution for the population of isolated dwarf galaxies in {\sc Romulus25} (black, solid). While UDGs do tend to have slightly higher spin values than non-UDGs, the difference is small and both are similar to the overall population of isolated dwarf galaxies.  The lower panels of Figure~\ref{lambda_tot} are scatter plots showing the relationship between the final effective radius and the spin parameter measured at $t_{50}$. Galaxy sizes tend to increase toward larger spin parameters, with a stronger relationship at higher masses, but the correlation is relatively weak as the halos do not have a large diversity of spin parameters within any mass bin. Controlling for spin parameter, cluster dwarf galaxies are still typically larger, lacking the  smaller dwarf galaxies present in isolation in all three mass bins. 

%There is virtually no difference in the distribution between UDGs, the total cluster dwarf population, and isolated dwarf galaxies. We also show the distribution of spin parameter for isolated dwarfs that are relatively large (R$_{eff} > 2$ kpc) and small (R$_{eff}$ < 1 kpc). The lower panels of Figure~\ref{lambda_tot} are scatter plots showing the relationship between the final effective radius and the spin parameter measured at $t_{50}$. Galaxy sizes do tend to increase toward larger spin parameters, with a stronger relationship at higher masses. Controlling for spin parameter we see that still cluster dwarf galaxies are typically larger, lacking the population of smaller dwarf galaxies present in isolation at all spin parameters. 

%split the population of isolated dwarfs based on their final effective radii, finding some preference at higher masses for larger spin parameters in galaxies that are bigger at $z=0$. The bottom panels are scatter plots of the effective radius as a function of $\lambda_{tot}$. The cluster galaxies follow similar relationships as isolated galaxies. Some dependence on $\lambda$ is apparent at higher masses, while low mass galaxies are essentially flat in size with respect to the spin parameter. 

\begin{figure*}
\centering
\includegraphics[trim=100mm 60mm 5mm 5mm, clip, width=190mm]{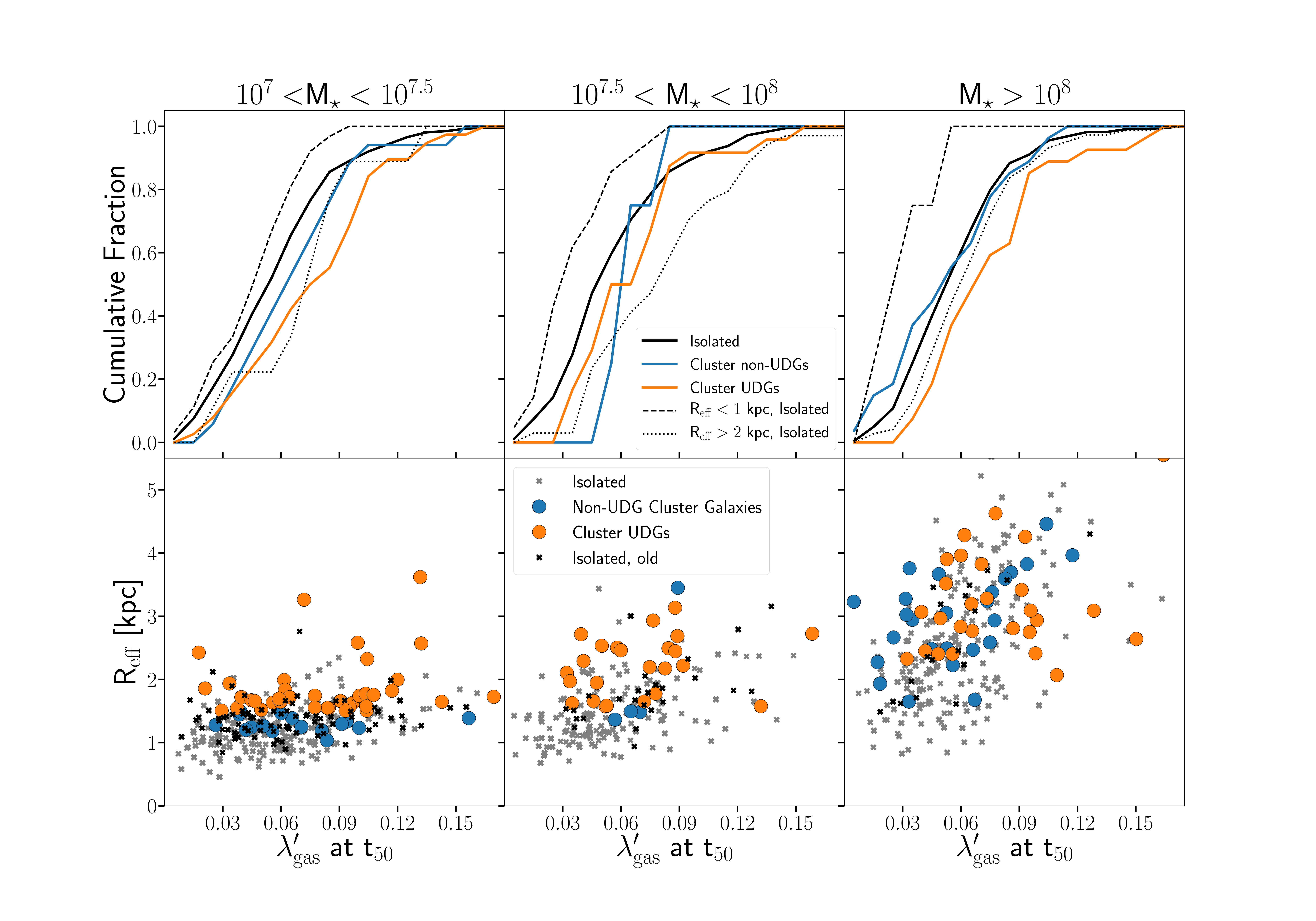}
\caption{{\sc Gas spin parameter at $t_{50}$}. The same as Figure~\ref{lambda_tot}, but for the spin just of the gas in each dwarf galaxy halo at $t_{50}$. In all mass bins UDGs are more likely to have high spin compared to non-UDG dwarf galaxies. In the two lowest mass bins, the non-UDG cluster dwarf galaxy population also has higher spin compared to  the overall isolated dwarf population in {\sc Romulus25}. The difference is more pronounced when compared to the spin distribution of compact (R$_{eff}<1$ kpc; dashed) isolated dwarf galaxies. Low mass galaxies with low angular momentum gas make up a substantial part of the compact isolated dwarf galaxy population and these galaxies are less likely to exist in the cluster environment. The bottom panels show a slight dependence of R$_{eff}$ at $z=0$ on $\lambda'_{gas}$ at $t_{50}$ becoming more important for higher mass dwarfs. The trend is more pronounced compared to $\lambda'_{tot}$ (Figure~\ref{lambda_tot}) and the halos have more diversity in gas spin compared to total spin. At a given $\lambda'_{gas}$ value the cluster environment still lacks compact dwarfs relative to the field. This difference goes away if we consider only isolated galaxies with old ($>7 Gyr$) stellar populations in their central 1 kpc (black points). Both gas angular momentum and a lack of recent star formation in their centers contribute to the lack of compact cluster dwarf galaxies in {\sc RomulusC}.}
\label{lambda_gas}
\end{figure*}

Figure~\ref{lambda_gas} is similar to Figure~\ref{lambda_tot} but for $\lambda'_{\mathrm{gas}}$, the spin parameter calculated only for gas particles, at t$_{50}$. Compared to total spin, there is a more significant difference in the gas spin distribution for isolated compact (R$_{\mathrm{eff}}<1$ kpc; dashed, black) and large (R$_{\mathrm{eff}}>2$ kpc; dotted, black) dwarf galaxies in all three mass bins. This can also be seen in the scatter plots on the bottom, where there is a more pronounced dependence of R$_{\mathrm{eff}}$ on $\lambda'_{\mathrm{gas}}$ (and overall more diversity among galaxies at all mass bins). Similar to total spin, the UDGs have a slight tendency to have higher gas spin values than non-UDGs in the cluster. At the two lower mass bins, even non-UDGs are more likely to have higher gas angular momentum compared to isolated dwarfs. This lack of low angular momentum gas at low masses can explain, in part, the lack of compact dwarf galaxies in {\sc RomulusC}. Galaxies with low spin parameters ($\lambda'_{\mathrm{gas}} \sim 0.03$) and M$_{\star} < 10^8$ M$_{\odot}$ make up a large portion of the most compact isolated dwarf galaxies. %This lack of low angular momentum gas while cluster dwarfs are still forming their stars can contribute to the lack of compact dwarfs in {\sc RomulusC} compared to {\sc Romulus25}.

%the lack of very low angular momentum gas in cluster dwarfs can contribute to the lack of compact dwarfs in {\sc RomulusC} compared to {\sc Romulus25}.

Still, the scatter plot on the bottom of Figure~\ref{lambda_gas} shows that even controlling for gas spin, isolated galaxies tend to be smaller than cluster galaxies. This difference goes away, however, if we control for the fact that cluster dwarfs are preferentially quenched with much older stellar populations. The black points on the scatter plot show that, for a given gas spin parameter at t$_{50}$, isolated dwarf galaxies with older stellar populations in their centers (a mass-weighted age of $>7.5$ Gyrs within the inner 1 kpc) have larger sizes, more comparable to cluster dwarf galaxies. While gas angular momentum may contribute to the lack of small galaxies, particularly at low mass, the quenching of star formation also results in dwarfs with larger effective radii. This is likely a combination of 1) younger central stellar populations being brighter and pushing the effective radius inward and 2) wind mass loss and dark matter tidal stripping decreasing the inner mass density over time and puffing the galaxy up, as we describe in \S4.1. This may also be partially due to a resolution effect which we discuss further in \S5.%, in particular why this is likely only a secondary effect.

\section{Discussion}

In this section we will review
%reiterate and discuss 
our results in light of other observational and theoretical work and discuss various numerical effects that may have influenced our results. %and examine the various ways in which our results are subject to both resolution effects and our exact choice of UDG selection.

\subsection{The role of the cluster environment in the formation of UDGs}

%Ultra diffuse galaxies
UDGs in the {\sc RomulusC} simulation are formed through interactions with their dense cluster environment. A combination of tidal stripping and heating result in a typical increase in effective radius of $\sim30\%$ after in-fall into the cluster, consistent with previous simulations presented in \citet{jiang19}, but only at the lowest masses does this increase in radius affect the final UDG classification. The cluster environment is crucial in shutting off star formation in low mass galaxies via ram pressure stripping (as seen in Figure~\ref{ram_pressure}). \citet{jiang19} also find that ram-pressure, rather than tidal stripping, is the main mechanism through which gas is removed from low mass galaxies in groups. Star formation in the more massive dwarfs often remains present longer ($\sim1-2$ Gyr) after in-fall, typically until first pericenter passage, but lower mass galaxies can quench even before cluster in-fall.
%\textbf{While tidal effects from the cluster potential do play a role in their final properties (particularly at high mass; \S4.2), contrary to previous work \citep[e.g.][]{ogiya18,carleton19} we do not find such tidal interactions to be the contributing factor for a galaxy to be classified as a UDG at $z=0$.} Rather, we find that the cluster environment is crucial in shutting off star formation in low mass galaxies via ram pressure stripping (as seen in Figure~\ref{ram_pressure}). \citet{jiang19} also find that ram-pressure, rather than tidal stripping, is the main mechanism through which gas is removed from low mass galaxies in groups. Star formation in the more massive dwarfs often remains present longer ($\sim1-2$ Gyr) after in-fall, \textbf{typically until first pericenter passage, but lower mass galaxies can quench even before cluster in-fall, as seen in Figure~\ref{pos_vel}}. 
In \citet{tremmel19} we find a significant population of galaxies that quench prior to in-fall and, of these, 71\% were once within the virial radius of another halo. Indeed observations have shown a significant population of quenched galaxies at large cluster-centric distances \citep{fujita04,haines15}. Simulations have shown that galaxies can also be quenched by the larger scale ICM without being a satellite prior to in-fall \citep{bahe13, zinger18}. It is possible that there may be other observational signatures in cluster dwarf galaxies such that their history of being pre-processed by another halo may be inferred, which we leave to future work. One limitation to this is that our zoom-in region only extends to $\sim2$R$_{200}$ at $z=0$. Future cosmological cluster simulations will be needed to better understand the formation of UDGs at larger cluster-centric distances.

\begin{figure}
    \centering
    \includegraphics[trim=5mm 0mm 10mm 30mm, clip, width=90mm]{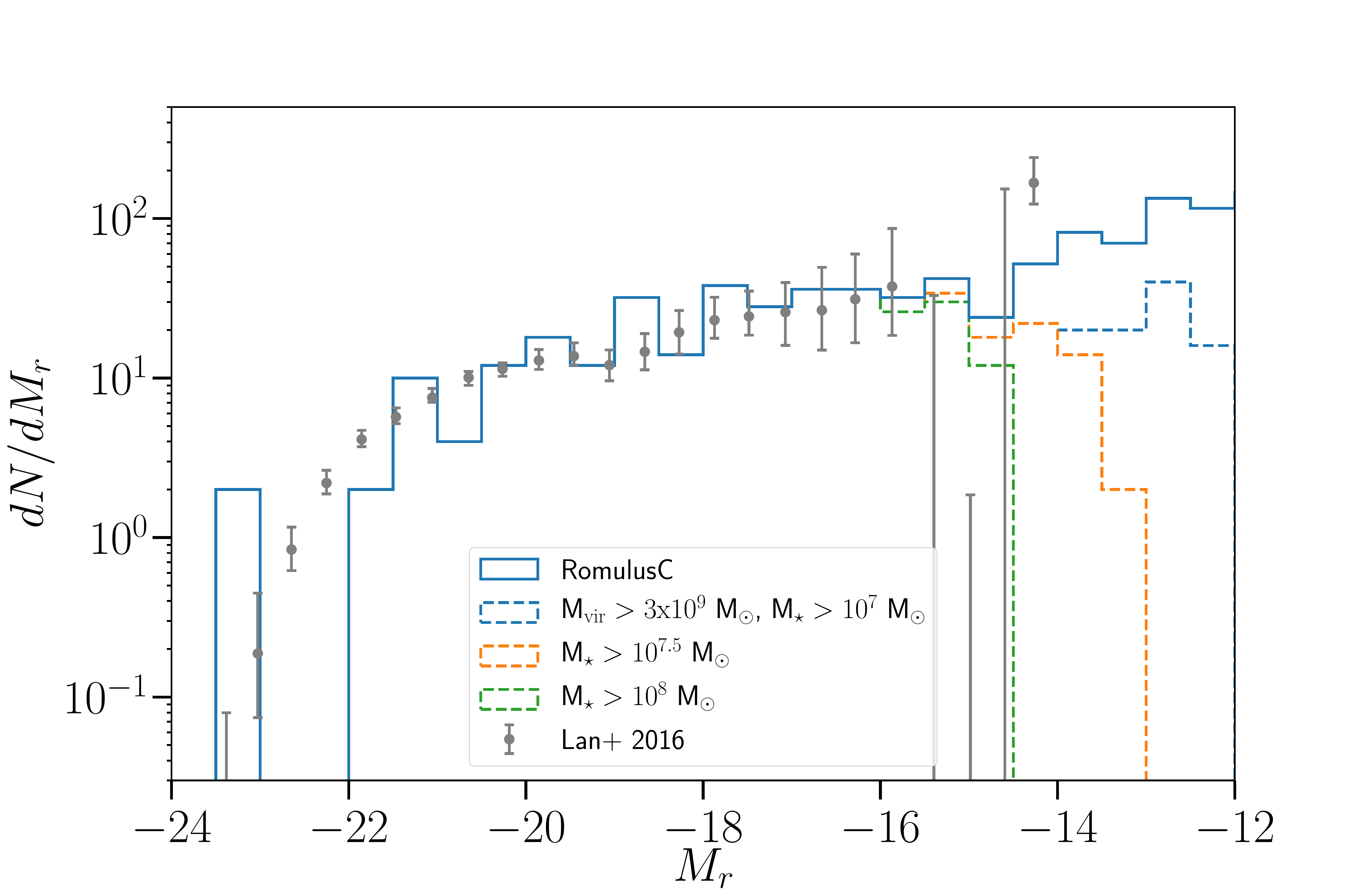}
    \caption{{\sc Luminosity function for RomulusC galaxies.} Here we plot the number of galaxies as a function of $M_r$, comparing with observations from \citet{lan16} for clusters of mass $\sim10^{14.2}$ M$_{\odot}$. The blue line represents all galaxies in the simulation with a dark matter halo mass above $3\times10^8$ M$_{\odot}$ (i.e. those with at least 1000 particles). Subsequent lines are representative of the various lower limits we have implemented when analyzing the data. Similar to observations, we see an up-turn at low luminosity, but not as steep. The majority of these low luminosity galaxies are excluded by our conservative halo mass cut and lowest stellar mass cut, indicating that in this regime we are pushing against our resolution limit. The fact that the blue line is below observations may indicate that at these low masses galaxies are disrupted too easily in the simulation.}
    \label{fig:lum_fun}
\end{figure}

%This is the first work to self-consistently predict the properties and evolution of UDGs within a cosmologically evolving cluster environment. Previous work has so far focused on lower mass halos, often at similar or lower resolution than {\sc RomulusC} \citep{jiang19,liao19}. Higher resolution simulations have been run for only isolated dwarf galaxies \citep[e.g.][]{dicintio17b,chan18}. 

We see an evolution in effective radius for both UDGs and non-UDGs that at high masses is associated with pericenter passage, in agreement with previous studies \citep[e.g.][]{jiang19,liao19}. The cause of such a sudden expansion is likely tidal heating. At low masses, the typical evolution in effective radius occurs more gradually as the stars respond to the loss of stellar mass via stellar winds and supernovae, a sudden loss of gas due to ram pressure, and the tidal stripping of dark matter in their cores, as seen in simulations of satellites in lower mass systems \citep{arraki14}. We find no clear difference in the size evolution of galaxies classified as UDGs and non-UDGs at $z=0$. Only for the lowest mass galaxies does this size evolution play a significant role in their final classification, as all of the higher mass galaxies have large enough effective radii to meet our criteria. We stress that while on average low mass galaxies do not experience the sudden increase in effective radii that would result from tidal heating, there are individual cases where this does occur. What we conclude, rather, is that rapid expansion due to tidal heating is not a requirement for UDG formation in {\sc RomulusC}, nor is it a unique occurrence in UDGs with respect to non-UDGs.

The importance of tidal heating may also depend on the mass of the host halo. On cluster scales, the magnitude of tidal heating seems to be similar to what is found at group scales \citep{jiang19}, but \citet{liao19} find significantly more size evolution in their Milky Way-mass simulations. With earlier formation times, it may be that satellite halos experience a different evolution on average. For example, if more satellites are accreted at earlier times, they will interact with a denser halo and may be more susceptible to tidal heating.

While size evolution matters at the lowest masses, passive evolution is the dominant mechanism that controls formation of UDGs in this simulation. None of our dwarf galaxies have low enough central surface brightness to be considered UDGs prior to quenching. Dwarf galaxies can easily be large enough to be considered UDGs while still relatively isolated (i.e. well before in-fall into the cluster) but the cluster environment is important for quenching star formation, leading to decreased central surface brightness and redder colors more consistent with observed cluster UDGs across all masses explored here.  This is in contrast to simulations that find isolated UDGs with low central surface brightnesses without quenching \citep{dicintio17,chan18}.  This may have to do with the different, more explosive feedback prescriptions used in the other works, as well as their higher resolution capable of capturing the influence of SN feedback on the stellar and dark matter mass distribution. However, this is not to say that UDGs can only form in cluster environments within the {\sc Romulus} framework.  In \citet{wright20} we show that UDGs can form in isolation in {\sc Romulus25}, and we will compare the formation channels in different environments. %The Sersic index does not evolve significantly after in-fall and the quenching of star formation, further evidence that passive evolution is the most important aspect to galaxy evolution post in-fall. Similar to the results from \citet{chan18}, we find that dwarf galaxies can easily be large enough \textbf{to be considered UDGs} while still relatively isolated (i.e. well before in-fall into the cluster) but that the cluster environment is important for quenching star formation, \textbf{leading} to decreased central surface brightness and redder colors more consistent with observed cluster UDGs \textbf{across all masses explored here. Contrary to simulations of isolated UDGs \citep{dicintio17,chan18},} none of our dwarf galaxies have low enough central surface brightness to be considered UDGs prior to quenching. \textbf{This may have to do with the different, more explosive feedback prescriptions used in the other works, as well as their higher resolution capable of capturing the influence of SN feedback on the stellar and dark matter mass distribution.}

Our prediction that the passive evolution of aging stellar populations is the critical mechanism to form UDGs agrees with  observations of UDGs in clusters. \citet{romantrujillo17b} find populations of both red UDGs and blue UDGs, with the latter having slightly higher surface brightness than what would be considered a UDG under our definition. They then predict the surface brightness profiles for the blue UDGs after 6 Gyrs of passive evolution and find similar properties to the red UDGs, specifically a much lower central and average surface brightness. This is essentially the exact scenario predicted by {\sc RomulusC}. %, where quenching results in a decrease in central surface brightness due to passive evolution. 
As in the observed sample in \citet{romantrujillo17b}, there is little evolution in morphology (as traced by the Sersic index) during this process. This scenario is also consistent with the fact that there is little evidence for tidal distortions in the stellar distributions of UDGs in the Coma cluster \citep{mowla17}.

On average UDGs have experienced very similar evolutionary histories compared to non-UDGs in {\sc RomulusC} and cannot be considered a fully separate population. \citet{mancerapina19} also find, using a large sample of UDGs spanning several galaxy clusters, that UDG properties are a continuation of the ambient dwarf galaxy population. One way in which UDGs and non-UDGs tend to differ is their in-fall time, with UDGs typically crossing R$_{200}$ earlier than non-UDGs.
A small sub-set of UDGs in Coma have been found to have metallicities, ages, and star formation histories consistent with normal dwarf galaxies, along with kinematics consistent with recent in-fall into the cluster \citep{alabi18,ferremateu18}. While we do find that there are plenty of UDGs that have recently fallen into the cluster, plenty more have fallen in much earlier ($z>0.5$). %Due to the importance of passive evolution, dwarf galaxies in {\sc RomulusC} that fall into the cluster earlier are more likely to become UDGs.

\subsection{Angular momentum and galaxy sizes}

We show in \S4.3 that cluster galaxies lack the population of dwarfs that form from low angular momentum gas. These are the galaxies that make up the most compact dwarf galaxies in isolation. Dwarf galaxies residing in very high spin dark matter halos has been proposed as a way of forming UDGs \citep{amoriscoloeb16,liao19}. We study this connection using the dimensionless gas spin parameter, $\lambda'_{gas}$ measured at $t_{50}$, the time at which half of the stars in the galaxy have formed. Ex-situ star formation makes this connection complicated, but we confirm that using $t_{20}$ and $t_{80}$ yield qualitatively similar results, although $t_{50}$ does show the most significant difference between cluster and field galaxies.

%While UDGs in {\sc RomulusC} do not form with drastically different gas angular momentum than non-UDGs or field galaxies, we find fewer dwarf galaxies forming from the lowest angular momentum gas in the cluster compared to the field. For isolated galaxies in {\sc Romulus25} there is a weak correlation between gas angular momentum at $t_{50}$ and galaxy sizes at $z=0$. 
The connection between gas angular momentum and galaxy size has been seen in other cosmological simulations, where star forming gas with high angular momentum remains at large distances. This results in lower surface brightness, larger galaxies \citep{dicintio19}.  The origin of the difference in gas angular momentum among cluster and isolated dwarf galaxies in {\sc Romulus} is still uncertain and we leave this to future work. It is possible that these galaxies have expelled their lower angular momentum gas \citep[e.g.][]{brook11,christensen16}, but that doesn't explain the environmental dependence. It is possible that the galaxies in cluster environments are more likely to accrete higher angular momentum gas. Recently, \citet{tadaki19} showed evidence for high rates of gas accretion in proto-cluster environments, but the star formation rates were comparable to field galaxies. This lower star formation efficiency could be due to gas accreting with higher angular momentum. Such gas would spend more time at the outskirts of the galaxy at lower densities rather than quickly falling to the center where densities would increase faster.

Looking at either total or gas spin parameter, we find that our results contrast those from \citet{amoriscoloeb16} who predict that UDGs form in halos with significantly higher spin compared to average dwarf galaxies.  Despite hints that gas angular momentum is important for creating larger galaxies in cluster environments, the values for both $\lambda'_{tot}$ and $\lambda'_{gas}$ are not significantly different from the average, isolated dwarf galaxy. This disconnect between gas and halo angular momentum is in agreement with other cosmological simulations \citep[e.g.][]{teklu15,jiang19b}, many of which also predict misalignment between galaxy and dark matter halo angular momenta \citep[e.g.][]{hahn10}. As discussed in \citet{jiang19b}, the connection between galaxy size and spin is probably stronger if one only considered the inner halo, although part of this would be driven by the dominance of baryons in the mass (and therefore angular momentum) budget in this region.

\subsection{Choice of UDG definition}
We chose to define our UDG population based on the definition used in \citet{PvD15} as well as several other works.
Using this criteria for UDG classification, we identify 80 UDGs within R$_{200}$ of the $10^{14}$ M$_{\odot}$ {\sc RomulusC} galaxy cluster, approximately twice as may as observations have found for halos of this mass \citep{romantrujillo17b, vdBurg17, mancerapina18}. It is important to note that our classifications are done assuming a face-on observer, where galaxies have the least surface brightness. Were they to be observed at different viewing angles, these galaxies may not be classified as UDGs so this number should be considered an upper limit. We also calculate surface brightness profiles using simple radial bins, rather than performing a full analysis with GalFit. It is unlikely that this would have a large effect on our results, but it could mean that we overestimate the effective radius for our galaxies with lower axis ratios. In future work we will both examine the effect of galaxy orientation and examine in more detail the morphology of UDGs, including their axis ratios (Van Nest et al., in prep).

There are other definitions that have been used in the literature for various bandpasses. All definitions generally require an effective radius of $\sim1.5$ kpc or similar. A common alternative definition involves using the mean surface brightness within R$_{eff}$, rather than the central surface brightness \citep{koda15, vdBurg16, ferremateu18}. At a fixed stellar mass, R$_{eff}$ and mean surface brightness are directly related to one another: larger galaxies have a lower mean surface brightness. In this way, the central surface brightness is a more orthogonal criterion to the effective radius which is why we choose to adopt it in this work. Following \citet{vdBurg16}, were we to define galaxies as UDGs using their r-band properties such that $24 <  \langle \mu(R_{\mathrm{eff}}) \rangle < 26$ ~\rm mag/arcsec$^2$, most of our high mass dwarf galaxies would be classified as UDGs while most of our low mass galaxies would lie below the allowed range of surface brightness. Were we to apply this criteria, we would get 51 UDGs within R$_{200}$, significantly closer to the observations. %Similarly, using our fiducial definition of UDG while requiring $\langle \mu_r(R_{\mathrm{eff}}) \rangle < 27$ ~\rm mag/arcsec$^2$ to approximately account for the unobserved systems, we would predict 50 UDGs. \amb{I'm confused.  You'd find 51 UDGs between 24-26, but 50 between 24 and 27?} 
The effective radius is similar between g and r-bands so the main effect of adopting different criteria is really the range of allowed mean surface brightness. Taking into account observational limitations in detecting objects with low mean surface brightness pushes the number of large, diffuse dwarf galaxies in {\sc RomulusC} closer to observations.

While the detailed accounting of the UDG population in {\sc RomulusC} does depend on the classification criteria, our results regarding formation are insensitive to the specific selection criteria adopted for UDGs. We show in Figures~\ref{sersic_evol},~\ref{reff_evol}, and~\ref{mu0_evol} that the evolution of UDGs and non-UDG dwarf galaxies are very similar. Figure~\ref{ram_pressure} demonstrates that the passive evolution of the quenched galaxies affects the mean surface brightness as well as $\mu_0$. Similarly there is little difference in the evolution in $R_{\mathrm{eff}}$ except that UDG progenitors tend to be larger than those of non-UDGs prior to cluster in-fall. When each galaxy does or does not cross the UDG threshold depends mostly on how long it has been quenched. The same qualitative results will hold true for UDG criteria that use the mean surface brightness rather than the central surface brightness.

\subsection{Important Caveats and Resolution Limitations}

While {\sc RomulusC} is state-of-the-art for cosmological simulations of halos of this mass, the spatial and mass resolution are still low compared to zoom-in cosmological simulations of isolated galaxies and it is important to understand how this might affect our results. 

\subsubsection{Artificial disruption of cluster galaxies}

Limitations from both resolution and halo finding technique can lead to us missing galaxies closest to the center of the cluster. Figure ~\ref{fig:lum_fun} plots the luminosity function for cluster galaxies in {\sc RomulusC} compared to similar mass observed clusters from \citet{lan16}. As in observations, there is an up-turn in the number of low luminosity galaxies at $M_r\sim-15$. This feature is common among a wide range of host halo masses, but in {\sc RomulusC} this effect is smaller than in observations. While the solid blue line in Figure~\ref{fig:lum_fun} is for all galaxies hosted in halos with at least 1000 dark matter particles, the other three lines are for cuts we make in this work. Many low luminosity galaxies are ignored in our analysis due to our cuts on stellar and dark matter mass, driven by a strict criteria for what we consider resolved in the simulation. Thus at low masses we are pushing against the resolution limits of the simulation, which is why it is not surprising we begin to see more discrepancy with observations in this regime. It is possible that this is caused by galaxies and their host halos becoming artificially disruption due to low particle counts and poor spatial resolution \citep{vdBosch18a,vdBosch18b}. In other words, our analysis may miss low mass galaxies that have experienced the strongest tidal interactions with the cluster potential and should survive to $z=0$. This may contribute to the lack of tidal heating we see in our low mass cluster dwarfs (\S 4.2) compared to higher mass galaxies as we may be selecting a biased sub-sample of the low mass dwarf galaxy population. However, the fact that we still form UDGs despite this bias indicates that strong tidal heating is not a strict requirement for low mass galaxies to attain UDG-like properties. We find that low mass galaxies are less likely to have experienced the highest background densities and therefore strongest tidal forces compared to more massive dwarfs. In part this is because massive galaxies with earlier in-fall times are more likely to survive until $z=0$ and are therefore more likely to have interacted with a denser (proto-)cluster environment. Controlling for in-fall time accounts for at least some of this difference, but not all. The results are similar if we control for the time of pericenter passage.

In a similar vein, for our time evolution analysis (\S4.1 and 4.2) we did not include galaxies that could not be traced back in time to before cluster in-fall (a population that encompasses $\sim30\%$ of our $z=0$ surviving galaxies). While artificial disruption described above will result in `over-merging' of substructure, effectively removing them from the $z=0$ population, halo finding algorithms also have known problems tracking substructure within the inner regions of a halo \citep[e.g.][]{joshi16}. This could hinder our ability to effectively trace a galaxy's full history through each simulation snapshot and would preferentially affect galaxies that come closer to cluster center.

\subsubsection{Resolution and galaxy sizes}

The {\sc RomulusC} simulation notably lacks compact dwarf galaxies (R$_{eff}< 1$ kpc) in the cluster environment, despite the fact that many such galaxies have been observed \citep{gavazzi05,eigenthaler18,aku19}. In \S4.3 and \S5.2 we discuss the role that angular momentum of gas plays in the difference between field and cluster dwarf galaxies. Indeed the field galaxies can attain more compact sizes, but the average size even for field dwarfs in {\sc Romulus25} plateaus around $\sim1$ kpc, or approximately four times the gravitational softening length of the simulation. Many simulations, including some at higher resolution, have had difficulty creating compact low mass galaxies in isolation or as satellites \citep[e.g.][]{elbadry16, lupi17, jiang19, chan18, santossantos18, GK19}. %\amb{Santos-Santos 2018 and Garrison-Kimmel 2019? are the references I would cite} 
While this can often be attributed to feedback and/or ISM sub-grid models, in {\sc Romulus} the issue is at least partially due to resolution. With a spline kernel softening length of 350 pc, structures well below 1 kpc are difficult to resolve. Further, the relatively low resolution and simple ISM physics means that star formation must be allowed to occur at relatively low densities. All together, this means that it is difficult to generate very dense structures in the simulation such that more than half of a galaxy's mass is within $<1$ kpc from the galaxy center. This lower limit to structure can be seen in the flattening of the size-mass relation (Figure~\ref{size_mass}) at stellar masses lower than $\sim10^{7.5}$ M$_{\odot}$. The flattening occurs just above 1 kpc and results in our lowest mass galaxies being a factor of $\sim2$ times larger than the empirical relation. If the resolution of {\sc Romulus} acts to create, on average, larger low mass galaxies it may mean that we form UDGs too easily because more galaxies are already more diffuse. However, we stress that the results we present here connecting quenching, passive evolution, and surface brightness evolution are robust to such resolution effects.

\subsubsection{Two body interactions and energy equipartition}

Galaxies resolved with only a few hundred star particles (like those in our lowest mass bin) are subject to effects related to two-body relaxation and energy equipartition among different mass particles \citep{ludlow19}. Even though dark matter particles have similar mass to gas particles in the {\sc Romulus} simulations, star particles form with only 30\% of their parent gas particle mass. It may be the case that the decrease in central stellar mass densities (see Figure~\ref{star_den_evol}) and increase in galaxy sizes (see Figure~\ref{reff_evol}) are results of this effect, which causes the puffing up of stars relative to the more massive dark matter particles, particularly in quiescent galaxies. However, the change we see in central density is only $\sim30\%$ on average and can mostly be explained by wind mass loss from aging stellar populations as well as the decreasing dark matter mass density (see Figure~\ref{dm_den_evol}). According to \citet{ludlow19}, energy equipartition should result in an increase in the sizes of galaxies over time. However, while \citet{ludlow19} predict a continuous puffing up of stars over time, the evolution we find at low masses slows after a few Gyr from when star formation was quenched. This is evidence that the majority of this evolution is directly related to passive stellar evolution and mass loss.

%, becoming more prominent in lower mass galaxies that are resolved with fewer star particles. However, we see just as much if not more evolution in effective radius for more massive dwarf galaxies (see Figure~\ref{reff_evol}) that are resolved with several thousand star particles as we do at our lowest mass bin. It is therefore unlikely that our simulated dwarf galaxies are suffering from this effect.

\subsubsection{Resolution and ram pressure stripping}

Dense, molecular gas would be more resistant to ram pressure in the cluster environment, but such dense gas is unresolved in the simulation. Therefore even though at this resolution ram pressure stripping itself is well resolved \citep{roedinger15}, the fact that star forming gas is less dense will make it artificially more efficient at being removed from galaxies interacting with the cluster environment. Further, feedback taking place in low density gas in low mass galaxies may be too efficient at disturbing the cold gas and make it even more susceptible to being stripped even in relatively low density environments \citep{eagle15, bahe17}. Using {\sc Romulus25} we confirm that in isolation we do not produce a significant population of quenched dwarf galaxies above a stellar mass of $10^{7.5}$M$_{\odot}$. Only at our lowest masses (M$_{\star}<10^{7.5}$M$_{\odot}$) do we begin to see significant evidence of such overquenching. However, we do not expect this to have a significant impact on our results, as the majority of UDGs fell into the cluster at early times. Even if they quench sooner than they should due to these resolution issues, they would have still had time to passively evolve for Gyrs, which in most cases is enough for galaxies to become UDGs at these low masses.

%Many of the UDGs have been within the cluster environment for a long time already and would likely have been stripped no matter what. At low masses the evolution in central surface brightness happens such that the galaxies are well below the threshold for being UDGs, so a delay in their quenching would only be a secondary effect. This may mean that the population of UDGs that have fallen in more recently is overestimated, but they are in the minority. More massive galaxies, which depend more closely on the evolution of $\mu_0$ for being considered UDGs, may be less likely to be considered UDGs at $z=0$ if they quench later, but this will not qualitatively affect our conclusions as to their origin or evolution.

\subsubsection{The lack of dark matter cores}

Feedback from supernovae can affect the distributions of both stars and dark matter in dwarf galaxies, but this requires simulations that can resolve individual, dense star-forming regions  \citep{G10,G12,PG12,PG13,dicintio14,elbadry16,dutton19}. As already discussed at length, the limited resolution attainable for such a massive simulation as {\sc RomulusC} means that the multiphase ISM cannot be resolved and stars must form in gas with lower density and higher temperature thresholds. This affects the environment in which supernovae explode and their spatial distribution throughout the galaxy. This low resolution model is incapable of forming cored dark matter profiles \citep[e.g.][]{genina18, bose19, dutton19}.Other simulations have shown that this same interaction is important for puffing up the stellar mass distribution in dwarf galaxies and is directly related to forming UDGs in isolation \citep{dicintio17b, chan18}. While the lack of dark matter cores is not a prediction of the simulation, but a result of its lower resolution, the fact that UDGs still form in {\sc RomulusC} means that the formation channels we predict in this work are unique compared to that of \citet{dicintio19} where bursty star formation, efficient feedback, and cored dark matter profiles are critical. In reality, it is likely that a combination of both processes are important for shaping the UDG population in both isolation and in clusters.

A cuspier dark matter profile may contribute to the lack of tidal heating of low mass dwarf galaxies in {\sc RomulusC} as there is more binding energy to hold the galaxies together  \citep{carleton19}. 
%It is possible that, with a heavily cored profile, cluster dwarfs experience more evolution in their central densities and effective radii due to various tidal interactions. 
Still, we show in this work that impulsive tidal interactions are not required to form UDGs in the first place. The lack of dark matter cores may also contribute to Sersic indices that are slightly larger than observed UDGs (see Figure~\ref{morph_compare}). More central mass concentration could mean more concentrated stellar distribution and therefore higher Sersic index.

\section{Summary and Conclusions}

We have selected galaxies from the {\sc RomulusC} cosmological zoom-in simulation of a galaxy cluster that fit the criteria of \citet{PvD15} for being considered ultra-diffuse (R$_{eff} > 1.5$ kpc, $\mu_0 > 24$ mag/arcsec$^2$). The UDGs in {\sc RomulusC} match well with the observed properties of cluster UDGs. Our simulated UDGs are all dwarf galaxies, so we compare the population of UDGs to the overall population of dwarf galaxies in the cluster (M$_{\star}<10^9$ M$_{\odot}$). We leverage the fact that we can trace all galaxies back in time to better understand the role of environment in forming the UDG population. We also compare to the population of isolated dwarf galaxies from the {\sc Romulus25} simulation.
\begin{itemize}

    \item UDGs in galaxy clusters form from dwarf galaxies that have their gas removed and star formation quenched due to ram pressure  as they travel through the hot, dense cluster environment. The subsequent passive stellar evolution is the cause of their low surface brightness at $z=0$.
    \\
     \item The evolution in central surface brightness is similar for UDGs and non-UDGs. Prior to quenching, UDG progenitors are 1-2 magnitudes brighter than what is required for UDG classification. This evolution cannot be accounted for by decreasing central stellar densities, which fall by typically only $\sim30\%$ over the course of the simulation.
    \\
    \item Passive evolution of stars results in a relationship between stellar age and central surface brightness that is the same in the cluster environment as it is for isolated galaxies. Dwarfs with older stellar populations have lower central surface brightness.
     \\
    \item The evolution of galaxy morphology is similar for UDGs and non-UDGs. UDG progenitors tend to be slightly larger at low masses. Dwarf galaxies that fall into the cluster earlier ($z > 0.5$) are approximately twice as likely ($>80\%$) to become UDGs as those that fall in later.
    \\
    \item The classification of lower mass dwarf galaxies (M$_{\star}<10^8$ M$_{\odot}$) as UDGs depends mostly on their effective radius, while for higher mass dwarfs it depends on their central surface brightness.
\\
    \item UDGs are not a separate population compared to non-UDG dwarf galaxies in terms of absolute magnitude, central surface brightness, and effective radius. Rather, they are part of a continuous distribution of dwarf morphologies. UDGs also have similar positions and velocities relative to the center of the cluster as non-UDGs and have similar Sersic indices compared to both non-UDG cluster dwarfs and isolated dwarfs.
    \\
    \item Both UDGs and non-UDG dwarf galaxies are, at the time of their maximum halo mass, consistent with the stellar mass-halo mass relation. UDGs have dark matter halos with peak virial mass below $10^{11}$ M$_{\odot}$. Tidal stripping results in decreased halo masses and central dark matter densities at $z=0$, but does not affect their final stellar mass.
    \\
    \item Both UDGs and non-UDGs have similar halo spin values at $t_{50}$, the time at which 50\% of the galaxy's final stellar mass has assembled. Halo spin is similar between cluster and isolated galaxies.% Cluster dwarf galaxies and isolated galaxies also have similar values for total halo spin at $t_{50}$. Cluster UDGs are  more likely to have slightly higher gas spin values at $t_{50}$ compared to isolated galaxies.
    \\
    \item The effective radius of all dwarf galaxies increases typically by $\sim30\%$ after quenching and in-fall into the cluster. For higher mass dwarfs this evolution is due to tidal heating at pericenter, consistent with previous results \citep{jiang19}. For lower mass galaxies this evolution is more gradual, likely due to adiabatic expansion in response to a changing mass distribution. Only at low masses does this expansion play a significant role in the $z=0$ classification of UDGs.
    \\
    %occurs quickly at pericenter, where the galaxies also typically quench, consistent with previous work by \citet{jiang19}. For these high mass galaxies this increase is likely due to tidal heating. Low mass galaxies, on the other hand, evolve more gradually in response to the loss of both dark matter and stellar mass from tidal stripping and stellar evolution (i.e. wind mass loss). Only for the lowest mass UDGs does this effective radius evolution play a role in determining their final classification.}
    %This evolution is the result of a decrease in central mass density due to the tidal stripping of dark matter, wind mass loss from stellar evolution, and ram pressure stripping of gas. At low masses, where the final effective radius strongly determines their classification, UDGs typically have larger effective radii compared to non-UDGs Gyrs before quenching or in-fall.
    
    %\\
    \item Cluster dwarf galaxies lack the population of more compact dwarfs (R$_{eff} < 1$ kpc) seen for isolated galaxies in {\sc Romulus25}. This can be partly attributed to the increasing size following in-fall and quenching due to tidal stripping, tidal heating, and passive evolution. There is also a dearth of dwarf galaxies forming from very low angular momentum gas in the cluster environment compared to the field.
        \\

These results show that UDGs do not experience unique evolution compared to other cluster dwarf galaxies. Rather, early in-fall and quenching followed by Gyrs of passive evolution naturally result in a population of low surface brightness galaxies. This is aided by tidal interactions with the cluster potential, such as tidal stripping and heating, which, in conjunction with passive stellar evolution and ram pressure, result in a modest expansion in galaxy effective radius. Despite, on average, having similar evolutionary paths, there are individual UDG cases where the evolution occurs more violently through both interaction with the cluster potential or close encounters with other cluster galaxies. We leave it to future work to examine individual cases in the {\sc RomulusC} simulation. In \citet{wright20}  we build on these results and examine the population of UDGs in lower mass halos and in isolation using the {\sc Romulus25} simulation in order to compare formation channels in different environments.

    %form in clusters through passive evolution of a population of quenched dwarf galaxies and the formation of a dark matter core, a high spin halo, \textbf{or tidal heating are} not requirements in their formation. They also support the notion that UDGs are, in general, just a sub-sample of the overall dwarf galaxy population, rather than a unique type of galaxy with a unique evolutionary path. Although the overall population of UDGs is not unique in terms of their evolution compared to non-UDGs, there may be individual cases where UDGs are created more violently through mergers, close interactions, or tidal heating from the cluster potential. We leave it to future work to examine individual cases in the {\sc RomulusC} simulation, though it is clear from Figures 10-14 that there are plenty of candidates for such unique evolutionary paths. In Wright et al. (in prep)  we will build on these results and examine the population of UDGs in lower mass halos and in isolation using the {\sc Romulus25} simulation in order to compare formation channels in different environments. 
    
     \section*{Acknowledgments}
   %  \dn{We thank the anonymous referee for constructive comments on the manuscript that have significantly improved the manuscript.}
     The authors thank the anonymous referee for their thorough reading and constructive comments which have significantly improved the manuscript.
     The authors also thank Pieter van Dokkum, Shany Danieli, Lamiya Mowla, Johnny Greco, Frank van den Bosch, Arianna di Cintio, Pavel Mancera Piña, 
Christopher Conselice, and Priya Natarajan for useful discussions related to this work. MT gratefully acknowledges support from the YCAA Prize Postdoctoral Fellowship. ACW is supported by an ACM SIGHPC/Intel Computational \& Data Science fellowship. AMB, FM, and DN  acknowledge the hospitality at the Aspen Center for Physics, which is supported by National Science Foundation grant PHY-1607611. TQ was partially supported by NSF award AST-1514868. This research is part of the Blue Waters sustained-petascale computing project, which is supported by the National Science Foundation (awards OCI-0725070 and ACI-1238993) and the state of Illinois. Blue Waters is a joint effort of the University of Illinois at Urbana-Champaign and its National Center for Supercomputing Applications. This work is also part of a PRAC allocation support by the National Science Foundation (award number OAC-1613674). The analysis presented in this Paper was done using the publicly available software packages, pynbody \citep{pynbody} and TANGOS \citep{tangos}, primarily with resources provided by the NASA High-End Computing (HEC) Program through the NASA Advanced Supercomputing (NAS) Division at Ames Research Center.

\section*{Data Availability}

The data for this work was generated from a proprietary branch of the {\sc ChaNGa} N-Body+SPH code \citep{changa15}. The public repository for {\sc ChaNGa} is available on github (https://github.com/N-BodyShop/changa). Analysis was conducted using the publicly available software pynbody \citep[][https://github.com/pynbody/pynbody]{pynbody} and TANGOS \citep[][https://github.com/pynbody/tangos]{tangos}. These results were generated from the RomulusC and Romulus25 cosmological simulations. The raw output from these simulations can be accessed upon request from Michael Tremmel (michael.tremmel@yale.edu), along with the TANGOS database files that were generated from these outputs and directly used for this analysis.

    %While there are important limitations to these simulations regarding their resolution and simple ISM models, the qualitative results we present in this Paper are robust to such effects. 
    
\end{itemize}

\end{document}

%% file: rgb.tex
  \definecolor{snow}{rgb}{1.000000,0.980392,0.980392}
  \definecolor{ghost white}{rgb}{0.972549,0.972549,1.000000}
  \definecolor{GhostWhite}{rgb}{0.972549,0.972549,1.000000}
  \definecolor{white smoke}{rgb}{0.960784,0.960784,0.960784}
  \definecolor{WhiteSmoke}{rgb}{0.960784,0.960784,0.960784}
  \definecolor{gainsboro}{rgb}{0.862745,0.862745,0.862745}
  \definecolor{floral white}{rgb}{1.000000,0.980392,0.941176}
  \definecolor{FloralWhite}{rgb}{1.000000,0.980392,0.941176}
  \definecolor{old lace}{rgb}{0.992157,0.960784,0.901961}
  \definecolor{OldLace}{rgb}{0.992157,0.960784,0.901961}
  \definecolor{linen}{rgb}{0.980392,0.941176,0.901961}
  \definecolor{antique white}{rgb}{0.980392,0.921569,0.843137}
  \definecolor{AntiqueWhite}{rgb}{0.980392,0.921569,0.843137}
  \definecolor{papaya whip}{rgb}{1.000000,0.937255,0.835294}
  \definecolor{PapayaWhip}{rgb}{1.000000,0.937255,0.835294}
  \definecolor{blanched almond}{rgb}{1.000000,0.921569,0.803922}
  \definecolor{BlanchedAlmond}{rgb}{1.000000,0.921569,0.803922}
  \definecolor{bisque}{rgb}{1.000000,0.894118,0.768627}
  \definecolor{peach puff}{rgb}{1.000000,0.854902,0.725490}
  \definecolor{PeachPuff}{rgb}{1.000000,0.854902,0.725490}
  \definecolor{navajo white}{rgb}{1.000000,0.870588,0.678431}
  \definecolor{NavajoWhite}{rgb}{1.000000,0.870588,0.678431}
  \definecolor{moccasin}{rgb}{1.000000,0.894118,0.709804}
  \definecolor{cornsilk}{rgb}{1.000000,0.972549,0.862745}
  \definecolor{ivory}{rgb}{1.000000,1.000000,0.941176}
  \definecolor{lemon chiffon}{rgb}{1.000000,0.980392,0.803922}
  \definecolor{LemonChiffon}{rgb}{1.000000,0.980392,0.803922}
  \definecolor{seashell}{rgb}{1.000000,0.960784,0.933333}
  \definecolor{honeydew}{rgb}{0.941176,1.000000,0.941176}
  \definecolor{mint cream}{rgb}{0.960784,1.000000,0.980392}
  \definecolor{MintCream}{rgb}{0.960784,1.000000,0.980392}
  \definecolor{azure}{rgb}{0.941176,1.000000,1.000000}
  \definecolor{alice blue}{rgb}{0.941176,0.972549,1.000000}
  \definecolor{AliceBlue}{rgb}{0.941176,0.972549,1.000000}
  \definecolor{lavender}{rgb}{0.901961,0.901961,0.980392}
  \definecolor{lavender blush}{rgb}{1.000000,0.941176,0.960784}
  \definecolor{LavenderBlush}{rgb}{1.000000,0.941176,0.960784}
  \definecolor{misty rose}{rgb}{1.000000,0.894118,0.882353}
  \definecolor{MistyRose}{rgb}{1.000000,0.894118,0.882353}
  \definecolor{white}{rgb}{1.000000,1.000000,1.000000}
  \definecolor{black}{rgb}{0.000000,0.000000,0.000000}
  \definecolor{dark slate gray}{rgb}{0.184314,0.309804,0.309804}
  \definecolor{DarkSlateGray}{rgb}{0.184314,0.309804,0.309804}
  \definecolor{dark slate grey}{rgb}{0.184314,0.309804,0.309804}
  \definecolor{DarkSlateGrey}{rgb}{0.184314,0.309804,0.309804}
  \definecolor{dim gray}{rgb}{0.411765,0.411765,0.411765}
  \definecolor{DimGray}{rgb}{0.411765,0.411765,0.411765}
  \definecolor{dim grey}{rgb}{0.411765,0.411765,0.411765}
  \definecolor{DimGrey}{rgb}{0.411765,0.411765,0.411765}
  \definecolor{slate gray}{rgb}{0.439216,0.501961,0.564706}
  \definecolor{SlateGray}{rgb}{0.439216,0.501961,0.564706}
  \definecolor{slate grey}{rgb}{0.439216,0.501961,0.564706}
  \definecolor{SlateGrey}{rgb}{0.439216,0.501961,0.564706}
  \definecolor{light slate gray}{rgb}{0.466667,0.533333,0.600000}
  \definecolor{LightSlateGray}{rgb}{0.466667,0.533333,0.600000}
  \definecolor{light slate grey}{rgb}{0.466667,0.533333,0.600000}
  \definecolor{LightSlateGrey}{rgb}{0.466667,0.533333,0.600000}
  \definecolor{gray}{rgb}{0.745098,0.745098,0.745098}
  \definecolor{grey}{rgb}{0.745098,0.745098,0.745098}
  \definecolor{light grey}{rgb}{0.827451,0.827451,0.827451}
  \definecolor{LightGrey}{rgb}{0.827451,0.827451,0.827451}
  \definecolor{light gray}{rgb}{0.827451,0.827451,0.827451}
  \definecolor{LightGray}{rgb}{0.827451,0.827451,0.827451}
  \definecolor{midnight blue}{rgb}{0.098039,0.098039,0.439216}
  \definecolor{MidnightBlue}{rgb}{0.098039,0.098039,0.439216}
  \definecolor{navy}{rgb}{0.000000,0.000000,0.501961}
  \definecolor{navy blue}{rgb}{0.000000,0.000000,0.501961}
  \definecolor{NavyBlue}{rgb}{0.000000,0.000000,0.501961}
  \definecolor{cornflower blue}{rgb}{0.392157,0.584314,0.929412}
  \definecolor{CornflowerBlue}{rgb}{0.392157,0.584314,0.929412}
  \definecolor{dark slate blue}{rgb}{0.282353,0.239216,0.545098}
  \definecolor{DarkSlateBlue}{rgb}{0.282353,0.239216,0.545098}
  \definecolor{slate blue}{rgb}{0.415686,0.352941,0.803922}
  \definecolor{SlateBlue}{rgb}{0.415686,0.352941,0.803922}
  \definecolor{medium slate blue}{rgb}{0.482353,0.407843,0.933333}
  \definecolor{MediumSlateBlue}{rgb}{0.482353,0.407843,0.933333}
  \definecolor{light slate blue}{rgb}{0.517647,0.439216,1.000000}
  \definecolor{LightSlateBlue}{rgb}{0.517647,0.439216,1.000000}
  \definecolor{medium blue}{rgb}{0.000000,0.000000,0.803922}
  \definecolor{MediumBlue}{rgb}{0.000000,0.000000,0.803922}
  \definecolor{royal blue}{rgb}{0.254902,0.411765,0.882353}
  \definecolor{RoyalBlue}{rgb}{0.254902,0.411765,0.882353}
  \definecolor{blue}{rgb}{0.000000,0.000000,1.000000}
  \definecolor{dodger blue}{rgb}{0.117647,0.564706,1.000000}
  \definecolor{DodgerBlue}{rgb}{0.117647,0.564706,1.000000}
  \definecolor{deep sky blue}{rgb}{0.000000,0.749020,1.000000}
  \definecolor{DeepSkyBlue}{rgb}{0.000000,0.749020,1.000000}
  \definecolor{sky blue}{rgb}{0.529412,0.807843,0.921569}
  \definecolor{SkyBlue}{rgb}{0.529412,0.807843,0.921569}
  \definecolor{light sky blue}{rgb}{0.529412,0.807843,0.980392}
  \definecolor{LightSkyBlue}{rgb}{0.529412,0.807843,0.980392}
  \definecolor{steel blue}{rgb}{0.274510,0.509804,0.705882}
  \definecolor{SteelBlue}{rgb}{0.274510,0.509804,0.705882}
  \definecolor{light steel blue}{rgb}{0.690196,0.768627,0.870588}
  \definecolor{LightSteelBlue}{rgb}{0.690196,0.768627,0.870588}
  \definecolor{light blue}{rgb}{0.678431,0.847059,0.901961}
  \definecolor{LightBlue}{rgb}{0.678431,0.847059,0.901961}
  \definecolor{powder blue}{rgb}{0.690196,0.878431,0.901961}
  \definecolor{PowderBlue}{rgb}{0.690196,0.878431,0.901961}
  \definecolor{pale turquoise}{rgb}{0.686275,0.933333,0.933333}
  \definecolor{PaleTurquoise}{rgb}{0.686275,0.933333,0.933333}
  \definecolor{dark turquoise}{rgb}{0.000000,0.807843,0.819608}
  \definecolor{DarkTurquoise}{rgb}{0.000000,0.807843,0.819608}
  \definecolor{medium turquoise}{rgb}{0.282353,0.819608,0.800000}
  \definecolor{MediumTurquoise}{rgb}{0.282353,0.819608,0.800000}
  \definecolor{turquoise}{rgb}{0.250980,0.878431,0.815686}
  \definecolor{cyan}{rgb}{0.000000,1.000000,1.000000}
  \definecolor{light cyan}{rgb}{0.878431,1.000000,1.000000}
  \definecolor{LightCyan}{rgb}{0.878431,1.000000,1.000000}
  \definecolor{cadet blue}{rgb}{0.372549,0.619608,0.627451}
  \definecolor{CadetBlue}{rgb}{0.372549,0.619608,0.627451}
  \definecolor{medium aquamarine}{rgb}{0.400000,0.803922,0.666667}
  \definecolor{MediumAquamarine}{rgb}{0.400000,0.803922,0.666667}
  \definecolor{aquamarine}{rgb}{0.498039,1.000000,0.831373}
  \definecolor{dark green}{rgb}{0.000000,0.392157,0.000000}
  \definecolor{DarkGreen}{rgb}{0.000000,0.392157,0.000000}
  \definecolor{dark olive green}{rgb}{0.333333,0.419608,0.184314}
  \definecolor{DarkOliveGreen}{rgb}{0.333333,0.419608,0.184314}
  \definecolor{dark sea green}{rgb}{0.560784,0.737255,0.560784}
  \definecolor{DarkSeaGreen}{rgb}{0.560784,0.737255,0.560784}
  \definecolor{sea green}{rgb}{0.180392,0.545098,0.341176}
  \definecolor{SeaGreen}{rgb}{0.180392,0.545098,0.341176}
  \definecolor{medium sea green}{rgb}{0.235294,0.701961,0.443137}
  \definecolor{MediumSeaGreen}{rgb}{0.235294,0.701961,0.443137}
  \definecolor{light sea green}{rgb}{0.125490,0.698039,0.666667}
  \definecolor{LightSeaGreen}{rgb}{0.125490,0.698039,0.666667}
  \definecolor{pale green}{rgb}{0.596078,0.984314,0.596078}
  \definecolor{PaleGreen}{rgb}{0.596078,0.984314,0.596078}
  \definecolor{spring green}{rgb}{0.000000,1.000000,0.498039}
  \definecolor{SpringGreen}{rgb}{0.000000,1.000000,0.498039}
  \definecolor{lawn green}{rgb}{0.486275,0.988235,0.000000}
  \definecolor{LawnGreen}{rgb}{0.486275,0.988235,0.000000}
  \definecolor{green}{rgb}{0.000000,1.000000,0.000000}
  \definecolor{chartreuse}{rgb}{0.498039,1.000000,0.000000}
  \definecolor{medium spring green}{rgb}{0.000000,0.980392,0.603922}
  \definecolor{MediumSpringGreen}{rgb}{0.000000,0.980392,0.603922}
  \definecolor{green yellow}{rgb}{0.678431,1.000000,0.184314}
  \definecolor{GreenYellow}{rgb}{0.678431,1.000000,0.184314}
  \definecolor{lime green}{rgb}{0.196078,0.803922,0.196078}
  \definecolor{LimeGreen}{rgb}{0.196078,0.803922,0.196078}
  \definecolor{yellow green}{rgb}{0.603922,0.803922,0.196078}
  \definecolor{YellowGreen}{rgb}{0.603922,0.803922,0.196078}
  \definecolor{forest green}{rgb}{0.133333,0.545098,0.133333}
  \definecolor{ForestGreen}{rgb}{0.133333,0.545098,0.133333}
  \definecolor{olive drab}{rgb}{0.419608,0.556863,0.137255}
  \definecolor{OliveDrab}{rgb}{0.419608,0.556863,0.137255}
  \definecolor{dark khaki}{rgb}{0.741176,0.717647,0.419608}
  \definecolor{DarkKhaki}{rgb}{0.741176,0.717647,0.419608}
  \definecolor{khaki}{rgb}{0.941176,0.901961,0.549020}
  \definecolor{pale goldenrod}{rgb}{0.933333,0.909804,0.666667}
  \definecolor{PaleGoldenrod}{rgb}{0.933333,0.909804,0.666667}
  \definecolor{light goldenrod yellow}{rgb}{0.980392,0.980392,0.823529}
  \definecolor{LightGoldenrodYellow}{rgb}{0.980392,0.980392,0.823529}
  \definecolor{light yellow}{rgb}{1.000000,1.000000,0.878431}
  \definecolor{LightYellow}{rgb}{1.000000,1.000000,0.878431}
  \definecolor{yellow}{rgb}{1.000000,1.000000,0.000000}
  \definecolor{gold}{rgb}{1.000000,0.843137,0.000000}
  \definecolor{light goldenrod}{rgb}{0.933333,0.866667,0.509804}
  \definecolor{LightGoldenrod}{rgb}{0.933333,0.866667,0.509804}
  \definecolor{goldenrod}{rgb}{0.854902,0.647059,0.125490}
  \definecolor{dark goldenrod}{rgb}{0.721569,0.525490,0.043137}
  \definecolor{DarkGoldenrod}{rgb}{0.721569,0.525490,0.043137}
  \definecolor{rosy brown}{rgb}{0.737255,0.560784,0.560784}
  \definecolor{RosyBrown}{rgb}{0.737255,0.560784,0.560784}
  \definecolor{indian red}{rgb}{0.803922,0.360784,0.360784}
  \definecolor{IndianRed}{rgb}{0.803922,0.360784,0.360784}
  \definecolor{saddle brown}{rgb}{0.545098,0.270588,0.074510}
  \definecolor{SaddleBrown}{rgb}{0.545098,0.270588,0.074510}
  \definecolor{sienna}{rgb}{0.627451,0.321569,0.176471}
  \definecolor{peru}{rgb}{0.803922,0.521569,0.247059}
  \definecolor{burlywood}{rgb}{0.870588,0.721569,0.529412}
  \definecolor{beige}{rgb}{0.960784,0.960784,0.862745}
  \definecolor{wheat}{rgb}{0.960784,0.870588,0.701961}
  \definecolor{sandy brown}{rgb}{0.956863,0.643137,0.376471}
  \definecolor{SandyBrown}{rgb}{0.956863,0.643137,0.376471}
  \definecolor{tan}{rgb}{0.823529,0.705882,0.549020}
  \definecolor{chocolate}{rgb}{0.823529,0.411765,0.117647}
  \definecolor{firebrick}{rgb}{0.698039,0.133333,0.133333}
  \definecolor{brown}{rgb}{0.647059,0.164706,0.164706}
  \definecolor{dark salmon}{rgb}{0.913725,0.588235,0.478431}
  \definecolor{DarkSalmon}{rgb}{0.913725,0.588235,0.478431}
  \definecolor{salmon}{rgb}{0.980392,0.501961,0.447059}
  \definecolor{light salmon}{rgb}{1.000000,0.627451,0.478431}
  \definecolor{LightSalmon}{rgb}{1.000000,0.627451,0.478431}
  \definecolor{orange}{rgb}{1.000000,0.647059,0.000000}
  \definecolor{dark orange}{rgb}{1.000000,0.549020,0.000000}
  \definecolor{DarkOrange}{rgb}{1.000000,0.549020,0.000000}
  \definecolor{coral}{rgb}{1.000000,0.498039,0.313726}
  \definecolor{light coral}{rgb}{0.941176,0.501961,0.501961}
  \definecolor{LightCoral}{rgb}{0.941176,0.501961,0.501961}
  \definecolor{tomato}{rgb}{1.000000,0.388235,0.278431}
  \definecolor{orange red}{rgb}{1.000000,0.270588,0.000000}
  \definecolor{OrangeRed}{rgb}{1.000000,0.270588,0.000000}
  \definecolor{red}{rgb}{1.000000,0.000000,0.000000}
  \definecolor{hot pink}{rgb}{1.000000,0.411765,0.705882}
  \definecolor{HotPink}{rgb}{1.000000,0.411765,0.705882}
  \definecolor{deep pink}{rgb}{1.000000,0.078431,0.576471}
  \definecolor{DeepPink}{rgb}{1.000000,0.078431,0.576471}
  \definecolor{pink}{rgb}{1.000000,0.752941,0.796078}
  \definecolor{light pink}{rgb}{1.000000,0.713726,0.756863}
  \definecolor{LightPink}{rgb}{1.000000,0.713726,0.756863}
  \definecolor{pale violet red}{rgb}{0.858824,0.439216,0.576471}
  \definecolor{PaleVioletRed}{rgb}{0.858824,0.439216,0.576471}
  \definecolor{maroon}{rgb}{0.690196,0.188235,0.376471}
  \definecolor{medium violet red}{rgb}{0.780392,0.082353,0.521569}
  \definecolor{MediumVioletRed}{rgb}{0.780392,0.082353,0.521569}
  \definecolor{violet red}{rgb}{0.815686,0.125490,0.564706}
  \definecolor{VioletRed}{rgb}{0.815686,0.125490,0.564706}
  \definecolor{magenta}{rgb}{1.000000,0.000000,1.000000}
  \definecolor{violet}{rgb}{0.933333,0.509804,0.933333}
  \definecolor{plum}{rgb}{0.866667,0.627451,0.866667}
  \definecolor{orchid}{rgb}{0.854902,0.439216,0.839216}
  \definecolor{medium orchid}{rgb}{0.729412,0.333333,0.827451}
  \definecolor{MediumOrchid}{rgb}{0.729412,0.333333,0.827451}
  \definecolor{dark orchid}{rgb}{0.600000,0.196078,0.800000}
  \definecolor{DarkOrchid}{rgb}{0.600000,0.196078,0.800000}
  \definecolor{dark violet}{rgb}{0.580392,0.000000,0.827451}
  \definecolor{DarkViolet}{rgb}{0.580392,0.000000,0.827451}
  \definecolor{blue violet}{rgb}{0.541176,0.168627,0.886275}
  \definecolor{BlueViolet}{rgb}{0.541176,0.168627,0.886275}
  \definecolor{purple}{rgb}{0.627451,0.125490,0.941176}
  \definecolor{medium purple}{rgb}{0.576471,0.439216,0.858824}
  \definecolor{MediumPurple}{rgb}{0.576471,0.439216,0.858824}
  \definecolor{thistle}{rgb}{0.847059,0.749020,0.847059}
  \definecolor{snow1}{rgb}{1.000000,0.980392,0.980392}
  \definecolor{snow2}{rgb}{0.933333,0.913725,0.913725}
  \definecolor{snow3}{rgb}{0.803922,0.788235,0.788235}
  \definecolor{snow4}{rgb}{0.545098,0.537255,0.537255}
  \definecolor{seashell1}{rgb}{1.000000,0.960784,0.933333}
  \definecolor{seashell2}{rgb}{0.933333,0.898039,0.870588}
  \definecolor{seashell3}{rgb}{0.803922,0.772549,0.749020}
  \definecolor{seashell4}{rgb}{0.545098,0.525490,0.509804}
  \definecolor{AntiqueWhite1}{rgb}{1.000000,0.937255,0.858824}
  \definecolor{AntiqueWhite2}{rgb}{0.933333,0.874510,0.800000}
  \definecolor{AntiqueWhite3}{rgb}{0.803922,0.752941,0.690196}
  \definecolor{AntiqueWhite4}{rgb}{0.545098,0.513726,0.470588}
  \definecolor{bisque1}{rgb}{1.000000,0.894118,0.768627}
  \definecolor{bisque2}{rgb}{0.933333,0.835294,0.717647}
  \definecolor{bisque3}{rgb}{0.803922,0.717647,0.619608}
  \definecolor{bisque4}{rgb}{0.545098,0.490196,0.419608}
  \definecolor{PeachPuff1}{rgb}{1.000000,0.854902,0.725490}
  \definecolor{PeachPuff2}{rgb}{0.933333,0.796078,0.678431}
  \definecolor{PeachPuff3}{rgb}{0.803922,0.686275,0.584314}
  \definecolor{PeachPuff4}{rgb}{0.545098,0.466667,0.396078}
  \definecolor{NavajoWhite1}{rgb}{1.000000,0.870588,0.678431}
  \definecolor{NavajoWhite2}{rgb}{0.933333,0.811765,0.631373}
  \definecolor{NavajoWhite3}{rgb}{0.803922,0.701961,0.545098}
  \definecolor{NavajoWhite4}{rgb}{0.545098,0.474510,0.368627}
  \definecolor{LemonChiffon1}{rgb}{1.000000,0.980392,0.803922}
  \definecolor{LemonChiffon2}{rgb}{0.933333,0.913725,0.749020}
  \definecolor{LemonChiffon3}{rgb}{0.803922,0.788235,0.647059}
  \definecolor{LemonChiffon4}{rgb}{0.545098,0.537255,0.439216}
  \definecolor{cornsilk1}{rgb}{1.000000,0.972549,0.862745}
  \definecolor{cornsilk2}{rgb}{0.933333,0.909804,0.803922}
  \definecolor{cornsilk3}{rgb}{0.803922,0.784314,0.694118}
  \definecolor{cornsilk4}{rgb}{0.545098,0.533333,0.470588}
  \definecolor{ivory1}{rgb}{1.000000,1.000000,0.941176}
  \definecolor{ivory2}{rgb}{0.933333,0.933333,0.878431}
  \definecolor{ivory3}{rgb}{0.803922,0.803922,0.756863}
  \definecolor{ivory4}{rgb}{0.545098,0.545098,0.513726}
  \definecolor{honeydew1}{rgb}{0.941176,1.000000,0.941176}
  \definecolor{honeydew2}{rgb}{0.878431,0.933333,0.878431}
  \definecolor{honeydew3}{rgb}{0.756863,0.803922,0.756863}
  \definecolor{honeydew4}{rgb}{0.513726,0.545098,0.513726}
  \definecolor{LavenderBlush1}{rgb}{1.000000,0.941176,0.960784}
  \definecolor{LavenderBlush2}{rgb}{0.933333,0.878431,0.898039}
  \definecolor{LavenderBlush3}{rgb}{0.803922,0.756863,0.772549}
  \definecolor{LavenderBlush4}{rgb}{0.545098,0.513726,0.525490}
  \definecolor{MistyRose1}{rgb}{1.000000,0.894118,0.882353}
  \definecolor{MistyRose2}{rgb}{0.933333,0.835294,0.823529}
  \definecolor{MistyRose3}{rgb}{0.803922,0.717647,0.709804}
  \definecolor{MistyRose4}{rgb}{0.545098,0.490196,0.482353}
  \definecolor{azure1}{rgb}{0.941176,1.000000,1.000000}
  \definecolor{azure2}{rgb}{0.878431,0.933333,0.933333}
  \definecolor{azure3}{rgb}{0.756863,0.803922,0.803922}
  \definecolor{azure4}{rgb}{0.513726,0.545098,0.545098}
  \definecolor{SlateBlue1}{rgb}{0.513726,0.435294,1.000000}
  \definecolor{SlateBlue2}{rgb}{0.478431,0.403922,0.933333}
  \definecolor{SlateBlue3}{rgb}{0.411765,0.349020,0.803922}
  \definecolor{SlateBlue4}{rgb}{0.278431,0.235294,0.545098}
  \definecolor{RoyalBlue1}{rgb}{0.282353,0.462745,1.000000}
  \definecolor{RoyalBlue2}{rgb}{0.262745,0.431373,0.933333}
  \definecolor{RoyalBlue3}{rgb}{0.227451,0.372549,0.803922}
  \definecolor{RoyalBlue4}{rgb}{0.152941,0.250980,0.545098}
  \definecolor{blue1}{rgb}{0.000000,0.000000,1.000000}
  \definecolor{blue2}{rgb}{0.000000,0.000000,0.933333}
  \definecolor{blue3}{rgb}{0.000000,0.000000,0.803922}
  \definecolor{blue4}{rgb}{0.000000,0.000000,0.545098}
  \definecolor{DodgerBlue1}{rgb}{0.117647,0.564706,1.000000}
  \definecolor{DodgerBlue2}{rgb}{0.109804,0.525490,0.933333}
  \definecolor{DodgerBlue3}{rgb}{0.094118,0.454902,0.803922}
  \definecolor{DodgerBlue4}{rgb}{0.062745,0.305882,0.545098}
  \definecolor{SteelBlue1}{rgb}{0.388235,0.721569,1.000000}
  \definecolor{SteelBlue2}{rgb}{0.360784,0.674510,0.933333}
  \definecolor{SteelBlue3}{rgb}{0.309804,0.580392,0.803922}
  \definecolor{SteelBlue4}{rgb}{0.211765,0.392157,0.545098}
  \definecolor{DeepSkyBlue1}{rgb}{0.000000,0.749020,1.000000}
  \definecolor{DeepSkyBlue2}{rgb}{0.000000,0.698039,0.933333}
  \definecolor{DeepSkyBlue3}{rgb}{0.000000,0.603922,0.803922}
  \definecolor{DeepSkyBlue4}{rgb}{0.000000,0.407843,0.545098}
  \definecolor{SkyBlue1}{rgb}{0.529412,0.807843,1.000000}
  \definecolor{SkyBlue2}{rgb}{0.494118,0.752941,0.933333}
  \definecolor{SkyBlue3}{rgb}{0.423529,0.650980,0.803922}
  \definecolor{SkyBlue4}{rgb}{0.290196,0.439216,0.545098}
  \definecolor{LightSkyBlue1}{rgb}{0.690196,0.886275,1.000000}
  \definecolor{LightSkyBlue2}{rgb}{0.643137,0.827451,0.933333}
  \definecolor{LightSkyBlue3}{rgb}{0.552941,0.713726,0.803922}
  \definecolor{LightSkyBlue4}{rgb}{0.376471,0.482353,0.545098}
  \definecolor{SlateGray1}{rgb}{0.776471,0.886275,1.000000}
  \definecolor{SlateGray2}{rgb}{0.725490,0.827451,0.933333}
  \definecolor{SlateGray3}{rgb}{0.623529,0.713726,0.803922}
  \definecolor{SlateGray4}{rgb}{0.423529,0.482353,0.545098}
  \definecolor{LightSteelBlue1}{rgb}{0.792157,0.882353,1.000000}
  \definecolor{LightSteelBlue2}{rgb}{0.737255,0.823529,0.933333}
  \definecolor{LightSteelBlue3}{rgb}{0.635294,0.709804,0.803922}
  \definecolor{LightSteelBlue4}{rgb}{0.431373,0.482353,0.545098}
  \definecolor{LightBlue1}{rgb}{0.749020,0.937255,1.000000}
  \definecolor{LightBlue2}{rgb}{0.698039,0.874510,0.933333}
  \definecolor{LightBlue3}{rgb}{0.603922,0.752941,0.803922}
  \definecolor{LightBlue4}{rgb}{0.407843,0.513726,0.545098}
  \definecolor{LightCyan1}{rgb}{0.878431,1.000000,1.000000}
  \definecolor{LightCyan2}{rgb}{0.819608,0.933333,0.933333}
  \definecolor{LightCyan3}{rgb}{0.705882,0.803922,0.803922}
  \definecolor{LightCyan4}{rgb}{0.478431,0.545098,0.545098}
  \definecolor{PaleTurquoise1}{rgb}{0.733333,1.000000,1.000000}
  \definecolor{PaleTurquoise2}{rgb}{0.682353,0.933333,0.933333}
  \definecolor{PaleTurquoise3}{rgb}{0.588235,0.803922,0.803922}
  \definecolor{PaleTurquoise4}{rgb}{0.400000,0.545098,0.545098}
  \definecolor{CadetBlue1}{rgb}{0.596078,0.960784,1.000000}
  \definecolor{CadetBlue2}{rgb}{0.556863,0.898039,0.933333}
  \definecolor{CadetBlue3}{rgb}{0.478431,0.772549,0.803922}
  \definecolor{CadetBlue4}{rgb}{0.325490,0.525490,0.545098}
  \definecolor{turquoise1}{rgb}{0.000000,0.960784,1.000000}
  \definecolor{turquoise2}{rgb}{0.000000,0.898039,0.933333}
  \definecolor{turquoise3}{rgb}{0.000000,0.772549,0.803922}
  \definecolor{turquoise4}{rgb}{0.000000,0.525490,0.545098}
  \definecolor{cyan1}{rgb}{0.000000,1.000000,1.000000}
  \definecolor{cyan2}{rgb}{0.000000,0.933333,0.933333}
  \definecolor{cyan3}{rgb}{0.000000,0.803922,0.803922}
  \definecolor{cyan4}{rgb}{0.000000,0.545098,0.545098}
  \definecolor{DarkSlateGray1}{rgb}{0.592157,1.000000,1.000000}
  \definecolor{DarkSlateGray2}{rgb}{0.552941,0.933333,0.933333}
  \definecolor{DarkSlateGray3}{rgb}{0.474510,0.803922,0.803922}
  \definecolor{DarkSlateGray4}{rgb}{0.321569,0.545098,0.545098}
  \definecolor{aquamarine1}{rgb}{0.498039,1.000000,0.831373}
  \definecolor{aquamarine2}{rgb}{0.462745,0.933333,0.776471}
  \definecolor{aquamarine3}{rgb}{0.400000,0.803922,0.666667}
  \definecolor{aquamarine4}{rgb}{0.270588,0.545098,0.454902}
  \definecolor{DarkSeaGreen1}{rgb}{0.756863,1.000000,0.756863}
  \definecolor{DarkSeaGreen2}{rgb}{0.705882,0.933333,0.705882}
  \definecolor{DarkSeaGreen3}{rgb}{0.607843,0.803922,0.607843}
  \definecolor{DarkSeaGreen4}{rgb}{0.411765,0.545098,0.411765}
  \definecolor{SeaGreen1}{rgb}{0.329412,1.000000,0.623529}
  \definecolor{SeaGreen2}{rgb}{0.305882,0.933333,0.580392}
  \definecolor{SeaGreen3}{rgb}{0.262745,0.803922,0.501961}
  \definecolor{SeaGreen4}{rgb}{0.180392,0.545098,0.341176}
  \definecolor{PaleGreen1}{rgb}{0.603922,1.000000,0.603922}
  \definecolor{PaleGreen2}{rgb}{0.564706,0.933333,0.564706}
  \definecolor{PaleGreen3}{rgb}{0.486275,0.803922,0.486275}
  \definecolor{PaleGreen4}{rgb}{0.329412,0.545098,0.329412}
  \definecolor{SpringGreen1}{rgb}{0.000000,1.000000,0.498039}
  \definecolor{SpringGreen2}{rgb}{0.000000,0.933333,0.462745}
  \definecolor{SpringGreen3}{rgb}{0.000000,0.803922,0.400000}
  \definecolor{SpringGreen4}{rgb}{0.000000,0.545098,0.270588}
  \definecolor{green1}{rgb}{0.000000,1.000000,0.000000}
  \definecolor{green2}{rgb}{0.000000,0.933333,0.000000}
  \definecolor{green3}{rgb}{0.000000,0.803922,0.000000}
  \definecolor{green4}{rgb}{0.000000,0.545098,0.000000}
  \definecolor{chartreuse1}{rgb}{0.498039,1.000000,0.000000}
  \definecolor{chartreuse2}{rgb}{0.462745,0.933333,0.000000}
  \definecolor{chartreuse3}{rgb}{0.400000,0.803922,0.000000}
  \definecolor{chartreuse4}{rgb}{0.270588,0.545098,0.000000}
  \definecolor{OliveDrab1}{rgb}{0.752941,1.000000,0.243137}
  \definecolor{OliveDrab2}{rgb}{0.701961,0.933333,0.227451}
  \definecolor{OliveDrab3}{rgb}{0.603922,0.803922,0.196078}
  \definecolor{OliveDrab4}{rgb}{0.411765,0.545098,0.133333}
  \definecolor{DarkOliveGreen1}{rgb}{0.792157,1.000000,0.439216}
  \definecolor{DarkOliveGreen2}{rgb}{0.737255,0.933333,0.407843}
  \definecolor{DarkOliveGreen3}{rgb}{0.635294,0.803922,0.352941}
  \definecolor{DarkOliveGreen4}{rgb}{0.431373,0.545098,0.239216}
  \definecolor{khaki1}{rgb}{1.000000,0.964706,0.560784}
  \definecolor{khaki2}{rgb}{0.933333,0.901961,0.521569}
  \definecolor{khaki3}{rgb}{0.803922,0.776471,0.450980}
  \definecolor{khaki4}{rgb}{0.545098,0.525490,0.305882}
  \definecolor{LightGoldenrod1}{rgb}{1.000000,0.925490,0.545098}
  \definecolor{LightGoldenrod2}{rgb}{0.933333,0.862745,0.509804}
  \definecolor{LightGoldenrod3}{rgb}{0.803922,0.745098,0.439216}
  \definecolor{LightGoldenrod4}{rgb}{0.545098,0.505882,0.298039}
  \definecolor{LightYellow1}{rgb}{1.000000,1.000000,0.878431}
  \definecolor{LightYellow2}{rgb}{0.933333,0.933333,0.819608}
  \definecolor{LightYellow3}{rgb}{0.803922,0.803922,0.705882}
  \definecolor{LightYellow4}{rgb}{0.545098,0.545098,0.478431}
  \definecolor{yellow1}{rgb}{1.000000,1.000000,0.000000}
  \definecolor{yellow2}{rgb}{0.933333,0.933333,0.000000}
  \definecolor{yellow3}{rgb}{0.803922,0.803922,0.000000}
  \definecolor{yellow4}{rgb}{0.545098,0.545098,0.000000}
  \definecolor{gold1}{rgb}{1.000000,0.843137,0.000000}
  \definecolor{gold2}{rgb}{0.933333,0.788235,0.000000}
  \definecolor{gold3}{rgb}{0.803922,0.678431,0.000000}
  \definecolor{gold4}{rgb}{0.545098,0.458824,0.000000}
  \definecolor{goldenrod1}{rgb}{1.000000,0.756863,0.145098}
  \definecolor{goldenrod2}{rgb}{0.933333,0.705882,0.133333}
  \definecolor{goldenrod3}{rgb}{0.803922,0.607843,0.113725}
  \definecolor{goldenrod4}{rgb}{0.545098,0.411765,0.078431}
  \definecolor{DarkGoldenrod1}{rgb}{1.000000,0.725490,0.058824}
  \definecolor{DarkGoldenrod2}{rgb}{0.933333,0.678431,0.054902}
  \definecolor{DarkGoldenrod3}{rgb}{0.803922,0.584314,0.047059}
  \definecolor{DarkGoldenrod4}{rgb}{0.545098,0.396078,0.031373}
  \definecolor{RosyBrown1}{rgb}{1.000000,0.756863,0.756863}
  \definecolor{RosyBrown2}{rgb}{0.933333,0.705882,0.705882}
  \definecolor{RosyBrown3}{rgb}{0.803922,0.607843,0.607843}
  \definecolor{RosyBrown4}{rgb}{0.545098,0.411765,0.411765}
  \definecolor{IndianRed1}{rgb}{1.000000,0.415686,0.415686}
  \definecolor{IndianRed2}{rgb}{0.933333,0.388235,0.388235}
  \definecolor{IndianRed3}{rgb}{0.803922,0.333333,0.333333}
  \definecolor{IndianRed4}{rgb}{0.545098,0.227451,0.227451}
  \definecolor{sienna1}{rgb}{1.000000,0.509804,0.278431}
  \definecolor{sienna2}{rgb}{0.933333,0.474510,0.258824}
  \definecolor{sienna3}{rgb}{0.803922,0.407843,0.223529}
  \definecolor{sienna4}{rgb}{0.545098,0.278431,0.149020}
  \definecolor{burlywood1}{rgb}{1.000000,0.827451,0.607843}
  \definecolor{burlywood2}{rgb}{0.933333,0.772549,0.568627}
  \definecolor{burlywood3}{rgb}{0.803922,0.666667,0.490196}
  \definecolor{burlywood4}{rgb}{0.545098,0.450980,0.333333}
  \definecolor{wheat1}{rgb}{1.000000,0.905882,0.729412}
  \definecolor{wheat2}{rgb}{0.933333,0.847059,0.682353}
  \definecolor{wheat3}{rgb}{0.803922,0.729412,0.588235}
  \definecolor{wheat4}{rgb}{0.545098,0.494118,0.400000}
  \definecolor{tan1}{rgb}{1.000000,0.647059,0.309804}
  \definecolor{tan2}{rgb}{0.933333,0.603922,0.286275}
  \definecolor{tan3}{rgb}{0.803922,0.521569,0.247059}
  \definecolor{tan4}{rgb}{0.545098,0.352941,0.168627}
  \definecolor{chocolate1}{rgb}{1.000000,0.498039,0.141176}
  \definecolor{chocolate2}{rgb}{0.933333,0.462745,0.129412}
  \definecolor{chocolate3}{rgb}{0.803922,0.400000,0.113725}
  \definecolor{chocolate4}{rgb}{0.545098,0.270588,0.074510}
  \definecolor{firebrick1}{rgb}{1.000000,0.188235,0.188235}
  \definecolor{firebrick2}{rgb}{0.933333,0.172549,0.172549}
  \definecolor{firebrick3}{rgb}{0.803922,0.149020,0.149020}
  \definecolor{firebrick4}{rgb}{0.545098,0.101961,0.101961}
  \definecolor{brown1}{rgb}{1.000000,0.250980,0.250980}
  \definecolor{brown2}{rgb}{0.933333,0.231373,0.231373}
  \definecolor{brown3}{rgb}{0.803922,0.200000,0.200000}
  \definecolor{brown4}{rgb}{0.545098,0.137255,0.137255}
  \definecolor{salmon1}{rgb}{1.000000,0.549020,0.411765}
  \definecolor{salmon2}{rgb}{0.933333,0.509804,0.384314}
  \definecolor{salmon3}{rgb}{0.803922,0.439216,0.329412}
  \definecolor{salmon4}{rgb}{0.545098,0.298039,0.223529}
  \definecolor{LightSalmon1}{rgb}{1.000000,0.627451,0.478431}
  \definecolor{LightSalmon2}{rgb}{0.933333,0.584314,0.447059}
  \definecolor{LightSalmon3}{rgb}{0.803922,0.505882,0.384314}
  \definecolor{LightSalmon4}{rgb}{0.545098,0.341176,0.258824}
  \definecolor{orange1}{rgb}{1.000000,0.647059,0.000000}
  \definecolor{orange2}{rgb}{0.933333,0.603922,0.000000}
  \definecolor{orange3}{rgb}{0.803922,0.521569,0.000000}
  \definecolor{orange4}{rgb}{0.545098,0.352941,0.000000}
  \definecolor{DarkOrange1}{rgb}{1.000000,0.498039,0.000000}
  \definecolor{DarkOrange2}{rgb}{0.933333,0.462745,0.000000}
  \definecolor{DarkOrange3}{rgb}{0.803922,0.400000,0.000000}
  \definecolor{DarkOrange4}{rgb}{0.545098,0.270588,0.000000}
  \definecolor{coral1}{rgb}{1.000000,0.447059,0.337255}
  \definecolor{coral2}{rgb}{0.933333,0.415686,0.313726}
  \definecolor{coral3}{rgb}{0.803922,0.356863,0.270588}
  \definecolor{coral4}{rgb}{0.545098,0.243137,0.184314}
  \definecolor{tomato1}{rgb}{1.000000,0.388235,0.278431}
  \definecolor{tomato2}{rgb}{0.933333,0.360784,0.258824}
  \definecolor{tomato3}{rgb}{0.803922,0.309804,0.223529}
  \definecolor{tomato4}{rgb}{0.545098,0.211765,0.149020}
  \definecolor{OrangeRed1}{rgb}{1.000000,0.270588,0.000000}
  \definecolor{OrangeRed2}{rgb}{0.933333,0.250980,0.000000}
  \definecolor{OrangeRed3}{rgb}{0.803922,0.215686,0.000000}
  \definecolor{OrangeRed4}{rgb}{0.545098,0.145098,0.000000}
  \definecolor{red1}{rgb}{1.000000,0.000000,0.000000}
  \definecolor{red2}{rgb}{0.933333,0.000000,0.000000}
  \definecolor{red3}{rgb}{0.803922,0.000000,0.000000}
  \definecolor{red4}{rgb}{0.545098,0.000000,0.000000}
  \definecolor{DeepPink1}{rgb}{1.000000,0.078431,0.576471}
  \definecolor{DeepPink2}{rgb}{0.933333,0.070588,0.537255}
  \definecolor{DeepPink3}{rgb}{0.803922,0.062745,0.462745}
  \definecolor{DeepPink4}{rgb}{0.545098,0.039216,0.313726}
  \definecolor{HotPink1}{rgb}{1.000000,0.431373,0.705882}
  \definecolor{HotPink2}{rgb}{0.933333,0.415686,0.654902}
  \definecolor{HotPink3}{rgb}{0.803922,0.376471,0.564706}
  \definecolor{HotPink4}{rgb}{0.545098,0.227451,0.384314}
  \definecolor{pink1}{rgb}{1.000000,0.709804,0.772549}
  \definecolor{pink2}{rgb}{0.933333,0.662745,0.721569}
  \definecolor{pink3}{rgb}{0.803922,0.568627,0.619608}
  \definecolor{pink4}{rgb}{0.545098,0.388235,0.423529}
  \definecolor{LightPink1}{rgb}{1.000000,0.682353,0.725490}
  \definecolor{LightPink2}{rgb}{0.933333,0.635294,0.678431}
  \definecolor{LightPink3}{rgb}{0.803922,0.549020,0.584314}
  \definecolor{LightPink4}{rgb}{0.545098,0.372549,0.396078}
  \definecolor{PaleVioletRed1}{rgb}{1.000000,0.509804,0.670588}
  \definecolor{PaleVioletRed2}{rgb}{0.933333,0.474510,0.623529}
  \definecolor{PaleVioletRed3}{rgb}{0.803922,0.407843,0.537255}
  \definecolor{PaleVioletRed4}{rgb}{0.545098,0.278431,0.364706}
  \definecolor{maroon1}{rgb}{1.000000,0.203922,0.701961}
  \definecolor{maroon2}{rgb}{0.933333,0.188235,0.654902}
  \definecolor{maroon3}{rgb}{0.803922,0.160784,0.564706}
  \definecolor{maroon4}{rgb}{0.545098,0.109804,0.384314}
  \definecolor{VioletRed1}{rgb}{1.000000,0.243137,0.588235}
  \definecolor{VioletRed2}{rgb}{0.933333,0.227451,0.549020}
  \definecolor{VioletRed3}{rgb}{0.803922,0.196078,0.470588}
  \definecolor{VioletRed4}{rgb}{0.545098,0.133333,0.321569}
  \definecolor{magenta1}{rgb}{1.000000,0.000000,1.000000}
  \definecolor{magenta2}{rgb}{0.933333,0.000000,0.933333}
  \definecolor{magenta3}{rgb}{0.803922,0.000000,0.803922}
  \definecolor{magenta4}{rgb}{0.545098,0.000000,0.545098}
  \definecolor{orchid1}{rgb}{1.000000,0.513726,0.980392}
  \definecolor{orchid2}{rgb}{0.933333,0.478431,0.913725}
  \definecolor{orchid3}{rgb}{0.803922,0.411765,0.788235}
  \definecolor{orchid4}{rgb}{0.545098,0.278431,0.537255}
  \definecolor{plum1}{rgb}{1.000000,0.733333,1.000000}
  \definecolor{plum2}{rgb}{0.933333,0.682353,0.933333}
  \definecolor{plum3}{rgb}{0.803922,0.588235,0.803922}
  \definecolor{plum4}{rgb}{0.545098,0.400000,0.545098}
  \definecolor{MediumOrchid1}{rgb}{0.878431,0.400000,1.000000}
  \definecolor{MediumOrchid2}{rgb}{0.819608,0.372549,0.933333}
  \definecolor{MediumOrchid3}{rgb}{0.705882,0.321569,0.803922}
  \definecolor{MediumOrchid4}{rgb}{0.478431,0.215686,0.545098}
  \definecolor{DarkOrchid1}{rgb}{0.749020,0.243137,1.000000}
  \definecolor{DarkOrchid2}{rgb}{0.698039,0.227451,0.933333}
  \definecolor{DarkOrchid3}{rgb}{0.603922,0.196078,0.803922}
  \definecolor{DarkOrchid4}{rgb}{0.407843,0.133333,0.545098}
  \definecolor{purple1}{rgb}{0.607843,0.188235,1.000000}
  \definecolor{purple2}{rgb}{0.568627,0.172549,0.933333}
  \definecolor{purple3}{rgb}{0.490196,0.149020,0.803922}
  \definecolor{purple4}{rgb}{0.333333,0.101961,0.545098}
  \definecolor{MediumPurple1}{rgb}{0.670588,0.509804,1.000000}
  \definecolor{MediumPurple2}{rgb}{0.623529,0.474510,0.933333}
  \definecolor{MediumPurple3}{rgb}{0.537255,0.407843,0.803922}
  \definecolor{MediumPurple4}{rgb}{0.364706,0.278431,0.545098}
  \definecolor{thistle1}{rgb}{1.000000,0.882353,1.000000}
  \definecolor{thistle2}{rgb}{0.933333,0.823529,0.933333}
  \definecolor{thistle3}{rgb}{0.803922,0.709804,0.803922}
  \definecolor{thistle4}{rgb}{0.545098,0.482353,0.545098}
  \definecolor{gray0}{rgb}{0.000000,0.000000,0.000000}
  \definecolor{grey0}{rgb}{0.000000,0.000000,0.000000}
  \definecolor{gray1}{rgb}{0.011765,0.011765,0.011765}
  \definecolor{grey1}{rgb}{0.011765,0.011765,0.011765}
  \definecolor{gray2}{rgb}{0.019608,0.019608,0.019608}
  \definecolor{grey2}{rgb}{0.019608,0.019608,0.019608}
  \definecolor{gray3}{rgb}{0.031373,0.031373,0.031373}
  \definecolor{grey3}{rgb}{0.031373,0.031373,0.031373}
  \definecolor{gray4}{rgb}{0.039216,0.039216,0.039216}
  \definecolor{grey4}{rgb}{0.039216,0.039216,0.039216}
  \definecolor{gray5}{rgb}{0.050980,0.050980,0.050980}
  \definecolor{grey5}{rgb}{0.050980,0.050980,0.050980}
  \definecolor{gray6}{rgb}{0.058824,0.058824,0.058824}
  \definecolor{grey6}{rgb}{0.058824,0.058824,0.058824}
  \definecolor{gray7}{rgb}{0.070588,0.070588,0.070588}
  \definecolor{grey7}{rgb}{0.070588,0.070588,0.070588}
  \definecolor{gray8}{rgb}{0.078431,0.078431,0.078431}
  \definecolor{grey8}{rgb}{0.078431,0.078431,0.078431}
  \definecolor{gray9}{rgb}{0.090196,0.090196,0.090196}
  \definecolor{grey9}{rgb}{0.090196,0.090196,0.090196}
  \definecolor{gray10}{rgb}{0.101961,0.101961,0.101961}
  \definecolor{grey10}{rgb}{0.101961,0.101961,0.101961}
  \definecolor{gray11}{rgb}{0.109804,0.109804,0.109804}
  \definecolor{grey11}{rgb}{0.109804,0.109804,0.109804}
  \definecolor{gray12}{rgb}{0.121569,0.121569,0.121569}
  \definecolor{grey12}{rgb}{0.121569,0.121569,0.121569}
  \definecolor{gray13}{rgb}{0.129412,0.129412,0.129412}
  \definecolor{grey13}{rgb}{0.129412,0.129412,0.129412}
  \definecolor{gray14}{rgb}{0.141176,0.141176,0.141176}
  \definecolor{grey14}{rgb}{0.141176,0.141176,0.141176}
  \definecolor{gray15}{rgb}{0.149020,0.149020,0.149020}
  \definecolor{grey15}{rgb}{0.149020,0.149020,0.149020}
  \definecolor{gray16}{rgb}{0.160784,0.160784,0.160784}
  \definecolor{grey16}{rgb}{0.160784,0.160784,0.160784}
  \definecolor{gray17}{rgb}{0.168627,0.168627,0.168627}
  \definecolor{grey17}{rgb}{0.168627,0.168627,0.168627}
  \definecolor{gray18}{rgb}{0.180392,0.180392,0.180392}
  \definecolor{grey18}{rgb}{0.180392,0.180392,0.180392}
  \definecolor{gray19}{rgb}{0.188235,0.188235,0.188235}
  \definecolor{grey19}{rgb}{0.188235,0.188235,0.188235}
  \definecolor{gray20}{rgb}{0.200000,0.200000,0.200000}
  \definecolor{grey20}{rgb}{0.200000,0.200000,0.200000}
  \definecolor{gray21}{rgb}{0.211765,0.211765,0.211765}
  \definecolor{grey21}{rgb}{0.211765,0.211765,0.211765}
  \definecolor{gray22}{rgb}{0.219608,0.219608,0.219608}
  \definecolor{grey22}{rgb}{0.219608,0.219608,0.219608}
  \definecolor{gray23}{rgb}{0.231373,0.231373,0.231373}
  \definecolor{grey23}{rgb}{0.231373,0.231373,0.231373}
  \definecolor{gray24}{rgb}{0.239216,0.239216,0.239216}
  \definecolor{grey24}{rgb}{0.239216,0.239216,0.239216}
  \definecolor{gray25}{rgb}{0.250980,0.250980,0.250980}
  \definecolor{grey25}{rgb}{0.250980,0.250980,0.250980}
  \definecolor{gray26}{rgb}{0.258824,0.258824,0.258824}
  \definecolor{grey26}{rgb}{0.258824,0.258824,0.258824}
  \definecolor{gray27}{rgb}{0.270588,0.270588,0.270588}
  \definecolor{grey27}{rgb}{0.270588,0.270588,0.270588}
  \definecolor{gray28}{rgb}{0.278431,0.278431,0.278431}
  \definecolor{grey28}{rgb}{0.278431,0.278431,0.278431}
  \definecolor{gray29}{rgb}{0.290196,0.290196,0.290196}
  \definecolor{grey29}{rgb}{0.290196,0.290196,0.290196}
  \definecolor{gray30}{rgb}{0.301961,0.301961,0.301961}
  \definecolor{grey30}{rgb}{0.301961,0.301961,0.301961}
  \definecolor{gray31}{rgb}{0.309804,0.309804,0.309804}
  \definecolor{grey31}{rgb}{0.309804,0.309804,0.309804}
  \definecolor{gray32}{rgb}{0.321569,0.321569,0.321569}
  \definecolor{grey32}{rgb}{0.321569,0.321569,0.321569}
  \definecolor{gray33}{rgb}{0.329412,0.329412,0.329412}
  \definecolor{grey33}{rgb}{0.329412,0.329412,0.329412}
  \definecolor{gray34}{rgb}{0.341176,0.341176,0.341176}
  \definecolor{grey34}{rgb}{0.341176,0.341176,0.341176}
  \definecolor{gray35}{rgb}{0.349020,0.349020,0.349020}
  \definecolor{grey35}{rgb}{0.349020,0.349020,0.349020}
  \definecolor{gray36}{rgb}{0.360784,0.360784,0.360784}
  \definecolor{grey36}{rgb}{0.360784,0.360784,0.360784}
  \definecolor{gray37}{rgb}{0.368627,0.368627,0.368627}
  \definecolor{grey37}{rgb}{0.368627,0.368627,0.368627}
  \definecolor{gray38}{rgb}{0.380392,0.380392,0.380392}
  \definecolor{grey38}{rgb}{0.380392,0.380392,0.380392}
  \definecolor{gray39}{rgb}{0.388235,0.388235,0.388235}
  \definecolor{grey39}{rgb}{0.388235,0.388235,0.388235}
  \definecolor{gray40}{rgb}{0.400000,0.400000,0.400000}
  \definecolor{grey40}{rgb}{0.400000,0.400000,0.400000}
  \definecolor{gray41}{rgb}{0.411765,0.411765,0.411765}
  \definecolor{grey41}{rgb}{0.411765,0.411765,0.411765}
  \definecolor{gray42}{rgb}{0.419608,0.419608,0.419608}
  \definecolor{grey42}{rgb}{0.419608,0.419608,0.419608}
  \definecolor{gray43}{rgb}{0.431373,0.431373,0.431373}
  \definecolor{grey43}{rgb}{0.431373,0.431373,0.431373}
  \definecolor{gray44}{rgb}{0.439216,0.439216,0.439216}
  \definecolor{grey44}{rgb}{0.439216,0.439216,0.439216}
  \definecolor{gray45}{rgb}{0.450980,0.450980,0.450980}
  \definecolor{grey45}{rgb}{0.450980,0.450980,0.450980}
  \definecolor{gray46}{rgb}{0.458824,0.458824,0.458824}
  \definecolor{grey46}{rgb}{0.458824,0.458824,0.458824}
  \definecolor{gray47}{rgb}{0.470588,0.470588,0.470588}
  \definecolor{grey47}{rgb}{0.470588,0.470588,0.470588}
  \definecolor{gray48}{rgb}{0.478431,0.478431,0.478431}
  \definecolor{grey48}{rgb}{0.478431,0.478431,0.478431}
  \definecolor{gray49}{rgb}{0.490196,0.490196,0.490196}
  \definecolor{grey49}{rgb}{0.490196,0.490196,0.490196}
  \definecolor{gray50}{rgb}{0.498039,0.498039,0.498039}
  \definecolor{grey50}{rgb}{0.498039,0.498039,0.498039}
  \definecolor{gray51}{rgb}{0.509804,0.509804,0.509804}
  \definecolor{grey51}{rgb}{0.509804,0.509804,0.509804}
  \definecolor{gray52}{rgb}{0.521569,0.521569,0.521569}
  \definecolor{grey52}{rgb}{0.521569,0.521569,0.521569}
  \definecolor{gray53}{rgb}{0.529412,0.529412,0.529412}
  \definecolor{grey53}{rgb}{0.529412,0.529412,0.529412}
  \definecolor{gray54}{rgb}{0.541176,0.541176,0.541176}
  \definecolor{grey54}{rgb}{0.541176,0.541176,0.541176}
  \definecolor{gray55}{rgb}{0.549020,0.549020,0.549020}
  \definecolor{grey55}{rgb}{0.549020,0.549020,0.549020}
  \definecolor{gray56}{rgb}{0.560784,0.560784,0.560784}
  \definecolor{grey56}{rgb}{0.560784,0.560784,0.560784}
  \definecolor{gray57}{rgb}{0.568627,0.568627,0.568627}
  \definecolor{grey57}{rgb}{0.568627,0.568627,0.568627}
  \definecolor{gray58}{rgb}{0.580392,0.580392,0.580392}
  \definecolor{grey58}{rgb}{0.580392,0.580392,0.580392}
  \definecolor{gray59}{rgb}{0.588235,0.588235,0.588235}
  \definecolor{grey59}{rgb}{0.588235,0.588235,0.588235}
  \definecolor{gray60}{rgb}{0.600000,0.600000,0.600000}
  \definecolor{grey60}{rgb}{0.600000,0.600000,0.600000}
  \definecolor{gray61}{rgb}{0.611765,0.611765,0.611765}
  \definecolor{grey61}{rgb}{0.611765,0.611765,0.611765}
  \definecolor{gray62}{rgb}{0.619608,0.619608,0.619608}
  \definecolor{grey62}{rgb}{0.619608,0.619608,0.619608}
  \definecolor{gray63}{rgb}{0.631373,0.631373,0.631373}
  \definecolor{grey63}{rgb}{0.631373,0.631373,0.631373}
  \definecolor{gray64}{rgb}{0.639216,0.639216,0.639216}
  \definecolor{grey64}{rgb}{0.639216,0.639216,0.639216}
  \definecolor{gray65}{rgb}{0.650980,0.650980,0.650980}
  \definecolor{grey65}{rgb}{0.650980,0.650980,0.650980}
  \definecolor{gray66}{rgb}{0.658824,0.658824,0.658824}
  \definecolor{grey66}{rgb}{0.658824,0.658824,0.658824}
  \definecolor{gray67}{rgb}{0.670588,0.670588,0.670588}
  \definecolor{grey67}{rgb}{0.670588,0.670588,0.670588}
  \definecolor{gray68}{rgb}{0.678431,0.678431,0.678431}
  \definecolor{grey68}{rgb}{0.678431,0.678431,0.678431}
  \definecolor{gray69}{rgb}{0.690196,0.690196,0.690196}
  \definecolor{grey69}{rgb}{0.690196,0.690196,0.690196}
  \definecolor{gray70}{rgb}{0.701961,0.701961,0.701961}
  \definecolor{grey70}{rgb}{0.701961,0.701961,0.701961}
  \definecolor{gray71}{rgb}{0.709804,0.709804,0.709804}
  \definecolor{grey71}{rgb}{0.709804,0.709804,0.709804}
  \definecolor{gray72}{rgb}{0.721569,0.721569,0.721569}
  \definecolor{grey72}{rgb}{0.721569,0.721569,0.721569}
  \definecolor{gray73}{rgb}{0.729412,0.729412,0.729412}
  \definecolor{grey73}{rgb}{0.729412,0.729412,0.729412}
  \definecolor{gray74}{rgb}{0.741176,0.741176,0.741176}
  \definecolor{grey74}{rgb}{0.741176,0.741176,0.741176}
  \definecolor{gray75}{rgb}{0.749020,0.749020,0.749020}
  \definecolor{grey75}{rgb}{0.749020,0.749020,0.749020}
  \definecolor{gray76}{rgb}{0.760784,0.760784,0.760784}
  \definecolor{grey76}{rgb}{0.760784,0.760784,0.760784}
  \definecolor{gray77}{rgb}{0.768627,0.768627,0.768627}
  \definecolor{grey77}{rgb}{0.768627,0.768627,0.768627}
  \definecolor{gray78}{rgb}{0.780392,0.780392,0.780392}
  \definecolor{grey78}{rgb}{0.780392,0.780392,0.780392}
  \definecolor{gray79}{rgb}{0.788235,0.788235,0.788235}
  \definecolor{grey79}{rgb}{0.788235,0.788235,0.788235}
  \definecolor{gray80}{rgb}{0.800000,0.800000,0.800000}
  \definecolor{grey80}{rgb}{0.800000,0.800000,0.800000}
  \definecolor{gray81}{rgb}{0.811765,0.811765,0.811765}
  \definecolor{grey81}{rgb}{0.811765,0.811765,0.811765}
  \definecolor{gray82}{rgb}{0.819608,0.819608,0.819608}
  \definecolor{grey82}{rgb}{0.819608,0.819608,0.819608}
  \definecolor{gray83}{rgb}{0.831373,0.831373,0.831373}
  \definecolor{grey83}{rgb}{0.831373,0.831373,0.831373}
  \definecolor{gray84}{rgb}{0.839216,0.839216,0.839216}
  \definecolor{grey84}{rgb}{0.839216,0.839216,0.839216}
  \definecolor{gray85}{rgb}{0.850980,0.850980,0.850980}
  \definecolor{grey85}{rgb}{0.850980,0.850980,0.850980}
  \definecolor{gray86}{rgb}{0.858824,0.858824,0.858824}
  \definecolor{grey86}{rgb}{0.858824,0.858824,0.858824}
  \definecolor{gray87}{rgb}{0.870588,0.870588,0.870588}
  \definecolor{grey87}{rgb}{0.870588,0.870588,0.870588}
  \definecolor{gray88}{rgb}{0.878431,0.878431,0.878431}
  \definecolor{grey88}{rgb}{0.878431,0.878431,0.878431}
  \definecolor{gray89}{rgb}{0.890196,0.890196,0.890196}
  \definecolor{grey89}{rgb}{0.890196,0.890196,0.890196}
  \definecolor{gray90}{rgb}{0.898039,0.898039,0.898039}
  \definecolor{grey90}{rgb}{0.898039,0.898039,0.898039}
  \definecolor{gray91}{rgb}{0.909804,0.909804,0.909804}
  \definecolor{grey91}{rgb}{0.909804,0.909804,0.909804}
  \definecolor{gray92}{rgb}{0.921569,0.921569,0.921569}
  \definecolor{grey92}{rgb}{0.921569,0.921569,0.921569}
  \definecolor{gray93}{rgb}{0.929412,0.929412,0.929412}
  \definecolor{grey93}{rgb}{0.929412,0.929412,0.929412}
  \definecolor{gray94}{rgb}{0.941176,0.941176,0.941176}
  \definecolor{grey94}{rgb}{0.941176,0.941176,0.941176}
  \definecolor{gray95}{rgb}{0.949020,0.949020,0.949020}
  \definecolor{grey95}{rgb}{0.949020,0.949020,0.949020}
  \definecolor{gray96}{rgb}{0.960784,0.960784,0.960784}
  \definecolor{grey96}{rgb}{0.960784,0.960784,0.960784}
  \definecolor{gray97}{rgb}{0.968627,0.968627,0.968627}
  \definecolor{grey97}{rgb}{0.968627,0.968627,0.968627}
  \definecolor{gray98}{rgb}{0.980392,0.980392,0.980392}
  \definecolor{grey98}{rgb}{0.980392,0.980392,0.980392}
  \definecolor{gray99}{rgb}{0.988235,0.988235,0.988235}
  \definecolor{grey99}{rgb}{0.988235,0.988235,0.988235}
  \definecolor{gray100}{rgb}{1.000000,1.000000,1.000000}
  \definecolor{grey100}{rgb}{1.000000,1.000000,1.000000}
  \definecolor{dark grey}{rgb}{0.662745,0.662745,0.662745}
  \definecolor{DarkGrey}{rgb}{0.662745,0.662745,0.662745}
  \definecolor{dark gray}{rgb}{0.662745,0.662745,0.662745}
  \definecolor{DarkGray}{rgb}{0.662745,0.662745,0.662745}
  \definecolor{dark blue}{rgb}{0.000000,0.000000,0.545098}
  \definecolor{DarkBlue}{rgb}{0.000000,0.000000,0.545098}
  \definecolor{dark cyan}{rgb}{0.000000,0.545098,0.545098}
  \definecolor{DarkCyan}{rgb}{0.000000,0.545098,0.545098}
  \definecolor{dark magenta}{rgb}{0.545098,0.000000,0.545098}
  \definecolor{DarkMagenta}{rgb}{0.545098,0.000000,0.545098}
  \definecolor{dark red}{rgb}{0.545098,0.000000,0.000000}
  \definecolor{DarkRed}{rgb}{0.545098,0.000000,0.000000}
  \definecolor{light green}{rgb}{0.564706,0.933333,0.564706}
  \definecolor{LightGreen}{rgb}{0.564706,0.933333,0.564706}